\documentclass[aps,pre,amsmath,amssymb]{revtex4-1}
 
\usepackage{graphicx}
\usepackage{bm}
\usepackage[usenames, dvipsnames]{color}

\usepackage[utf8]{inputenc}
\usepackage[T1]{fontenc}
\usepackage{mathptmx}
\usepackage{hyperref}
\usepackage{subcaption}
\usepackage{caption,setspace}
\usepackage{multirow}

\usepackage{booktabs}

\newcommand{\angstrom}{\text{\normalfont\AA}}

\renewcommand{\thefigure}{ Fig. \arabic{figure}}

\usepackage[toc,title,page]{appendix}

\begin{document}

\title{Automated characterization of spatial and dynamical heterogeneity in supercooled liquids via implementation of Machine Learning}

\author{Viet Nguyen}
\author{Xueyu Song}
\email{xsong@iastate.edu}
\affiliation{Ames Laboratory and Department of Chemistry, Iowa State University, Ames, IA, USA}
\date{\today}

\begin{abstract}
A computational approach by an implementation of the Principle Component Analysis (PCA) with K-means and Gaussian Mixture (GM) clustering methods from Machine Learning (ML) algorithms to identify structural and dynamical heterogeneities of supercooled liquids is developed. In this method, a collection of the average weighted coordination numbers ($\overline{WCNs}$) of particles calculated from particles' positions are used as an order parameter to build a low-dimensional representation of feature (structural) space for K-means clustering to sort the particles in the system into few meso-states using PCA. Nano-domains or aggregated clusters are also formed in configurational (real) space from a direct mapping using associated meso-states' particle identities with some misclassified interfacial particles. These classification uncertainties can be improved by a co-learning strategy which utilizes the probabilistic GM clustering and the information transfer between the structural space and configurational space iteratively until convergence. A final classification of meso-states in structural space and domains in configurational space are stable over long times and measured to have dynamical heterogeneities. Armed with such a classification protocol, various studies over the thermodynamic and dynamical properties of these domains indicate that the observed heterogeneity is the result of liquid-liquid phase separation after quenching to a supercooled state.

\end{abstract}
\maketitle

\section{Introduction}

Glass plays a central role in nature and our daily lives. It is essential in food processing, preservation of wildlife animals under extreme cold ~\cite{Crowe}. Ordinary window glass, mostly made of sand (SiO$_2$), lime (CaCO$_3$) and soda (Na$_2$CO$_3$) is a best known manufactured amorphous solid product ~\cite{Debenedetti}.  Optical wave guides use pure amorphous silica while silicon in photovoltaic cell is amorphous. In principle, glassy state is attained by supercooling a liquid below its melting temperature fast enough to avoid crystallization. Under such rapid cooling, the supercooled liquid attains mesoscopic structural disorder with "complex dynamics" such as non-exponential relaxation, breakdown of Stokes-Einstein relation. Although these heterogeneities are well-known for decades ~\cite{Sastry,Andersen,Walter_Kob,WGotze,Cubuk,STILLINGER,Smessaert,Candelier,Kawasaki,Yang}, there is no direct evidence to consistently classify  and correlate these heterogeneities both structurally and dynamically. These following questions remain a puzzle: What cause these heterogeneities to arise? What is the spatial order of magnitude of the domains?  How much do dynamics vary among these domains? Answers to those questions could significantly impact our practical applications of glass-forming materials. 

Observation of heterogeneous dynamics is directly linked to the onset of cage effect~\cite{B.Doliwa} where particles become trapped in local cages by their neighboring particles to prevent them from moving around as a normal liquid. The cage effect is manifested as a plateau in the self intermediate scattering function $F(k,t)$ or the mean squared displacement of particles and could be explained as following: If we take an instant snapshot of the system, we see no impressive structure change close to $T_{g}$. Let's consider two different snapshots taken at two instants of time separated by a time interval $\it t$. We can now capture how particles move during this interval $\it t$. If the interval $\it t$ is too short, the system is still in ballistic regime, there is not a significant variations of particles mobility because interaction has not kicked in to make things interesting. Meanwhile, if  $\it t$ is too long, larger than the relaxation time $\tau_{r}$ (the longest relaxation process), time average is equivalent to ensemble average, hence all particles are statistically the same and each particle will have the same mobility.  $\it t$ is selected such that it is long enough to capture particles interaction but short enough to avoid statistical homogeneity to observe the difference of high or low mobility of particles. Hence, such intermediate time $\it t$ value is closely related to the plateau of the $\beta$-relaxation regime where particles become transiently trapped in cages and $F(k,t)$ remains constant. Only at sufficiently long times will particles break free and full relaxation takes place ($\alpha$-relaxation). Particles mobility can vary several orders of magnitude~\cite{GLOTZER2000342,SILLESCU199981,Ediger}. In addition, particles with one mobility tends to form a cluster or a domain such that the system are filled with different domains of particles. In other words, particles move in cooperatively manner  as a dynamically correlated mesoscopic domains with long relexation time scales.~\cite{C.PatrickRoyall,AndreaCavagna,ClaudioDonati,Adam_Gibbs_1965,E.VidalRussell,Adam_Gibbs_1965}. A dynamical length-scale $\xi$ can be associated with the increasing dynamic heterogeneities because it measures the size of mesoscopic domains as equivalently to size of growing cooperative motion of particles~\cite{Berthier1,Kirkpatrick,ViotPascal,Garrahan,Hurley,Bennemann,DONATI2002215,Whitelam,Berthier:2004vo}.

Several theories of glass transition have been developed to seek a fundamental understanding of these spatial domains: such as the energy landscape picture~\cite{Goldstein_1969,Berthier}, Adam-Gibbs theory~\cite{Adam_Gibbs_1965,GibbsJulianH,Bouchaud20047347}, and random first-order transition theories (RFOT)~\cite{Kirkpatrick}, to name a few. These theories present  various pictures of domains thermodynamically. Although these thermodynamic descriptions provide a simple and intuitive framework related to dynamics and spatial structures of supercooled liquids, it lacks a consistent classification protocol to characterize the structure of these mesoscopic domains. The lack of a clear characterization of these domain structures in supercooled liquids has hindered the formulation of a general theory for glass transition. Unlike crystalline solids whose structures can be easily detected due to its periodicity, no general classification scheme  has been formulated for supercooled liquids to the best of our knowledge. 

Meanwhile, several classification schemes are developed to identify structures in amorphous systems. The first kind of approaches include Voronoi polyhedra~\cite{BERNAL:1959vs,BERNAL:1960td,Finney1970,Anikeenko:2007uy,Anikeenko:2008vp}, bond-orientational order parameters~\cite{Steinhardt:1983uh,Lechner}, the common-neighbour analysis~\cite{TSUZUKI2007518,FAKEN1994279,Honeycutt} and topological cluster classification~\cite{Williams2007TopologicalCO,Malins} which are based on identification of a bond network among particles. However, these methods require some specific structural information $\it a$ $\it priori$ which is unknown in general except for few systems under certain situations. Other general ‘‘order-agnostic’’  approaches ~\cite{C.PatrickRoyall,Royall_2015} which rely not on a specific structure but on some general properties have been developed. One of them using mutual information based on Shannon entropy ~\cite{Shannon}, to determine structural length-scale. Structure in one part of the system can influence structure in another via mutual information, hence mutual information between two regions can be computed as a function of distance~\cite{Dunleavy:2012vi}, which does not require $\it a$ $\it priori$ knowledge of the structure. Another method is to seek networks among domains. Each domain is considered as a non-interacting isolated community. By minimizing the length-scale of these communities, it minimizes the interaction among communities~\cite{Ronhovde:2012tx}, hence leads to identification of clusters which are not specified beforehand. Another type of methods is to introduce an external, static perturbation in the form of an affine deformation of coordinate data. A drawback of these approaches is that the nature of structures identified is not as clear as the first kind of approaches because it lacks microscopic details of particles like bond network and coordination number. 

Given the significance of structural classification in supercooled liquids, we developed a new strategy to classify a supercooled liquid into nano-domains using some algorithms from machine learning (ML) such as the Principle Component Analysis (PCA), K-means and Gaussian Mixture (GM) clustering~\cite{G.James,F.Noe-pyemma,BishopChristopherM2006Pram,Murphy} both in structural and configurational spaces. This classification protocol shows improvement over discussed methods in previous paragraph because it is similar to ‘‘order-agnostic’’  where the emergence of domains requires no prior knowledge in one hand and at the same time contains information of microscopic details as the first kind of approaches(Voronoi polyhedra, bond-orientational order parameters, etc). 

The nano-domains from our approach agree with the picture in the Adam-Gibbs and RFOT theories. Based upon our classification, nature of these spatially distinct domains are clearly characterized and each of these domains is correlated with different diffusion constant distributions within a domain, hence the spatially heterogeneous dynamics naturally falls into two categories: the diffusion within various domains and the domain rearrangement dynamics which reflect the slow relaxation of the system. Structural evolution of these nano-domains is identified as coarsening kinetics from the liquid-liquid phase separation after rapid cooling or quenching.  Furthermore,  temperature dependence and other properties of nano-domains are also studied to support this picture. A well studied binary Lennard-Jones model system, the Kob-Andersen model~\cite{kob_andersen_PhysRevE.51.4626,WalterKob-scaling,Thomas}, is used  to to demonstrate the capability of our classification scheme for supercooled liquids since it is known that the model system does not crystallize when it is supercooled well below the melting temperature.

The paper is organized as follows. Section \ref{sec:methods} presents a detailed presentation of the proposed method. This is followed by an extensive result presentation with discussions. Some concluding remarks are given in the final section.

\section{Classification Scheme} \label{sec:methods}

\subsection{Simulation Details}

In this work, simulations are done with $NPT$ ensemble (where $N$ is the number of particles, $P$ is pressure and $T$ is temperature) using the molecular dynamics (MD) simulation package, LAMMPS ~\cite{LAMMPS}. Noose-Hoover thermostats are employed to control both external pressure (pressure is set to 0) and temperature. The atomic interaction potential used in our work is the well-known Kob-Andersen binary Lennard-Jones (LJ) model ~\cite{kob_andersen_PhysRevE.51.4626,WalterKob-scaling,Thomas}. The standard form of the LJ potential can be expressed as :

\begin{equation}
V(r)= 
\begin{cases}
 4\epsilon_{A,B} \left[ \left( \frac{\sigma_{A,B}}{r} \right)^{12} - \left( \frac{\sigma_{A,B}}{r} \right)^6 \right] & \text{for } (r\leq r_c)\\
 0&  \text{for }  (r > r_c),
\end{cases}
\end{equation} where the parameter $\epsilon$ is the potential well depth, $\sigma$ is the characteristic atomic diameter and the cutting distance  $r_c$ is set to $2.5\sigma_{A,B}$. The parameters for solid Ar are adopted~\cite{Vega,X.Bai-1.2184315} : $\sigma = 0.3405{\angstrom}$ and $\frac{\epsilon}{k_B}=119.8K$ where $k_B$ is the Boltzmann constant and particle mass m = 6.69 x $10^{-26}$kg. The conventional reduced unit for LJ system is used: the mass unit is set to the weight of one Ar atom while the length unit in $\sigma$, energy unit in $\epsilon$ , the time unit in term of $\tau=t\sqrt{m\sigma^2\over\epsilon}$ and reduced temperature is defined by $T^* = T(\frac{\epsilon}{k_B})$. The system consists of 80$\%$ of A and 20$\%$ of B particles with $\epsilon_{AA}=1$, $\sigma_{AA}=1$, $\epsilon_{AB}=1$ , $\sigma_{AB}=0.8$, $\epsilon_{BB}= 0.5$ and  $\sigma_{BB}=0.88$ while $m_A = m_B = 1$. Periodic boundary conditions are applied to all directions. The time step is set to $0.005\tau$ which is about 10 femtoseconds. Number of particles of the systems studied are 5000, 16000 and 50000. 
To prepare the liquid at supercooled conditions, we first heated up the system to a high temperature to obtain a liquid state. After a short period of equilibration and relaxation, the system is quenched to three different target temperatures  $T^*$ = $\{ 0.37, 0.3, 0.2 \}$. The cooling process has been done by linearly decreasing temperature via re-scaling atomic velocity: $T = T_0 -\gamma n$, where $\gamma$  is the cooling rate (3.3 x $10^{10}$ K/s if taking Ar parameters) and $n$ is the number of MD steps. These temperatures are reasonably selected because: $T^*$ = $\{ 0.37, 0.3 \}$ are below the mode-coupling temperature $T^*_c \approx 0.435$ predicted by mode-coupling theory~\cite{Walter_Kob,Janssen} but above glass transition temperature ($T^*_{g}=0.25$)~\cite{Andersen} to observe any change of dynamics~\cite{Schroder} while $T^*$ = 0.2 is below the $T^*_{g}$ to study the trend of structural heterogeneity for temperature dependence.  After two million time steps equilibration, the system is run for another 3 million time steps, saving configurations every 100 steps or $0.05\tau$. The average number density $\rho^*$ = $\{1.14, 1.17, 1.19\}$.

\subsection{Radial Distribution Function (rdf) and Weighted Coordination Numbers (WCNs)} \label{sec:rdf}

\begin{figure}
\resizebox{\columnwidth}{!}
{
\begin{subfigure}{0.35\textwidth}
\includegraphics[width=2.1in]{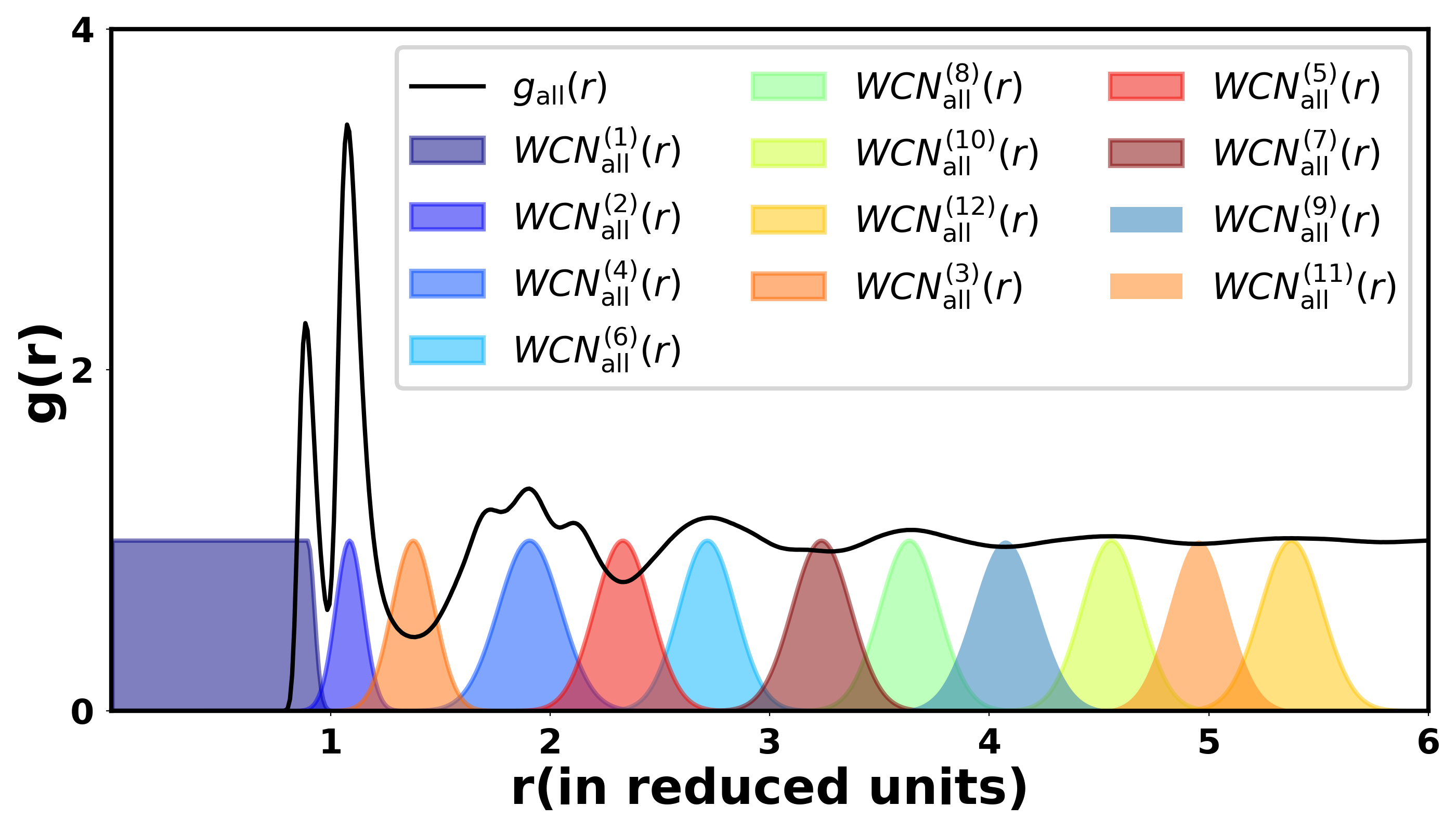}
\caption{WCNs based on rdf}
   \label{fig:wcn}
\end{subfigure}
\hfill
\begin{subfigure}{0.35\textwidth}
\includegraphics[width=2.1in]{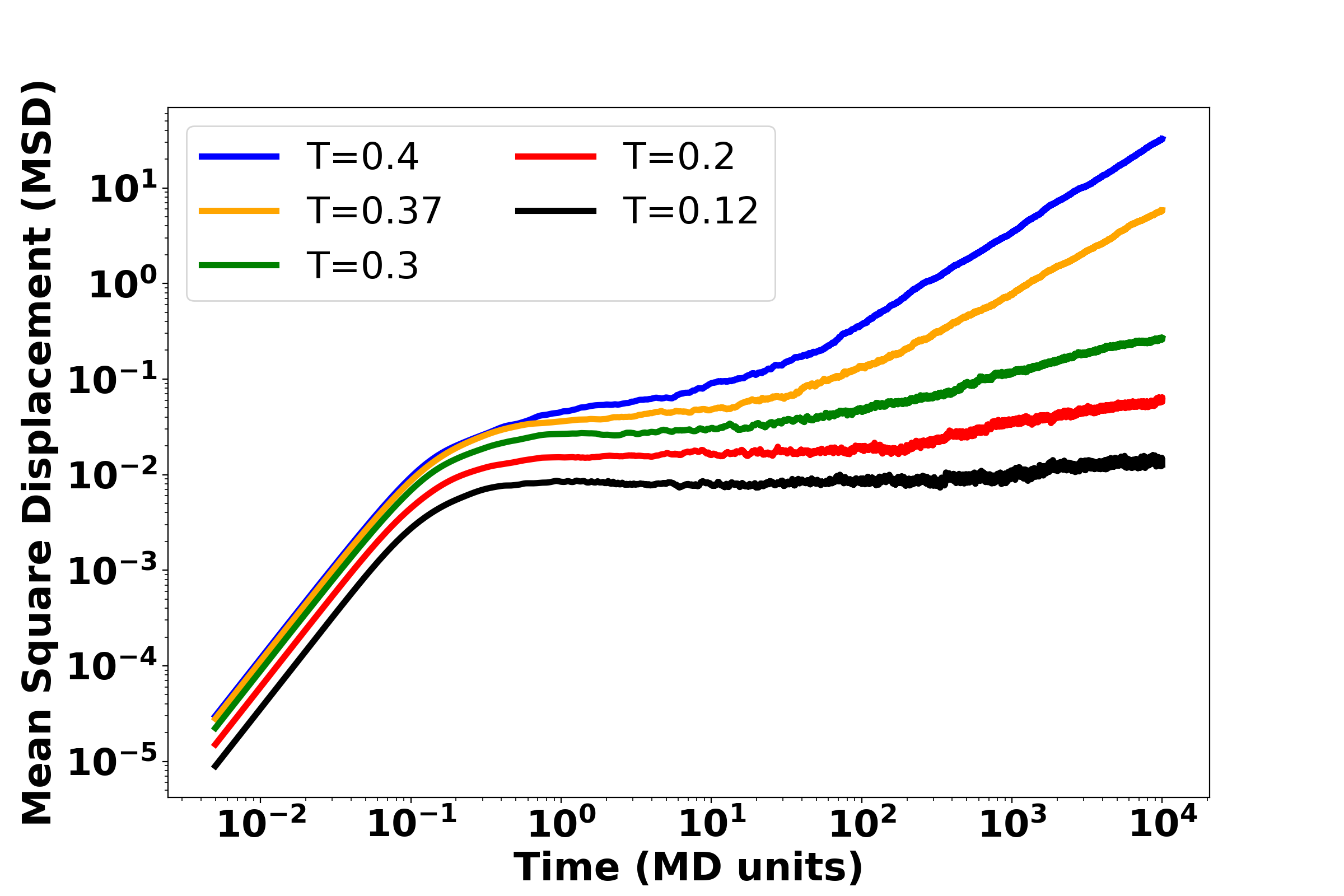}
\caption{MSD trajectories}
   \label{fig:msd}
\end{subfigure}
}
\caption{ (a) Radial distribution function (rdf) employed by the weighted Gaussian functions for  the Kob-Andersen binary LJ system at $T^*=0.2$, regardless of particles identities. The color and superscript indicate the WCN's positions and numbers along the rdf. (b) All-particle MSDs as a function of time at various temperatures regardless of their identities}  
\end{figure}

To investigate structural heterogeneities of a disordered system, radial distribution function {\it g(r)} is commonly employed to describe spatial local environments by means of collecting averaged coordination numbers (CNs) which describe the relative number of neighboring particles in a particular surrounding spherical shell of a particle, which is the same for all particles. However,  this highly averaged CNs representation of the system lacks the details to provide realistic features of the spatial heterogeneity of a supercooled liquid. Meanwhile, for a particular configuration of the system either by a snapshot from a molecular simulation or an experimental image of supercooled colloidal system from confocal microscopy, local structures for each of an $\textbf M$ particles system can be characterized with its local coordination shell structure. Naturally, a middle ground is an order parameter that can classify these local structures of the system into a few meso-states which is useful to describe the heterogenous structure of the system. In addition, aggregated clusters or domains, whose particles from the same meso-states should be formed in the configurational space, together tile up the whole system to make classification scheme work both structurally and configurationally. Furthermore,  meso-states in the structural space and domains in the configurational space should live long enough to afford further analysis. For example these meso-states and domains can directly relate to the onset of caging effect which is attributed to plateau region of mean-square displacement trajectories in \ref{fig:msd}). In this study, the timescale for this analysis is from 5x$10^1$ to 2x$10^4$ MD units or converted to 0.1 to 40ns which associates with the plateau region at different temperatures.

Using molecular dynamics simulations of this model system, the CN of a particle can be calculated. In this study, the A/B identity of the particles is ignored, which can be thought as the supercooled liquid state being generated from an effective one-component system. 
However, CN-based features suffer a strict cut-off value to determine whether a neighbor particle is counted as in or out of the shell. To avoid this hard assignment, weighted coordination numbers (WCNs)~\cite{rudzinski_1.5064808}, which utilize the normalized Gaussian distribution based on the shell structure of the system {\it g(r)}(\ref{fig:wcn}) to weight the contribution of each neighboring particle based on the particle's distance to the center one. Using the relevant solvation shell features as  identified maxima and minima along the radial distribution function, the normalized Gaussian distribution functions are placed at the center of these shell features as shown in \ref{fig:wcn}. The width of the Gaussian functions depends on the area that the solvation feature covers and neighboring Gaussians such that the value of the intersection is assigned to 0 or roughly 0.25 depending on whether or not the two shells are largely overlapping. However, width and the size of the overlapping areas of the Gaussians do not change the consistency of the final results. The WCNs smooth out transitions between solvation shells by counting the particles sitting at the center of the features as one while the one further away from the center feature is counted as a fraction based on the Gaussian distribution function. For each configuration, employing this WCN implementation, each component of a particle's WCNs vector is determined by summing the weight from all surrounding particles within that shell and the dimension of the WCN vector is determined by the number of shells reasonably covering the main features of the {\it g(r)}, $N=12$ in \ref{fig:wcn}; other numbers of shells tested yield consistent results.

For a single configuration of the simulation, WCNs of all particles are collected from the particles' coordination numbers smoothed using the {\it g(r)}, hence the features data for the entire system is represented by a matrix ${\widetilde{\bf X}}$ of {\textbf {M{\rm x}N}} which is obtained from {\textbf N} WCNs for each of {\textbf M} particles system. WCNs are noisy and complicated in a disordered system, hence require a further step to remove some of these noises. Instead of WCNs, averaged WCNs is used which has a form: $\overline{WCN}_i=\frac{1}{N_b}{\sum_{j}^{N_b}WCN_j}$, where $N_b$ is the number of neighboring particles in each shell plus the particle $i$ itself.

\subsection{Dimensionality reduction and clustering }

\begin{figure}
\resizebox{\columnwidth}{!}
{
\begin{subfigure}{0.35\textwidth}
\includegraphics[width=2.3in]{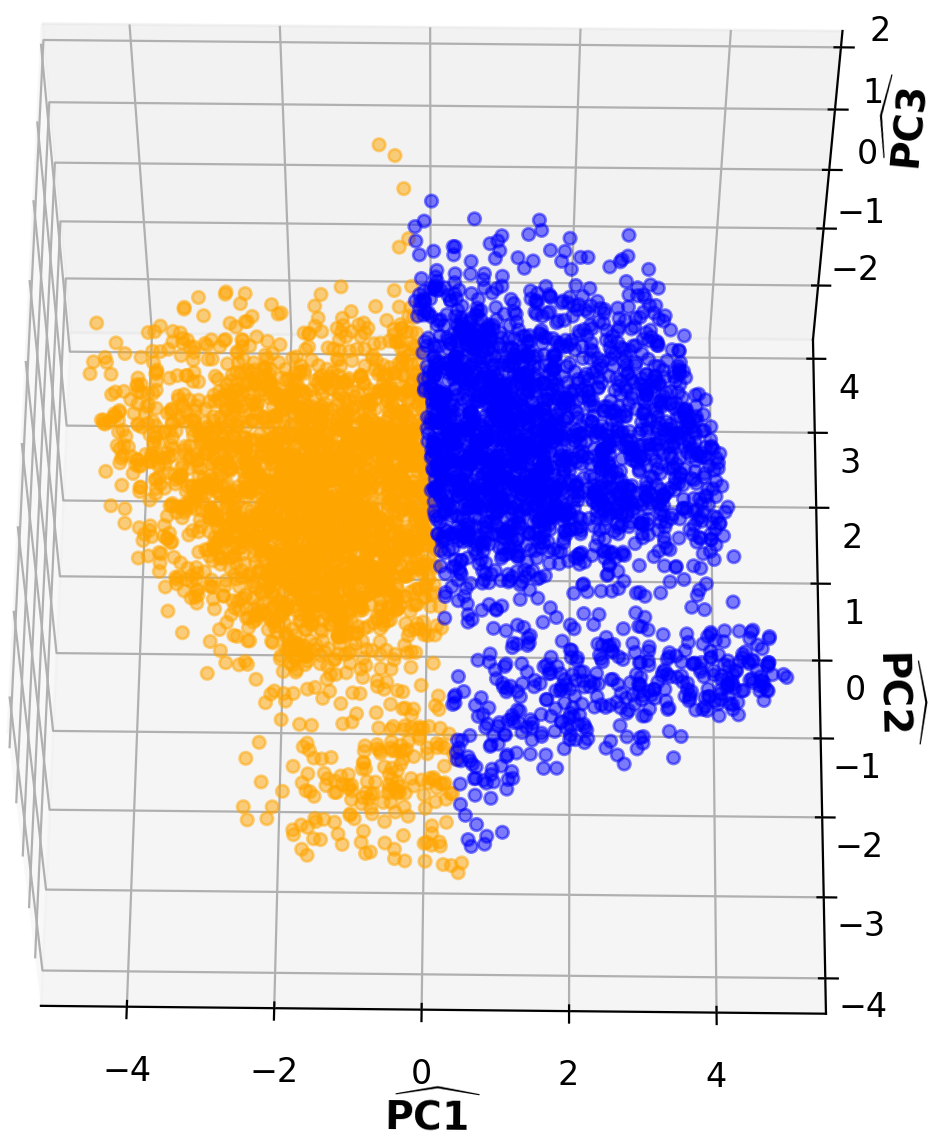}
\caption{PCA - initial Kmeans}
   \label{fig:init_PCA}
\end{subfigure}
\hfill
\begin{subfigure}{0.35\textwidth}
\includegraphics[width=2.3in]{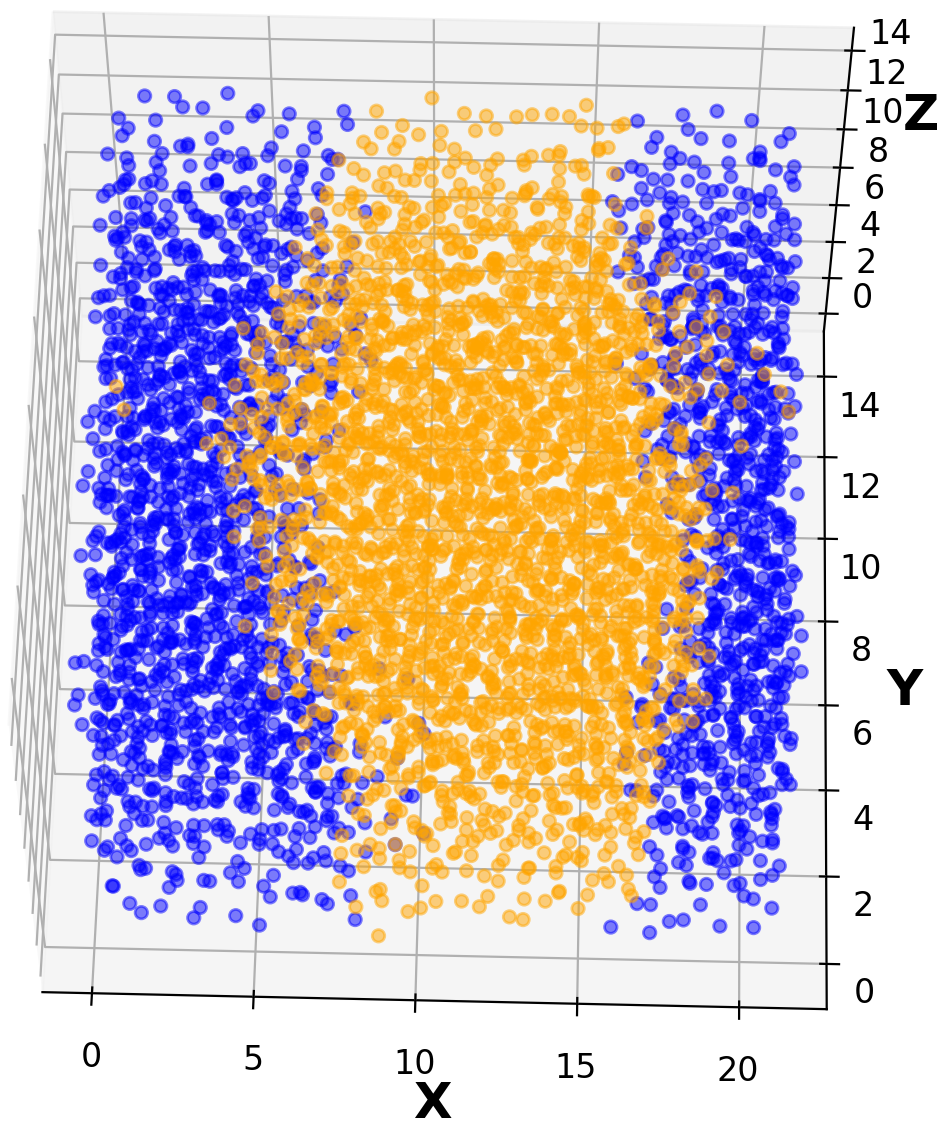}
\caption{xyz - initial Kmeans}
   \label{fig:init_xyz}
\end{subfigure}
\hfill
\begin{subfigure}{0.35\textwidth}
\includegraphics[width=2.3in,height=2.3in]{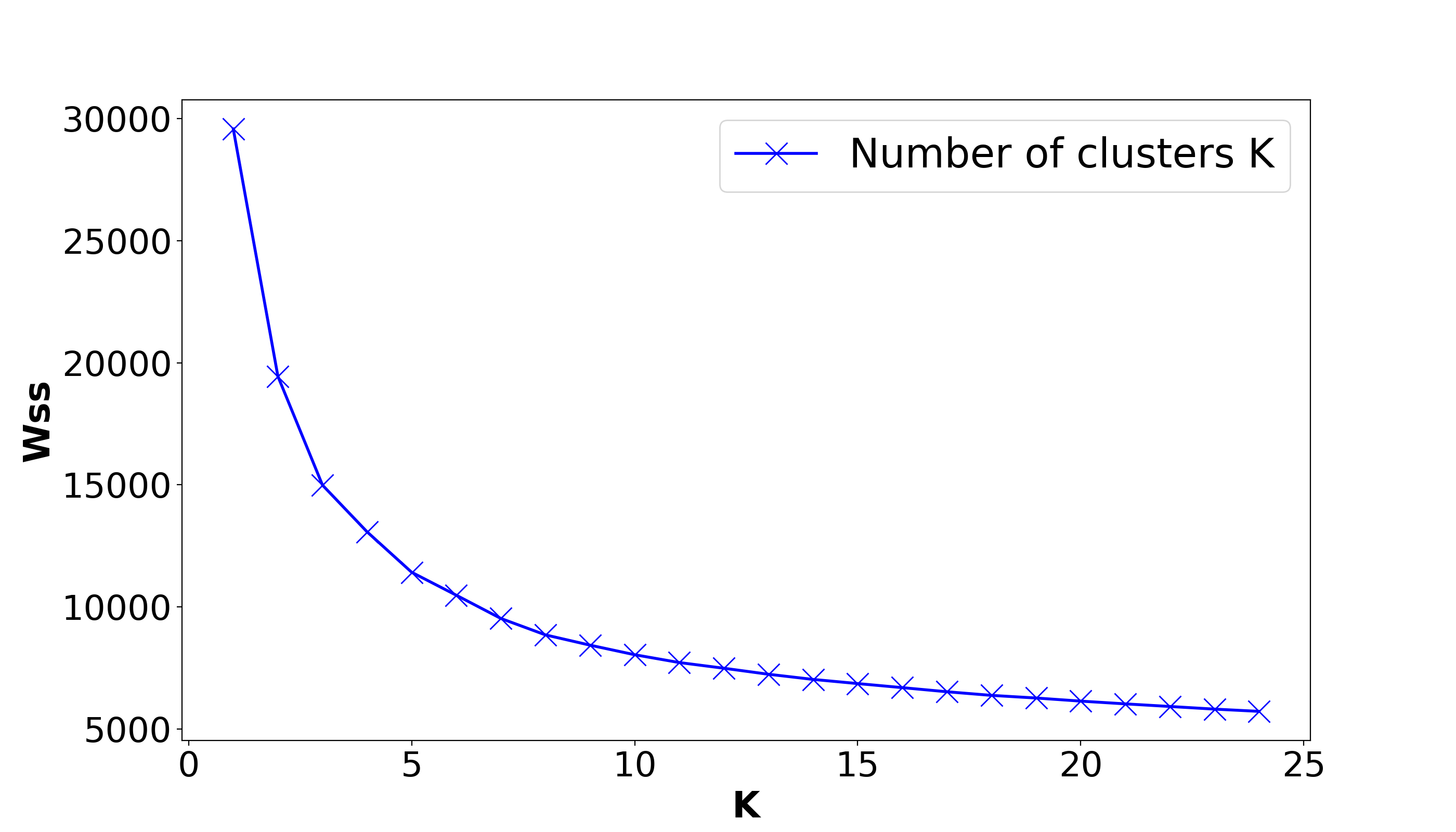}
\caption{Elbow test}
   \label{fig:Elbow}
\end{subfigure}
\hfill
}
\caption{A 3D snapshot of meso-states of 5000 particles system at $T^*=0.2$ from direct mapping and the Elbow test. Panels a and b are for the PC-space and configurational space while panel c displays the Elbow convergence test. The hat symbol for labelling axes of the PC space represents the inner product of a particle's $\overline{WCN}$s with the PC basis, in this case for the first three PC components. Blue and orange particles belong to meso-state 1 and  2, respectively. X,Y, Z are atomic coordinates in reduced units.}  
\label{fig:init}
\end{figure}

\begin{figure}
\resizebox{\columnwidth}{!}
{
\begin{subfigure}{0.35\textwidth}
\includegraphics[width=1.5in]{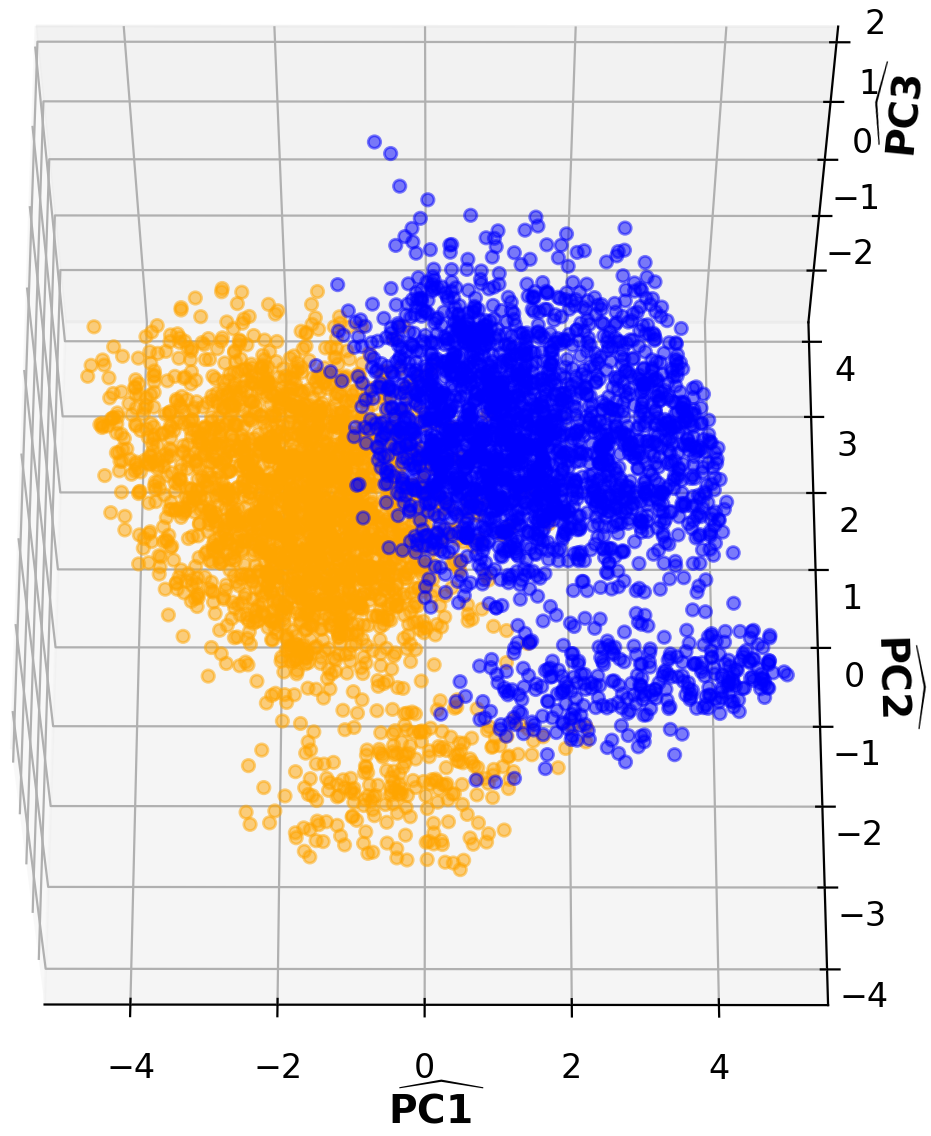}
\caption{ PCA - iteration 2}
   \label{fig: iter2_PCA}
\end{subfigure}
\hfill
\begin{subfigure}{0.35\textwidth}
\includegraphics[width=1.5in]{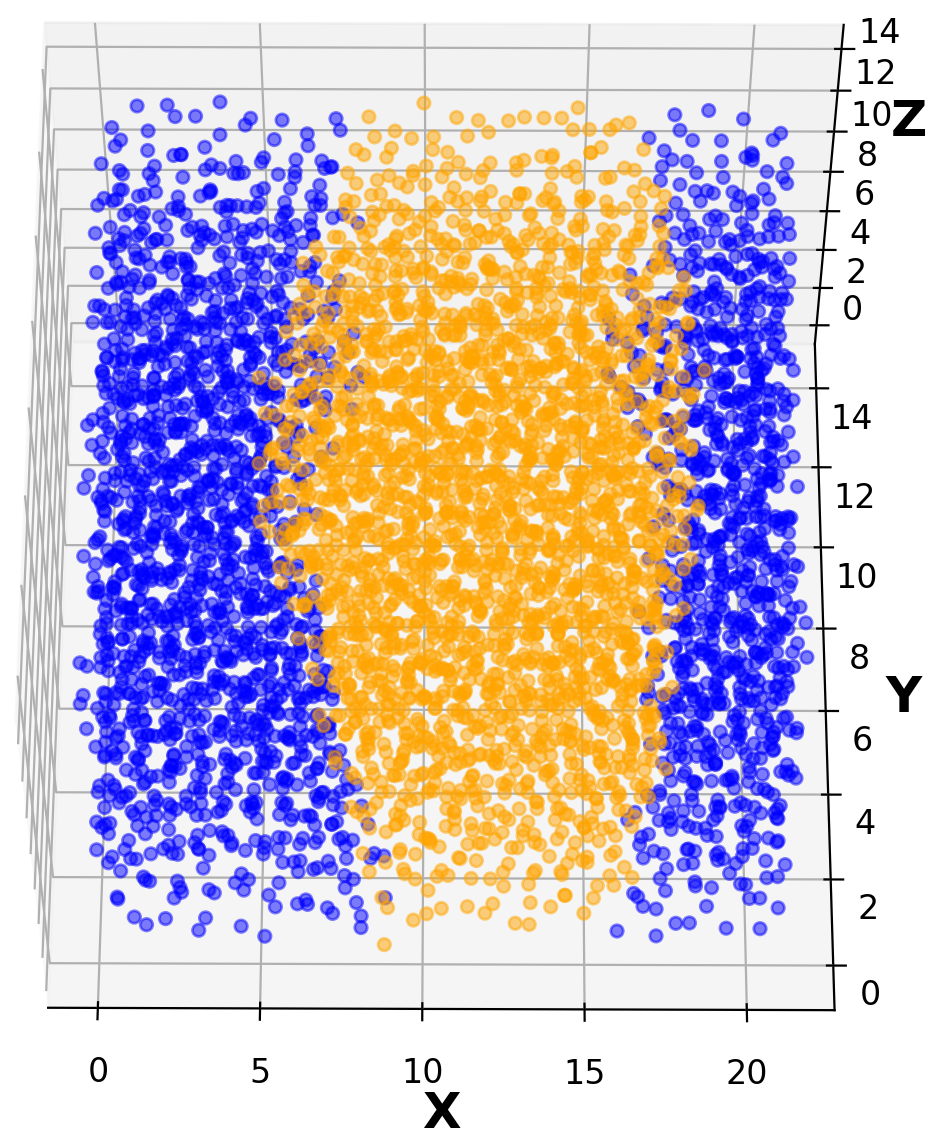}
\caption{xyz - iteration 2}
   \label{fig: iter2_xyz}
\end{subfigure}
\hfill
}

\resizebox{\columnwidth}{!}
{
\begin{subfigure}{0.35\textwidth}
\includegraphics[width=1.5in]{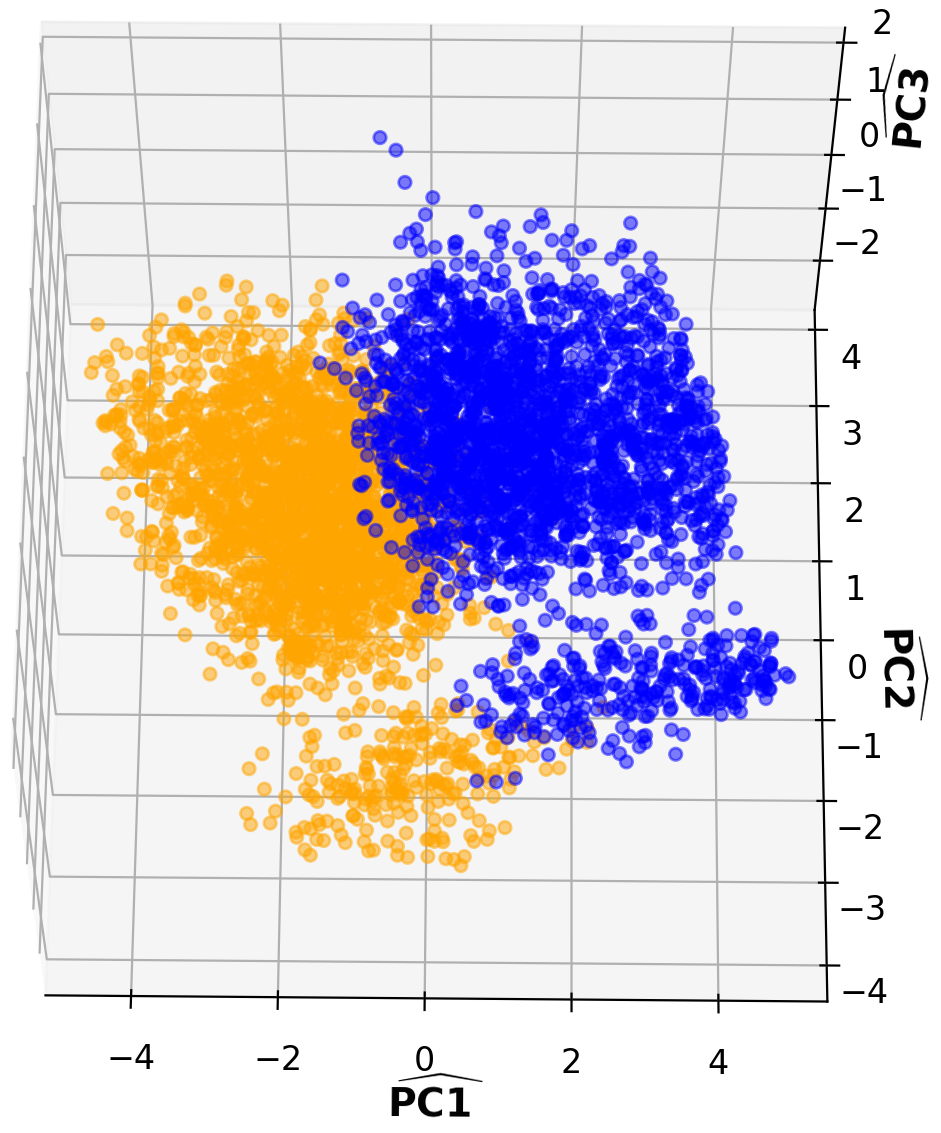}
\caption{$T^*=0.2$, PCA - iteration 3 }
   \label{fig: iter3_PCA}
\end{subfigure}
\hfill
\begin{subfigure}{0.35\textwidth}
\includegraphics[width=1.5in]{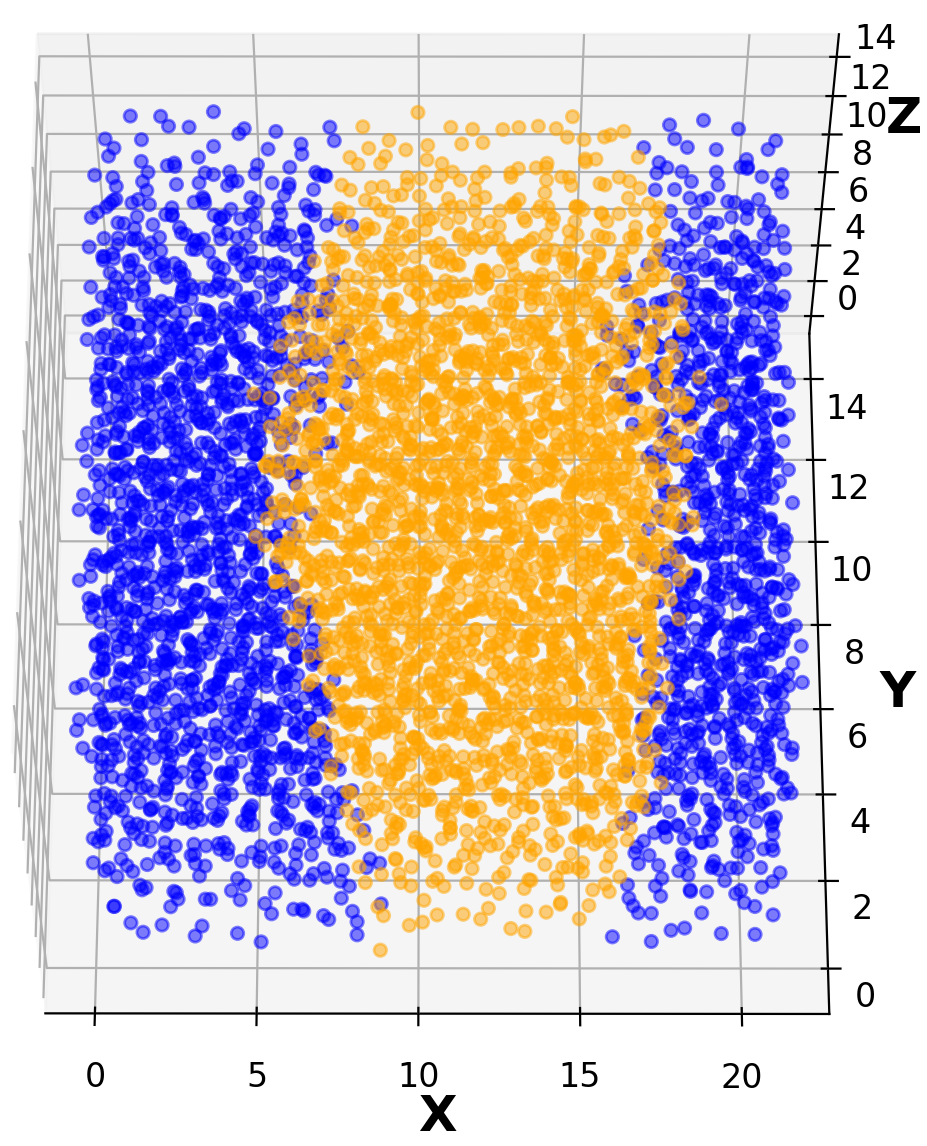}
\caption{$T^*=0.2$,  xyz - iteration 3}
   \label{fig: iter3_xyz}
\end{subfigure}
}
\caption{ A 3D snapshot of 2 meso-states of 5000 particles system at $T^*=0.2$ from our co-learning method. Panels a and c are for the PC-space and configurational space for iteration 2 while panels b) and d) represent clusters formed in the PC-space and configurational space for iteration 3 }  
\label{fig:nano-domain_co-learning_T0.2}
\end{figure}

For each particle, each of $N$ features in the $\overline{WCNs}$ matrix is constructed separately to describe its own local solvation shell environment with respect to its surrounding particles, it is disconnected from each other to form a proper feature space. To resolve this issue, Principal Component Analysis (PCA) ~\cite{J.Shlens} is used for dimension reduction, namely to linearly transform original $\overline{WCNs}$ matrix into a new feature space that reduce $N$ particles' features to a few correlated ones. Mathematically, PCA can be done through the following three steps:
\begin{itemize}
\item Obtaining the mean-free data $\bold{X = \widetilde{X} - \langle \widetilde{X} \rangle}$ where the average is over ${M}$ particles for each component of WCNs.
\item Forming the correlation matrix $\bold{C = X^\intercal X}$, which is $N\times N$.
\item The principle components $\bold{u_i}$ are obtained after solving the eigenvalue problem: $\bold{Cu_i = {\bm{\sigma_i}}^2u_i }$. The eigenvalue $\bold{\bm{\sigma_i^2}}$ measures the variance of the data along each principle component(PC) $i$. PCA is optimal in term of seeking small numbers of PCs but maximizing cumulative proportion of variance explained (PVE) $\bold{\bm{\sigma_i^2}}$ by each principle component. In other words, the numbers of retained PCs depend on their total PVE such that the total PVE is $\ge$ 95$\%$ of total variances presented in ${\widetilde{\bf X}}$.
\end{itemize}
The new complete basis composes of all PCs: $\bold{U = [u_1,u_2,..u_N]}$ where each $\bold{u_i}$ is a collective coordinate with $N$ components corresponding to the number of features in the data input. In our study, the first three PCs retains about 85-90\%, so 6-7 PCs are sufficient enough to form PC's basis whose PVE could be $\ge$ 95$\%$ of total variances presented in ${\widetilde{\bf X}}$. The new coordinates (PC representation) are generated from an inner product of original $\overline{WCNs}$ matrix with the PC's basis (PC-space), mathematically, $\bold{Y= U^\intercal \widetilde{\bf X}}$. We then use K-means clustering method to decipher hidden structures of the PC representation by classifying particles into distinct clusters called meso-states. K-means clustering is chosen because it is an unsupervised standard technique that geometrically separate particles into clusters that aggregated together because of certain similarities. 
 
However, the K-means requires prior knowledge of the number of existing clusters K in the data structure to work effectively, which is generally unknown in most cases. An implementation of the Elbow convergent test could provides a reasonable prediction of the K values. The Elbow test permits the number of clusters K being varied freely and computes the Within-Cluster Sum of Square Distance (Wss) which is the sum of square distances between each data point and the centroid within a cluster. As the number of clusters K increase, the Wss will start to decrease and eventually become roughly constant regardless of further increasing K. The Elbow plot of the Wss against K looks like an Elbow shape where the Elbow point normally corresponds to an initial guess of K used in K-means clustering. In many cases, the Elbow plot has a clear Elbow point which indicates a good guess for K-means. In our case, initial K remains uncertain because the Elbow shape is poor to single out an Elbow point, thus we can only narrow down a possible range of K values (K = 2 to 5) (\ref{fig:Elbow}). After a careful trial-and-error process with help of the co-learning strategy,  K = 2 is selected; details of the process is discussed in the Appendix \ref{sec:kmeans}. Given K = 2, particles in the PC-space are classified into 2 distinct meso-states (\ref{fig:init_PCA}), then a direct mapping using the identities of particles in each meso-state in the PC-space also forms aggregated clusters in the configurational space as shown in \ref{fig:init_xyz}; different projected angles of \ref{fig:init_PCA} and \ref{fig:init_xyz} to confirm the clustering structures both in PC and real space are in Appendix \ref{sec:angles}. Naturally, each meso-state have mixing A and B particles. On the other hand, each type of (A or B) particles itself appears as two distinct aggregated domains in the configurational space. This is clearly demonstrated in the Appendix \ref{sec:AB}. 

Although domains are generated in the real space by a simple mapping of particles' identities in the PC-space after K-means clustering, there are two issues needed to be addressed. Firstly, the principle of K-means clustering relies on assigning a particle to a cluster where its Euclidean distance (E-dist) to the centroid of that cluster is the closest among others. In other words, assignment of a particle depends on the E-dist measure sensitively which becomes robust for core particles of each meso-state because the difference of their distances from one state to another is well-defined. However, the E-dist criterion becomes an issue to assign interfacial particles due to the small differences in their distances to either states, so it could lead to misclassification. Secondly, even though identities of clusters are preserved from the PC-space to the configurational space the inverse transfer of the knowledge is not clear, but physically the transfer of  knowledge should be bi-directional. Thus, a co-learning strategy is developed as the following:

\begin{enumerate}
\item Perform K-means clustering in the PC-space.
\item Use the initial knowledge of the clustering  from the PC-space to perform a Gaussian Mixture (GM) classification in the configurational space to soften the hard assignment from the K-means: 
\begin{enumerate}
	\item do a direct mapping of particles identities in the PC space to identify distinct nano-domains in the real space.
	\item build a mixture model of multivariate Gaussian distributions of domains, then assignment of a particle belonging to a domain is determined by maximizing Gaussian probability among different domains.
\end{enumerate}
\item Similar to step 2, perform GM in the PC-space from the clustering knowledge in the configurational space.
\item Iteratively perform GM classification in both spaces until convergence. 
\end{enumerate}

Classification of interfacial particles by the co-learning strategy converges quickly in both PC-space(\ref{fig: iter2_PCA},\ref{fig: iter3_PCA}) and configurational space (\ref{fig: iter2_xyz},\ref{fig: iter3_xyz}) after few iterations (3-5 runs on average) as shown in the \ref{fig:nano-domain_co-learning_T0.2}. The co-learning strategy shows improvement over K-means as it generalizes and fills the missing information from a direct information transfer from the PC-space to the configurational space. In other words, it allows a bi-directional information transfer. Firstly, correct classification of interfacial particles comes from using probabilistic clustering like GM to avoid sensitivities of E-dist criterion of the K-means. This GM clustering allows assignment of interfacial particles to  two states and the decision is made by the maximum-likelihood of the Gaussian probability, which creates a boundary region of meso-states in the PC-space as shown in \ref{fig: iter3_PCA}. In the configurational space, we also find that the core particles in both domains are still the same, only interfacial particles are properly re-assigned to make the final results consistent with the direct mapping (\ref{fig:init_xyz}) and co-learning strategy (\ref{fig: iter3_xyz}). Secondly, the classification scheme utilizes the information from both spaces in a self-consistent manner.   

\section{Results and Discussion}
\label{sec:results}
\subsection{Nature of nano-domains: Statics} 
\label{sec:nature}

\begin{figure}
\resizebox{\columnwidth}{!}
{
\begin{subfigure}{0.35\textwidth}
\includegraphics[width=2.1in]{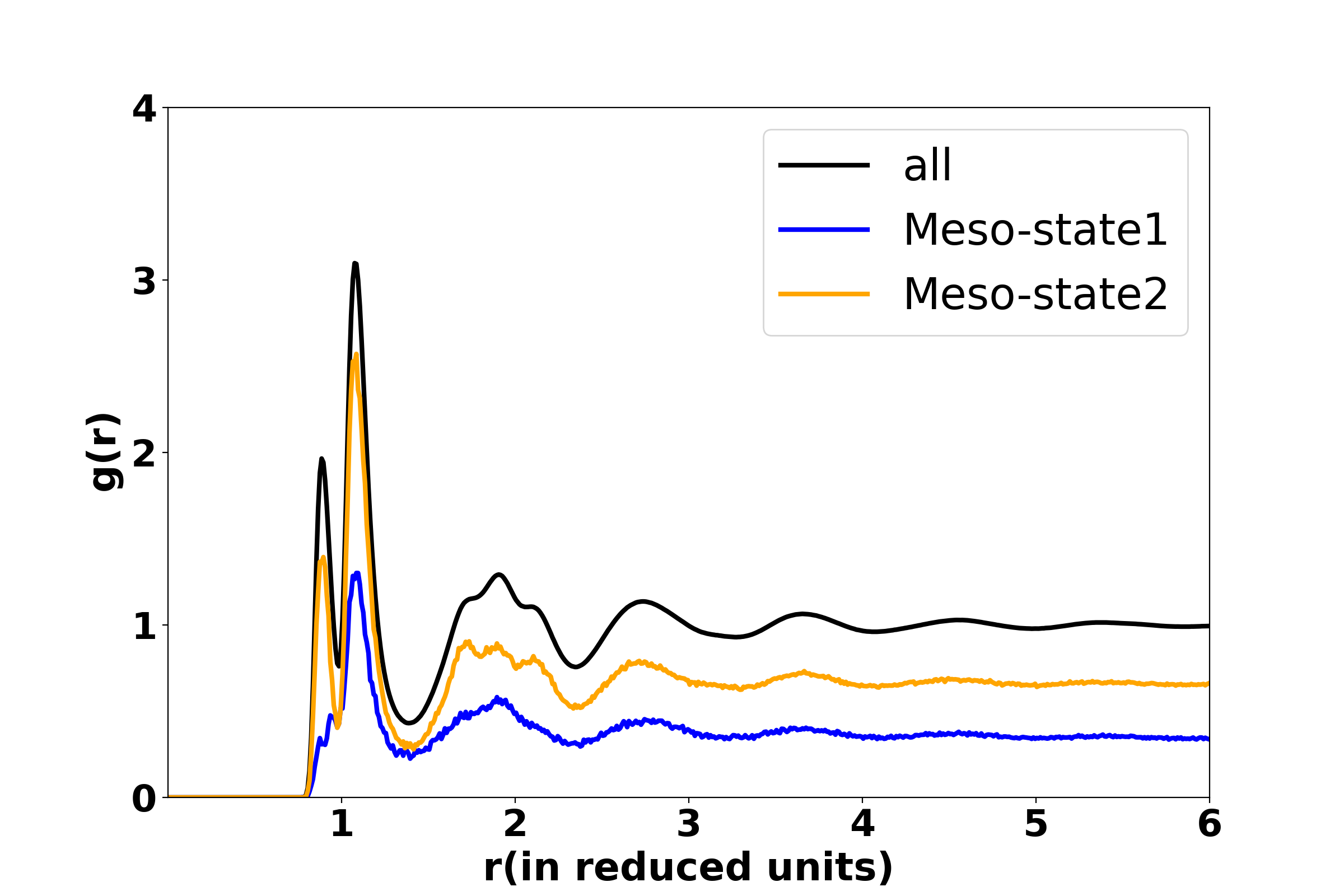}
\caption{{\it g(r)}s at $T^*=0.3$}
   \label{fig:1}
\end{subfigure}
\hfill
\begin{subfigure}{0.35\textwidth}
\includegraphics[width=2.1in]{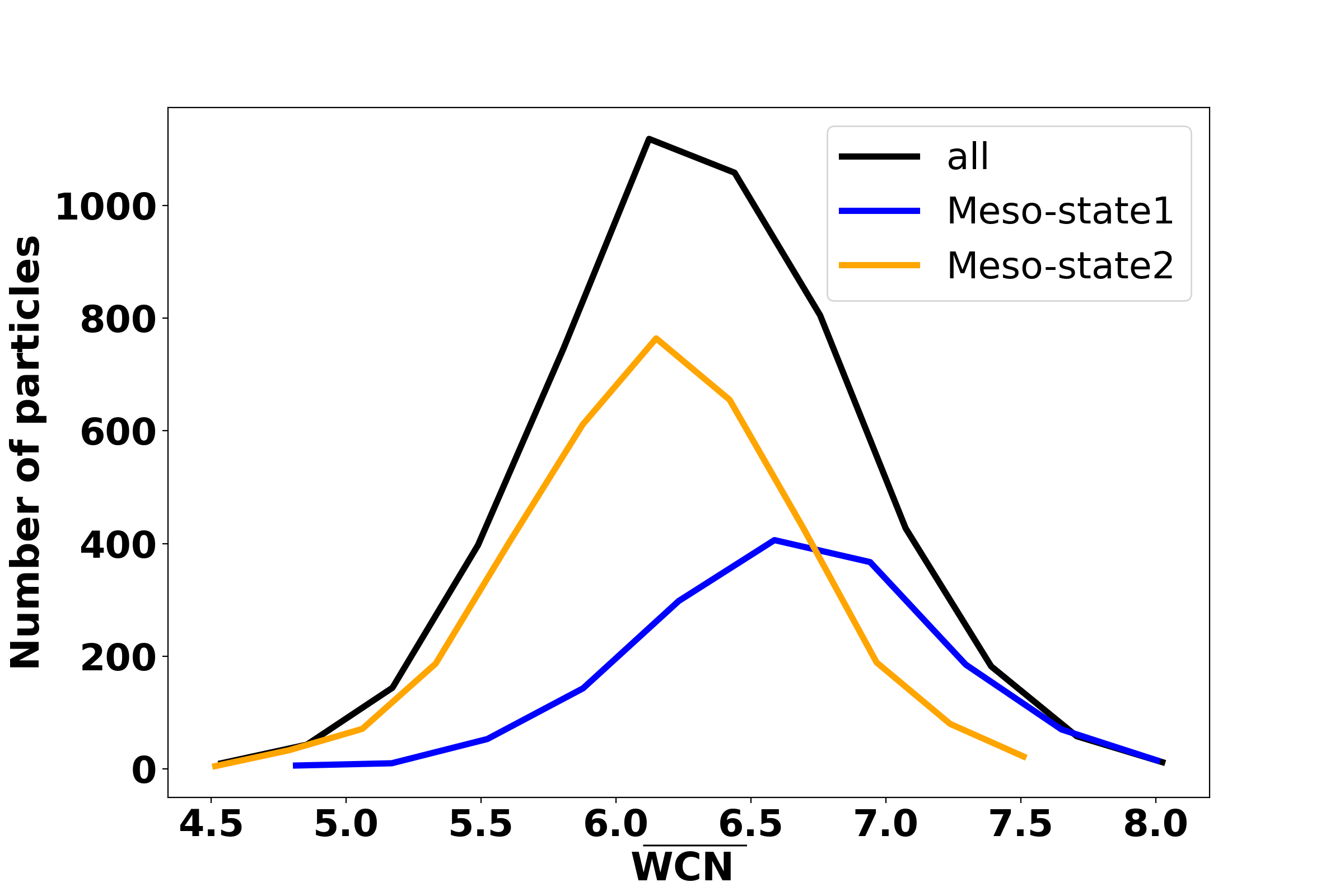}
\caption{ss1}
   \label{fig:2}
\end{subfigure}
\hfill
}
\resizebox{\columnwidth}{!}
{
\begin{subfigure}{0.35\textwidth}
\includegraphics[width=2.1in]{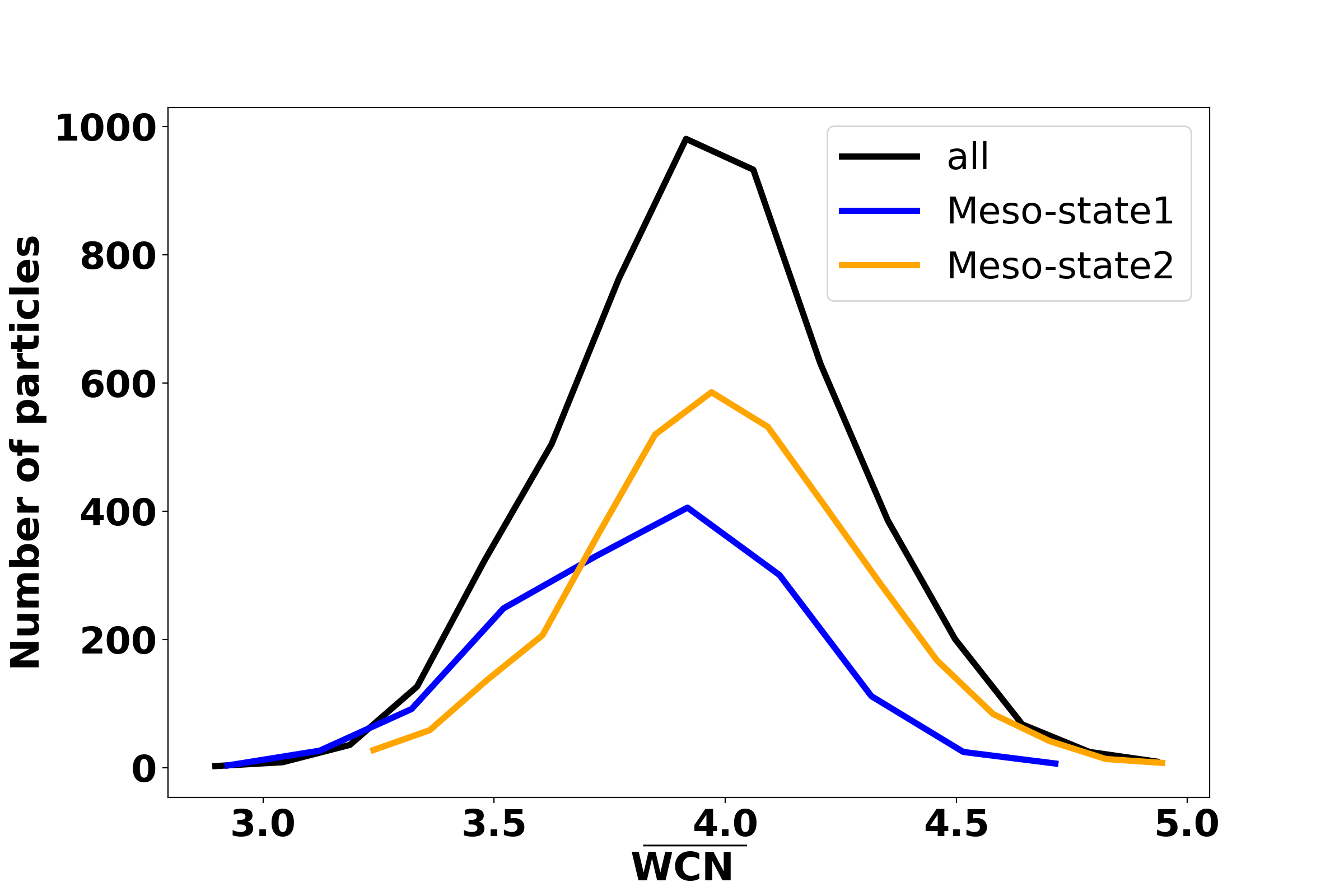}
\caption{ss2}
   \label{fig:3}
\end{subfigure}
\hfill
\begin{subfigure}{0.35\textwidth}
\includegraphics[width=2.1in]{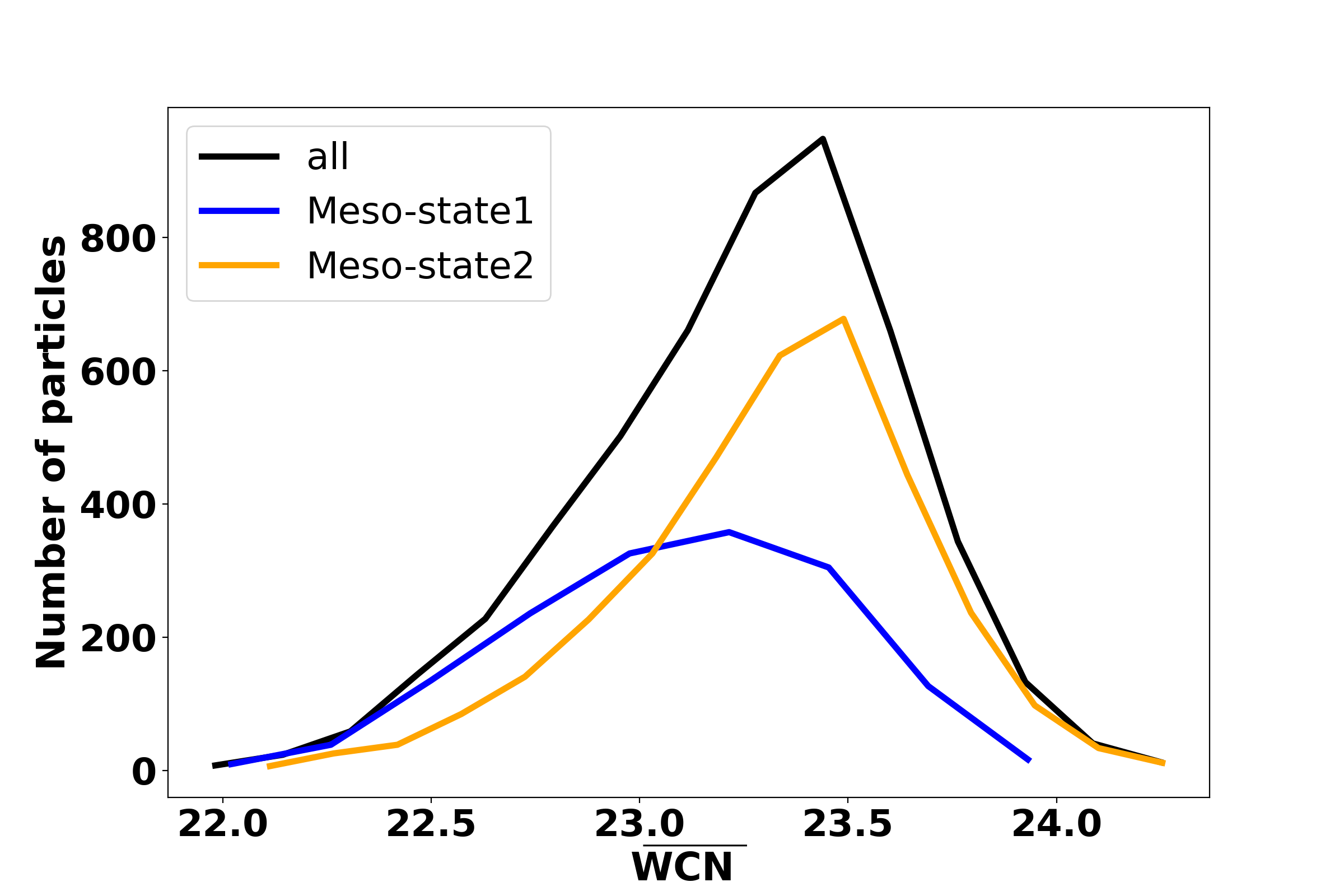}
\caption{ss3}
   \label{fig:4}
\end{subfigure}
\hfill
}
\resizebox{\columnwidth}{!}
{
\begin{subfigure}{0.35\textwidth}
\includegraphics[width=2.1in]{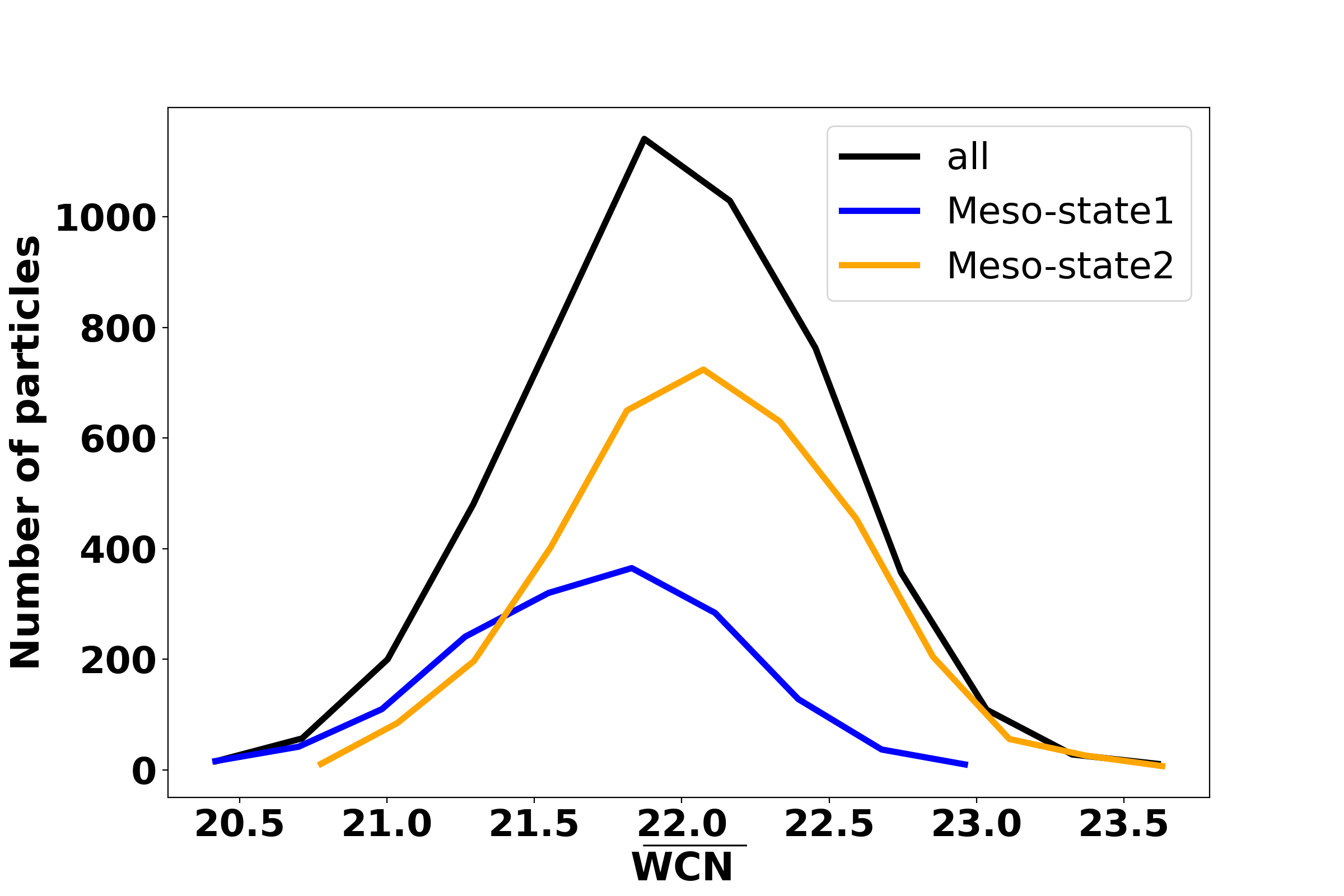}
\caption{ss4}
   \label{fig:5}
\end{subfigure}
\hfill
\begin{subfigure}{0.35\textwidth}
\includegraphics[width=2.1in]{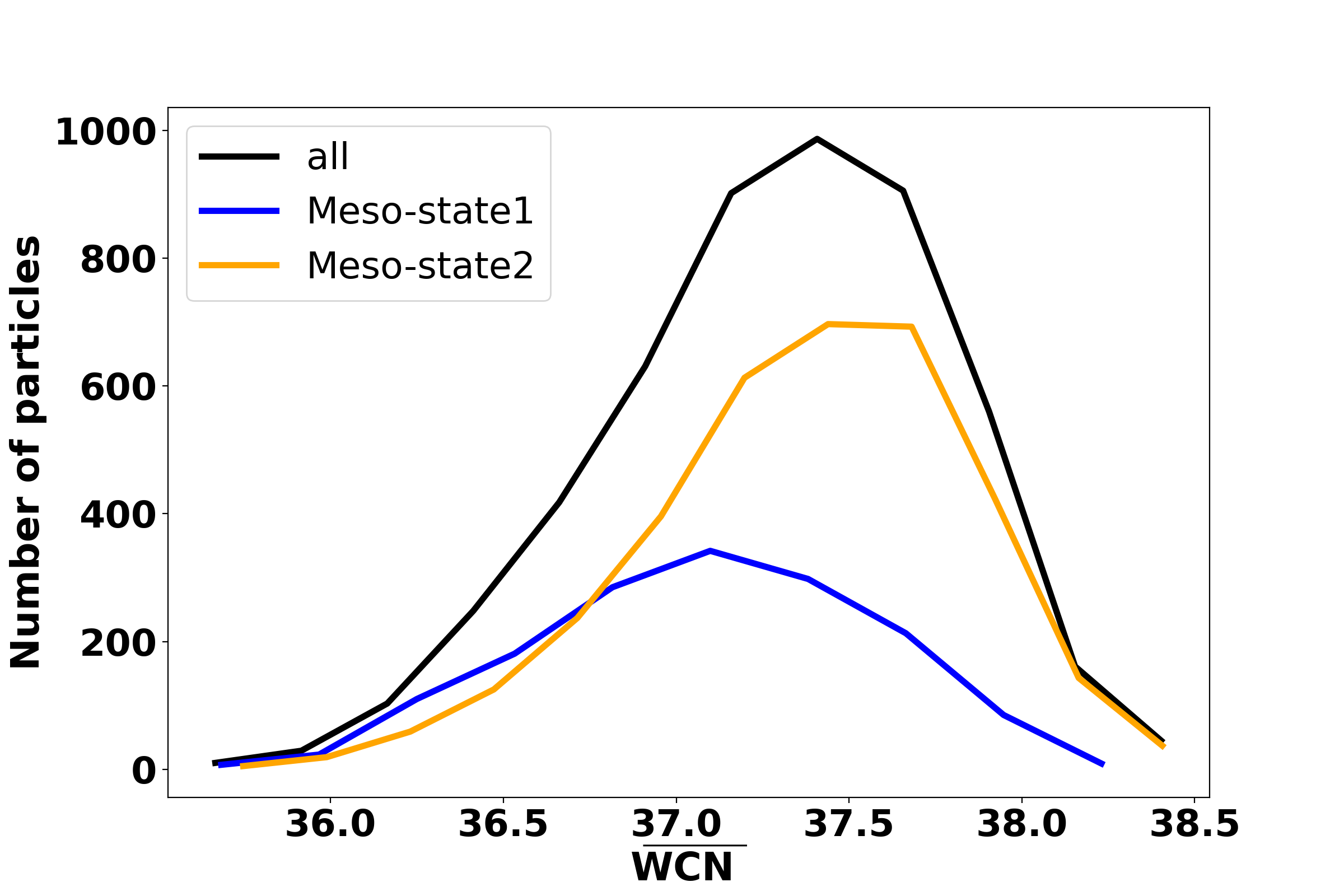}
\caption{ss5}
   \label{fig:6}
\end{subfigure}
}
\caption{ a) Weighted(according to the number of particles in each meso-state) partial {\it g(r)} decomposition of each meso-state at $T^* =0.3$. Panel b-f are the weighted $\overline{WCNs}$ distribution for the first five solvation shells of each meso-state and the whole system, respectively(ss stands for solvation shell). Given the classification of particles identities, the average WCN distributions along each solvation shell for each meso-state is obtained by  averaging over 5ns. The partial {\it g(r)}s in the \ref{fig:1} is constructed based on the particle's identities of each state and scaled by the weight obtained from $\overline{WCNs}$ bimodality along each solvation shell by averaging over 5ns. }
\label{fig:wcn_dist}
\end{figure}

In the previous section, a picture of structural and configurational  heterogeneity is revealed by the classification of the system into meso-states (in PC-space) or nano-domains (in real space). 
In order to clarify the physical interpretation of these nano-domains, it is observed that in the PC-space bimodality of $\overline{WCNs}$ distribution along each solvation shell. \ref{fig:wcn_dist}b-f provide clear evidence of two meso-states in the PC space, for example the total $\overline{WCNs}$ distributions along first five solvation shells of the system are decomposed into distributions of each individual meso-state as there is a co-existence of two meso-states with different unique local structures. Furthermore,  the bimodal distributions of the $\overline{WCNs}$ along all shells in the PC-space can be transformed into a construction of partial {\it g(r)}s in the configurational space as shown in \ref{fig:1}. The total $\it g(r)$ of the whole system is the summation of the weighted partial $\it g(r)$s (blue and orange curves) representing two meso-states. In other words, the classification scheme provides a method to decompose the total $\it g(r)$ of the system into two partial $\it g(r)$s, which represent two different meso-structures whose particles form various domains that tile up the whole configurational space.

\begin{figure}
\resizebox{\columnwidth}{!}
{
\begin{subfigure}{0.35\textwidth}
\includegraphics[width=2.1in]{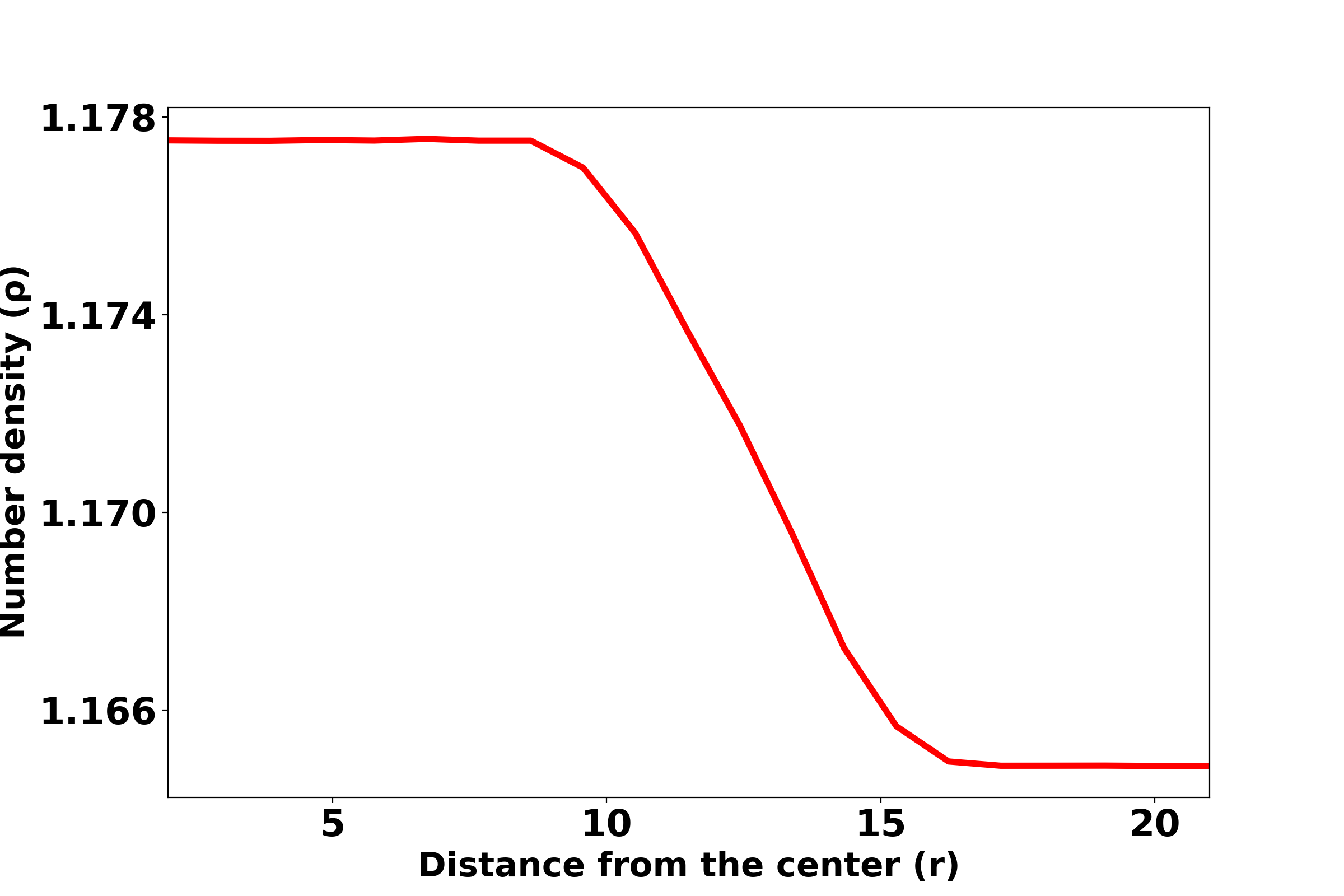}
\caption{Density profile}
   \label{fig:density}
\end{subfigure}
\hfill
\begin{subfigure}{0.35\textwidth}
\includegraphics[width=2.1in]{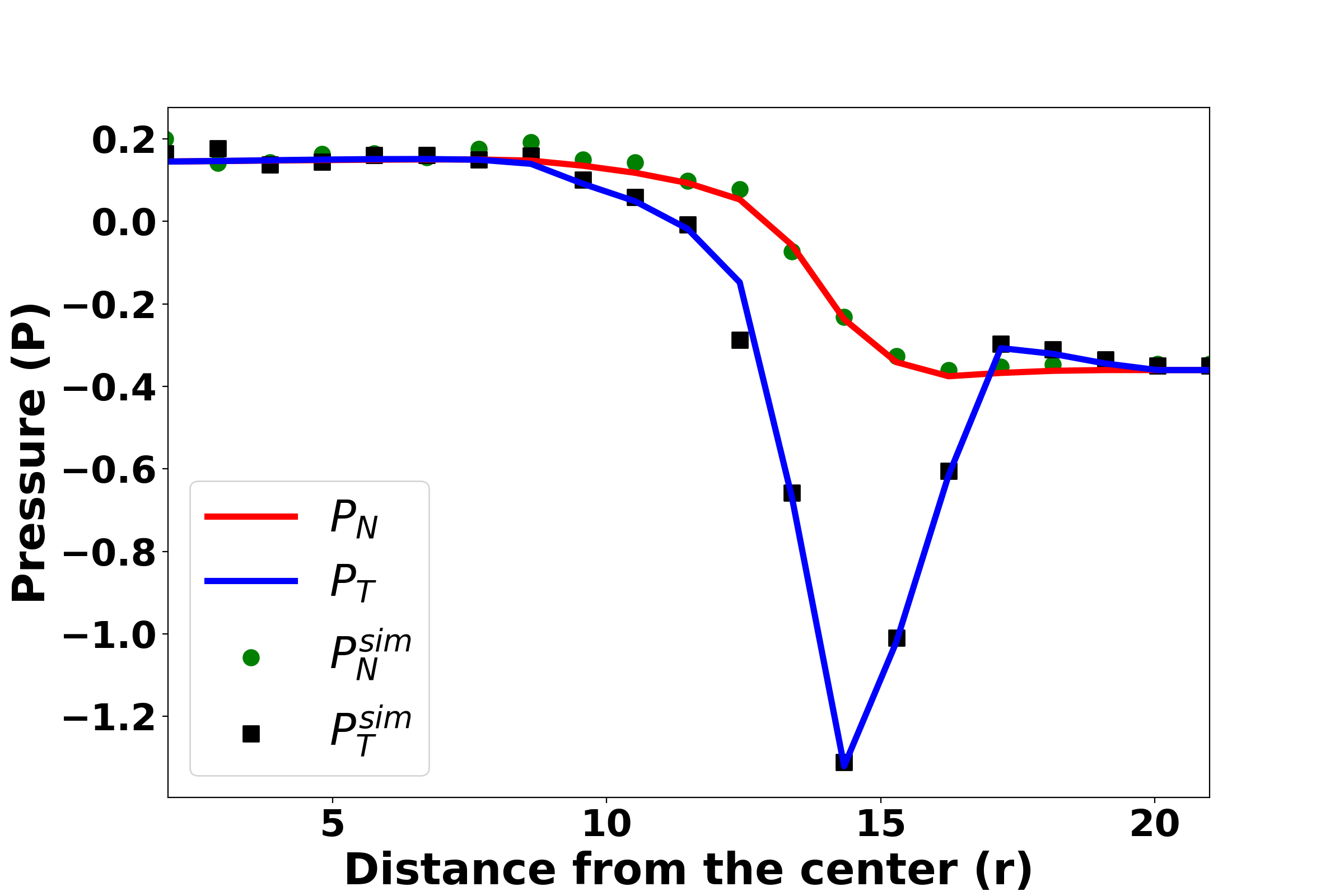}
\caption{Pressure profile}
   \label{fig:press}
\end{subfigure}
}
\caption{ (a) Density profile varying with distance from the center of a domain of meso-state 1  at $T^*=0.3$;  (b) Normal and tangential components of the pressure tensor varying with the distance from the center of a domain of meso-state 1. The green spheres and black squares are normal ($P_N^{sim}$) and tangential ($P_T^{sim}$) pressures obtained in the simulation, respectively. The red line is the fit to the normal pressure ($P_N$) and the blue line is the tangential pressure ($P_T$) obtained to verify mechanical equilibrium}  
\label{fig:density_press}
\end{figure}

Another quantitative measure of these distinct meso-states is to compute density and pressure profiles of the domains. Because the shapes of the domains are irregular, the thermodynamic properties of the two meso-states were calculated using a spherical region inside a domain of meso-state 1 and a thin shell in the outermost meso-state 2 region. \ref{fig:density} shows the radial distribution of the atomic number density from the center of meso-state 1. Meanwhile, the six  components of the pressure tensor ($\it p_{xx}$,  $\it p_{yy}$, $\it p_{zz}$, $\it p_{xy}$,$\it p_{yx}$,$\it p_{xz}$ and $\it p_{zx}$) for each atom are computed in the Cartesian coordinate. The pressure tensor is then transformed into polar coordinate representation whose corresponding components will be ($\it p_{rr}$,  $\it p_{\theta\theta}$, $\it p_{\phi\phi}$, $\it p_{r\theta}$,  $\it p_{\theta\phi}$ and $\it p_{\phi r}$). It is noted that the magnitude of the off-diagonal terms is negligible compared to diagonal terms, thus the pressure tensor can be expressed as ~\cite{Gunawardana,rowlinson2013molecular}:
\begin{equation} \label{eq:press_comp}
	P(r) = P_N (r)\textbf e_r\textbf e_r + P_T (r)(\textbf e_{\theta}\textbf e_{\theta}+ \textbf e_{\phi}\textbf e_{\phi} ),
\end{equation}
where $\textbf e_r$, $\textbf e_{\theta}$ and $\textbf e_{\phi}$ are unit vectors, $\it P_N$ and $\it P_T$ are the radial or normal and transverse components of the pressure tensor, respectively. The radial profiles of the components $\it P_N(r)$ and $\it P_T(r)$ are obtained by integrating out the angular degrees of freedom over thin spherical shells extending outwards from the origin. \ref{fig:press} shows the normal ($\it P_N$) and tangential ($\it P_T$) pressure profiles approximated as a spherical interface within the solid angle of the calculation. It is verified that the normal and tangential profiles statisfy the mechanical equilibrium,$\bm{\nabla } \cdot \bm{P} = 0$, which in spherical coordinates is given by~\cite{Ballal,rowlinson2013molecular}: 
\begin{equation} \label{eq:press_mech_eq}
	P_T (r) = P_N (r) + \frac{r}{2} \frac {dP_N (r)} {dr},
\end{equation}
where the second term is the derivative of the normal pressure with respect to distance from the center of meso-state 1. 

The formation of local spherical interfaces from density and pressure profiles in \ref{fig:density_press} signifies a strong indication for the co-existence of two local distinct meso-states
in supercooled states.

\subsection{ Nature of the nano-domains: Dynamics}

\begin{figure}
\resizebox{\columnwidth}{!}
{
\begin{subfigure}{0.35\textwidth}
\includegraphics[width=2.1in]{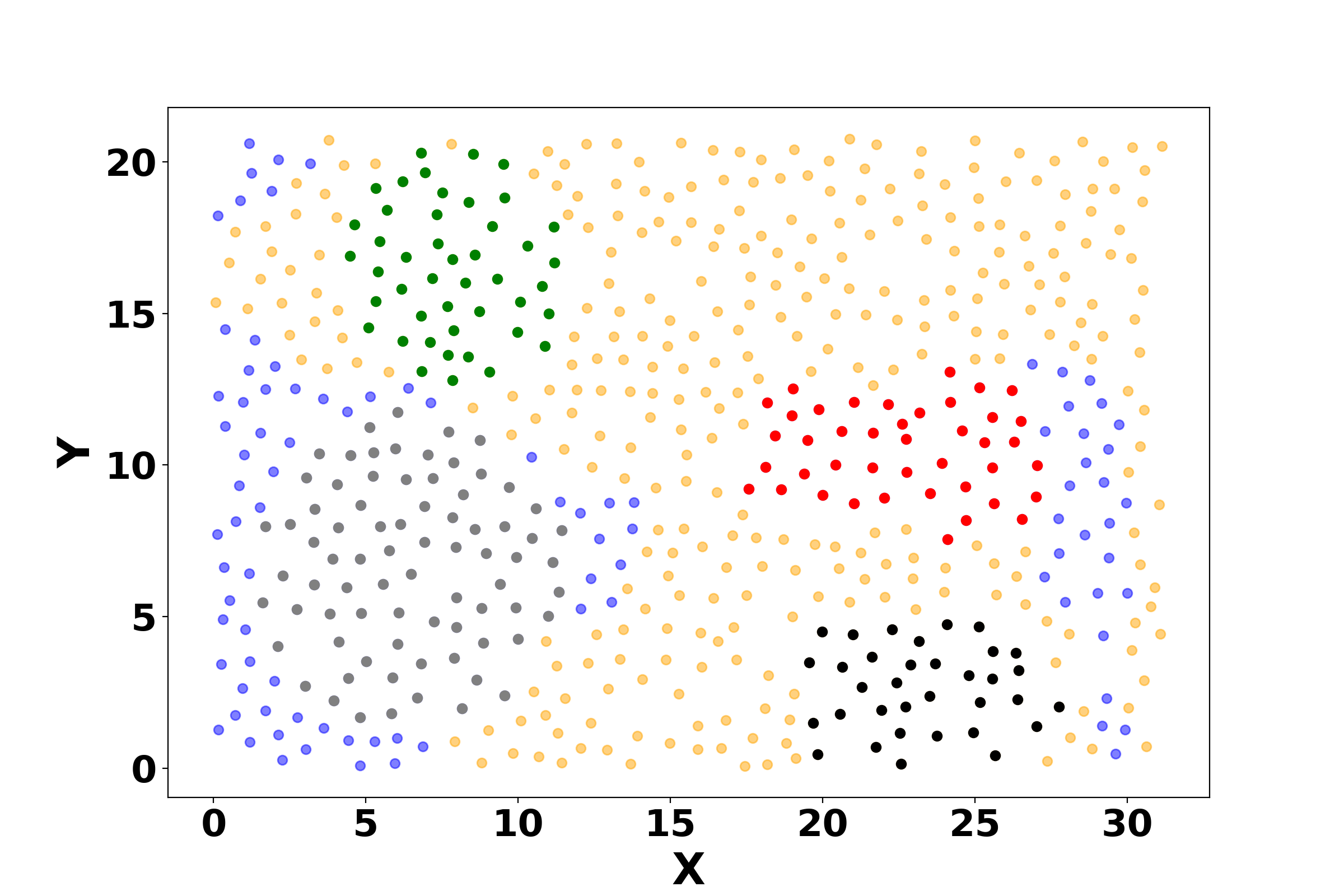}
\caption{Cross section 1}
   \label{fig:core1}
\end{subfigure}
\hfill
\begin{subfigure}{0.35\textwidth}
\includegraphics[width=2.1in]{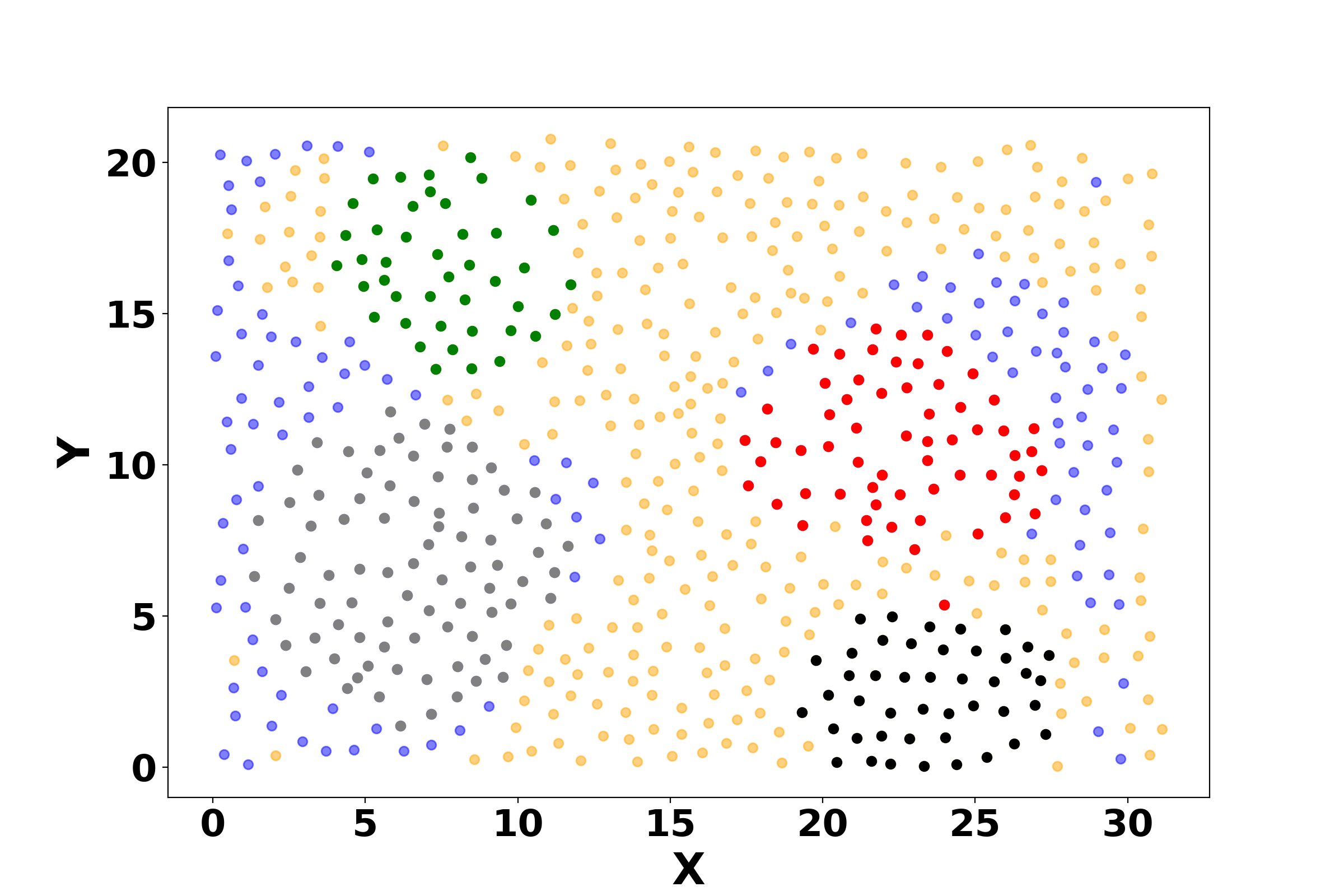}
\caption{Cross section 2}
   \label{fig:core2}
\end{subfigure}
\hfill
}
\resizebox{\columnwidth}{!}
{
\begin{subfigure}{0.35\textwidth}
\includegraphics[width=2.1in]{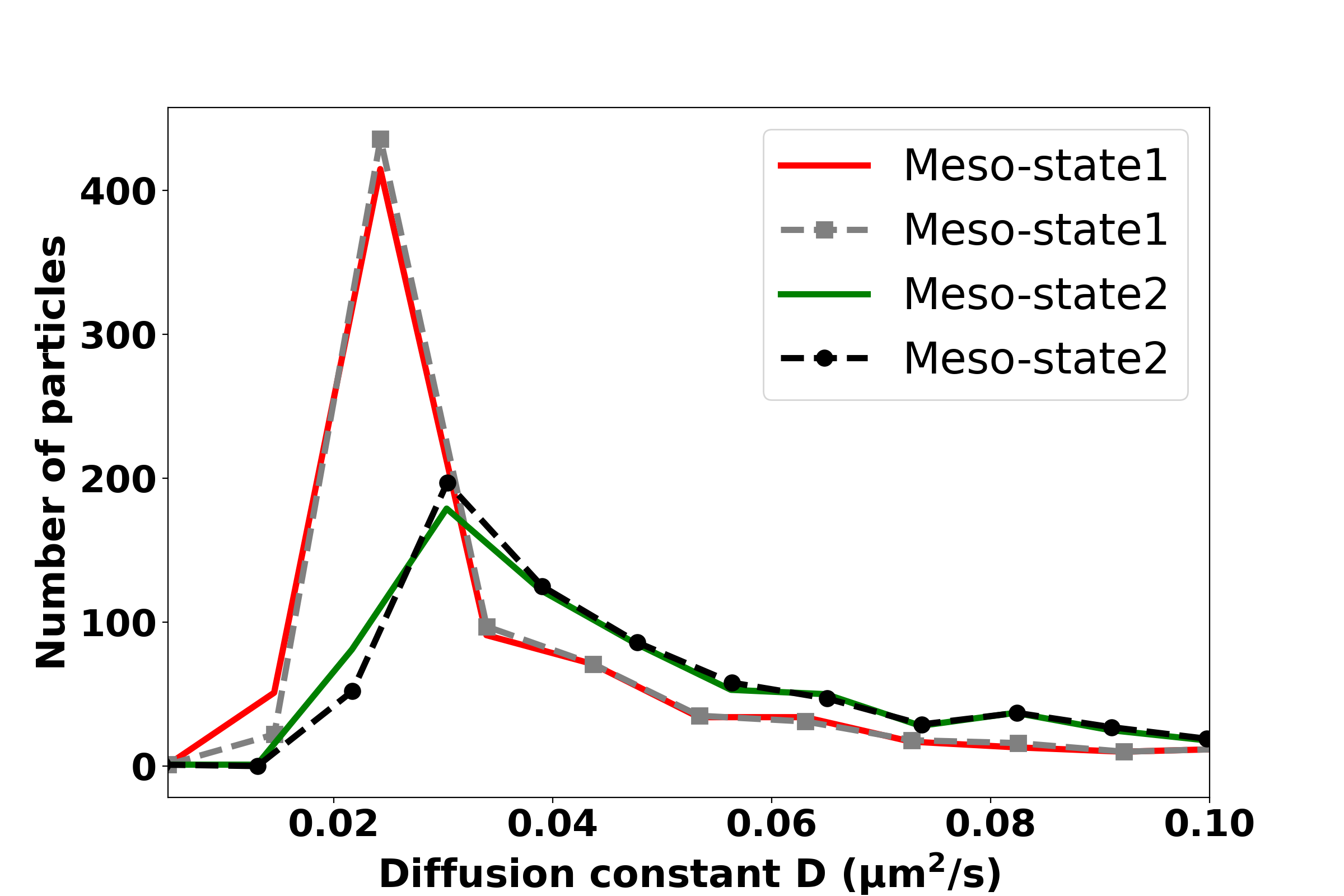}
\caption{$T^*$ = 0.37}
   \label{fig:T0.37_D_dist}
\end{subfigure}
\hfill
\begin{subfigure}{0.35\textwidth}
\includegraphics[width=2.1in]{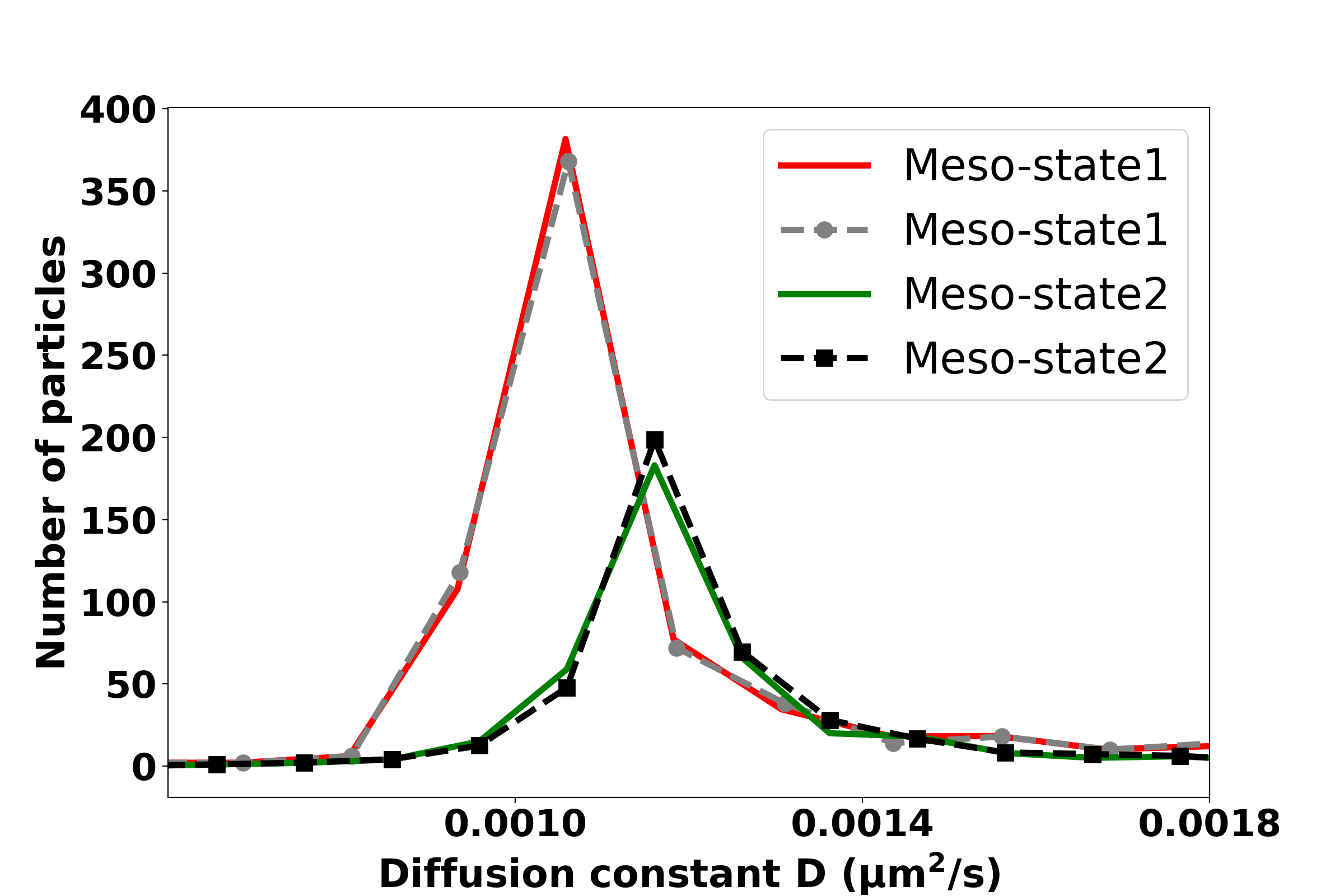}
\caption{$T^*$ = 0.3}
   \label{fig:T0.3_D_dist}
\end{subfigure}
\hfill
}
\caption{ a-b) 2D representation of sorted cores particles of meso-states in bifurcated domains of a 16000 particles system at $T^*$ = 0.3; c-d) Distribution of diffusion constants of core particles at different temperatures}  
\label{fig:D_dist}
\end{figure}

With our classification scheme, the bimodal decomposition of the {\it g(r)}, the density and pressure profiles seem to indicate an coexistence of two phases with domain structures after quenching, where similar liquid-liquid phase separation is also observed in a model 2D system with such classification scheme~\cite{viet}. To further check the validity of such a picture, 
 some dynamical signatures of a liquid-liquid phase separation are evaluated.

First of all, there will be two well separated relaxation time scales in such a scenario. The particles within the domains that belong to the same meso-states should have the same diffusion behavior as they are in the same thermodynamic state.  After finding the nano-domains in the configurational space, core particles, the particles stay in that domain during the whole simulation time, in each domain are sorted.  Core particles are colored as red and grey for meso-state 1 (blue) and black and green for meso-state 2 (orange) as shown in \ref{fig:D_dist}a,b. 2D cross section of core particles is taken for the purpose of visualization. Collected core particles from each domain are then used to compute mean-square displacements to get diffusion constant by Einstein relation. \ref{fig:D_dist}c,d show different diffusion constant distribution of different domains at different temperatures, hence supports the picture that the domains that belong to the same meso-state have the same diffusion behavior. 

\begin{figure}
\resizebox{\columnwidth}{!}
{
\begin{subfigure}{0.35\textwidth}
\includegraphics[width=2.1in]{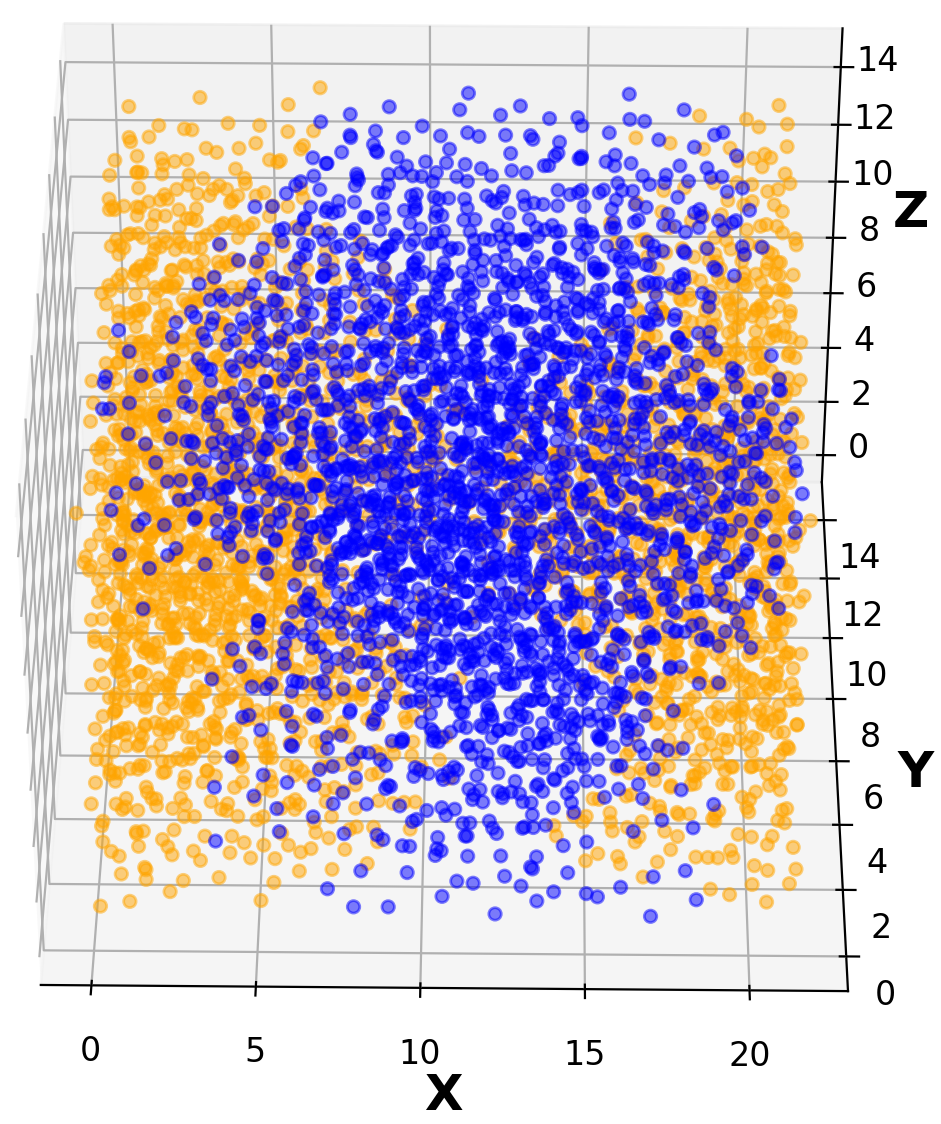}
\caption{$T^*$ = 0.37, Snapshot 1}
   \label{fig:T0.37_10ns}
\end{subfigure}
\hfill
\begin{subfigure}{0.35\textwidth}
\includegraphics[width=2.1in]{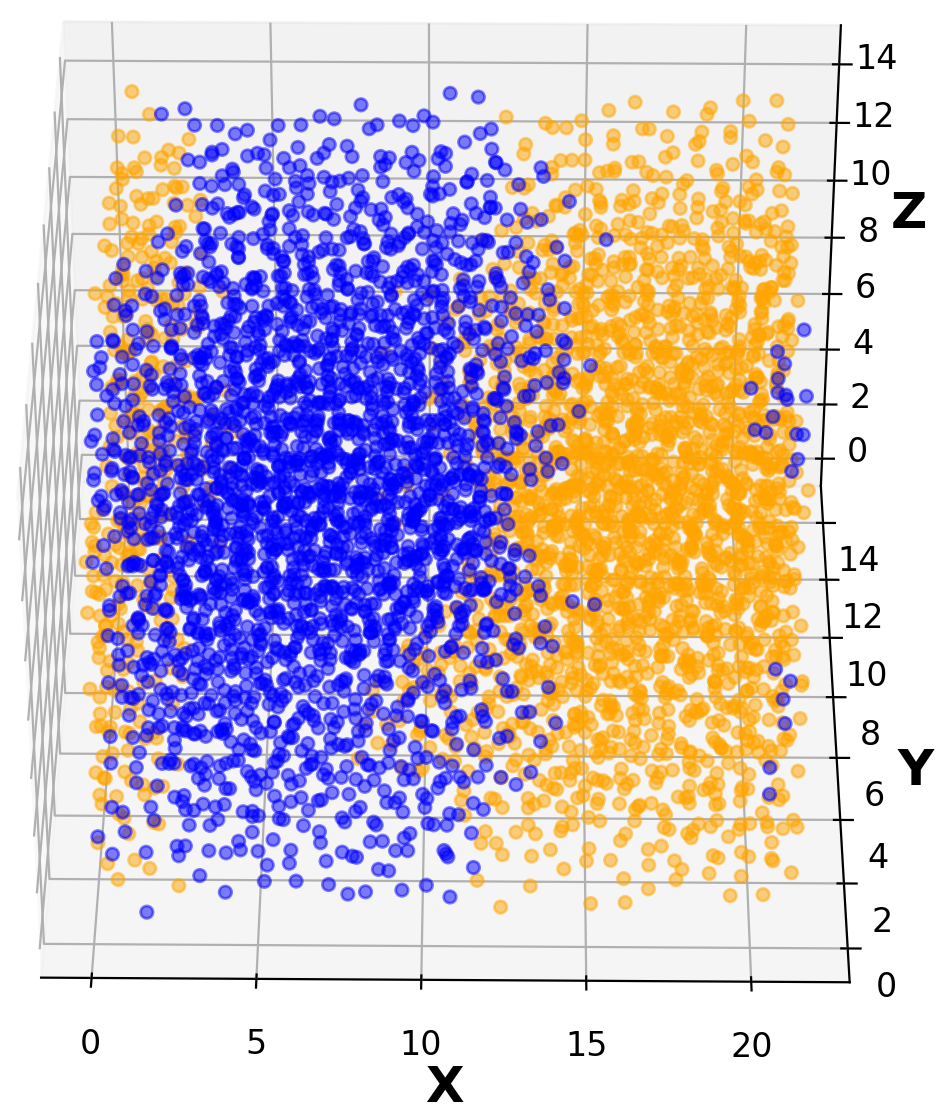}
\caption{$T^*$ = 0.37, Snapshot 2}
   \label{fig:T0.37_20ns}
\end{subfigure}
\hfill
\begin{subfigure}{0.35\textwidth}
\includegraphics[width=2.1in]{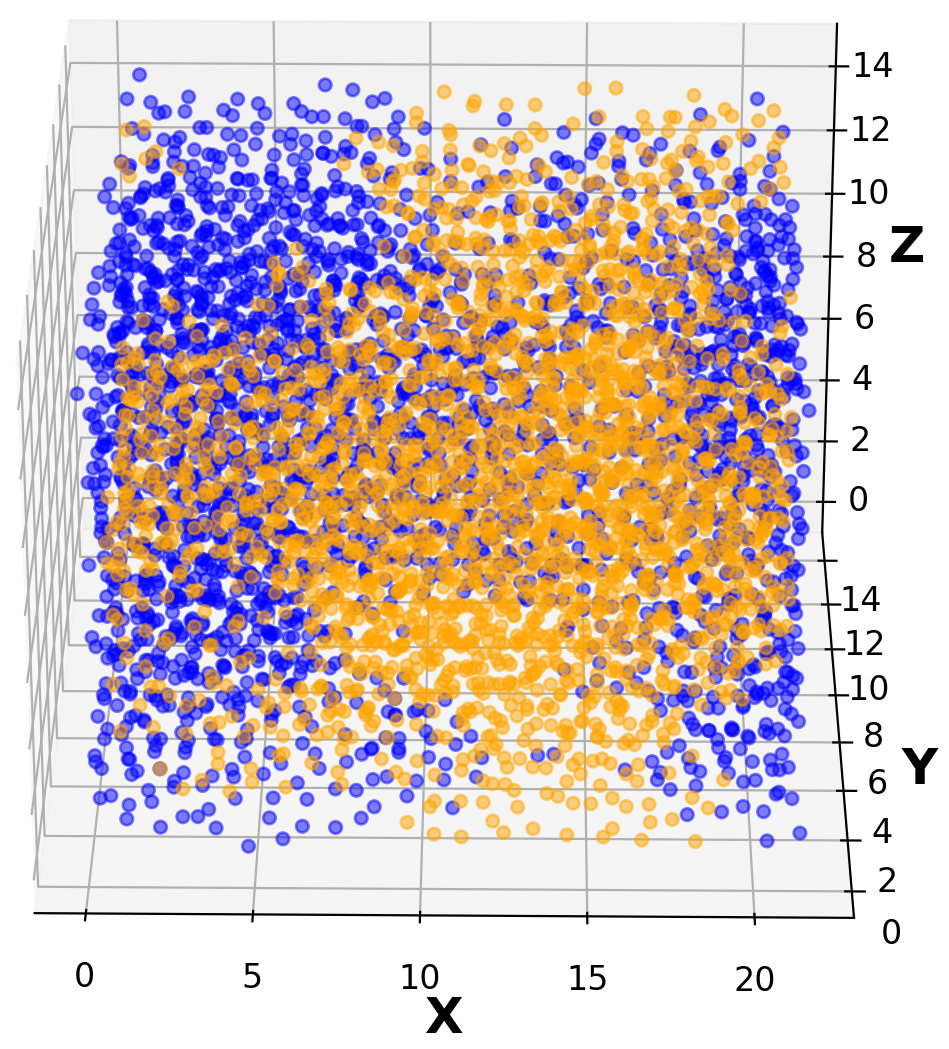}
\caption{$T^*$ = 0.37, Snapshot 3}
   \label{fig:T0.37_30ns}
\end{subfigure}
\hfill
}
\resizebox{\columnwidth}{!}
{
\begin{subfigure}{0.35\textwidth}
\includegraphics[width=2.1in]{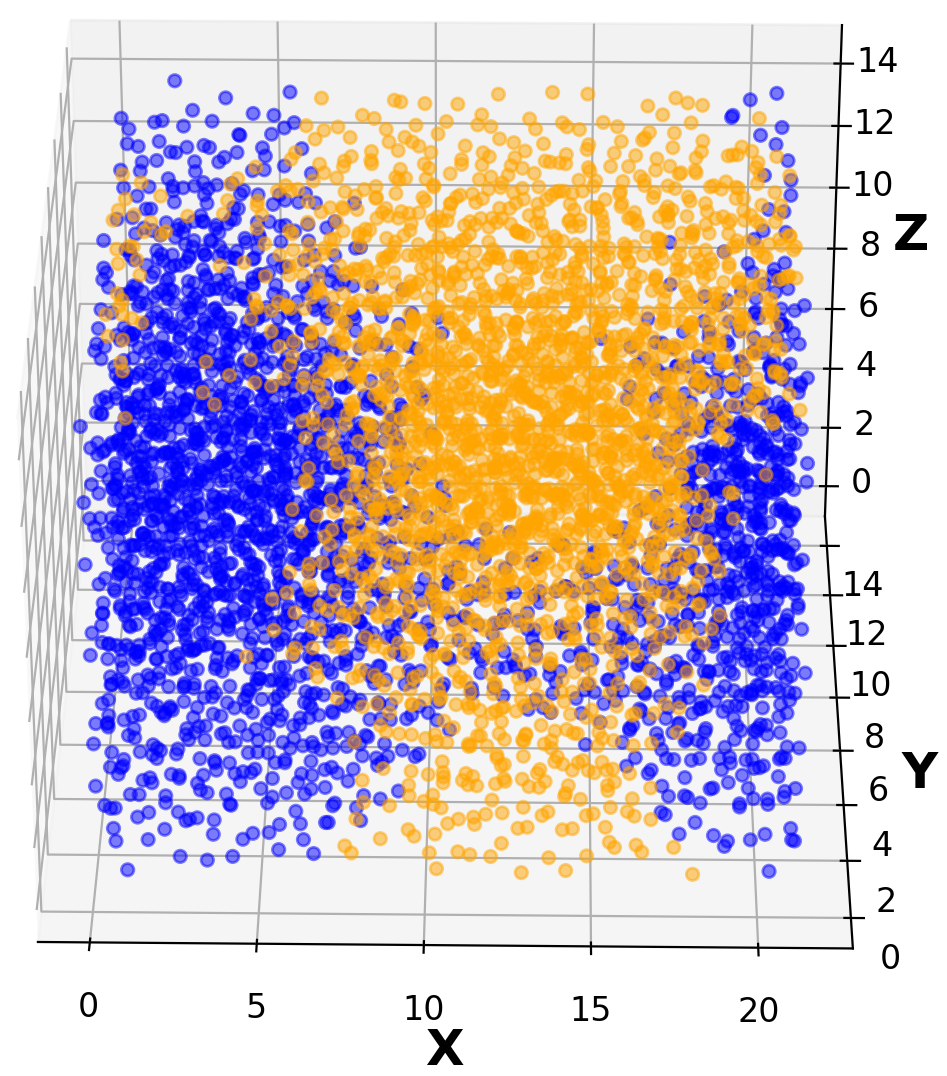}
\caption{$T^*$ = 0.3, Snapshot 1}
   \label{$T^*$ = 0.3,fig:T0.3_0ns}
\end{subfigure}
\hfill
\begin{subfigure}{0.35\textwidth}
\includegraphics[width=2.1in]{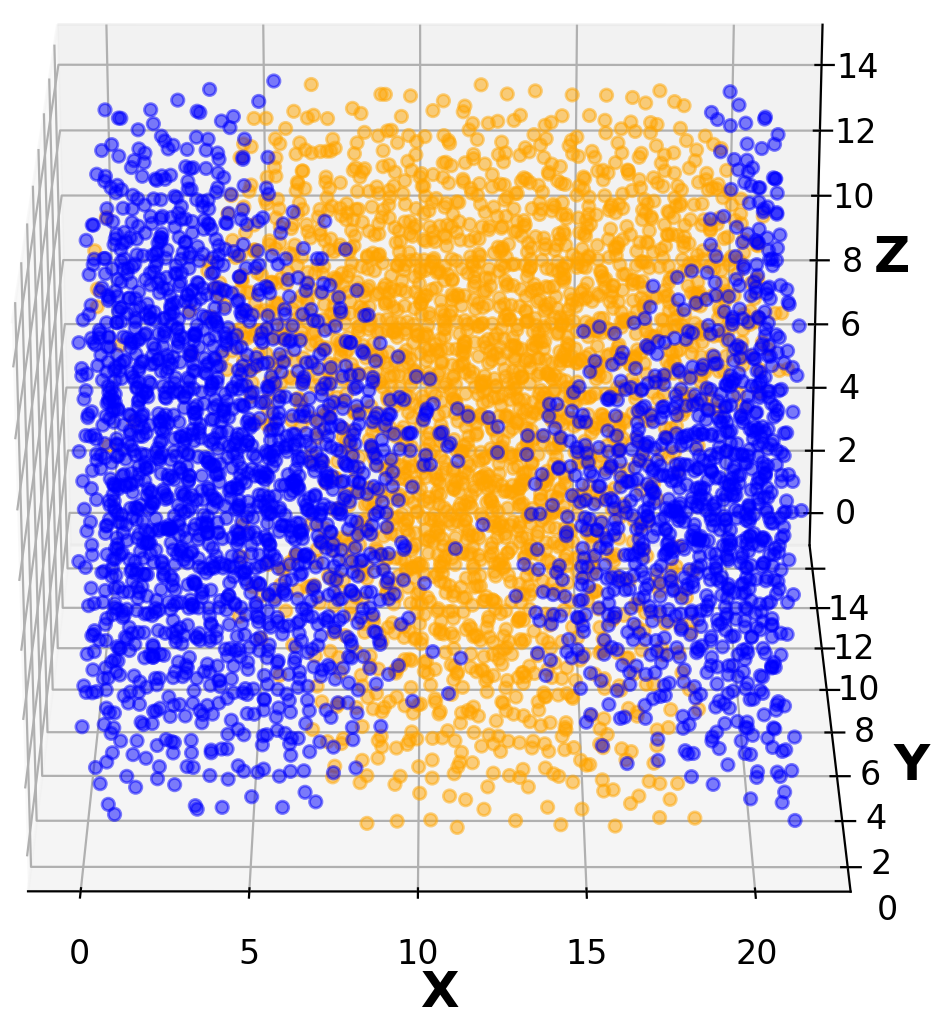}
\caption{$T^*$ = 0.3, Snapshot 2}
   \label{fig:T0.3_10ns}
\end{subfigure}
\hfill
\begin{subfigure}{0.35\textwidth}
\includegraphics[width=2.1in]{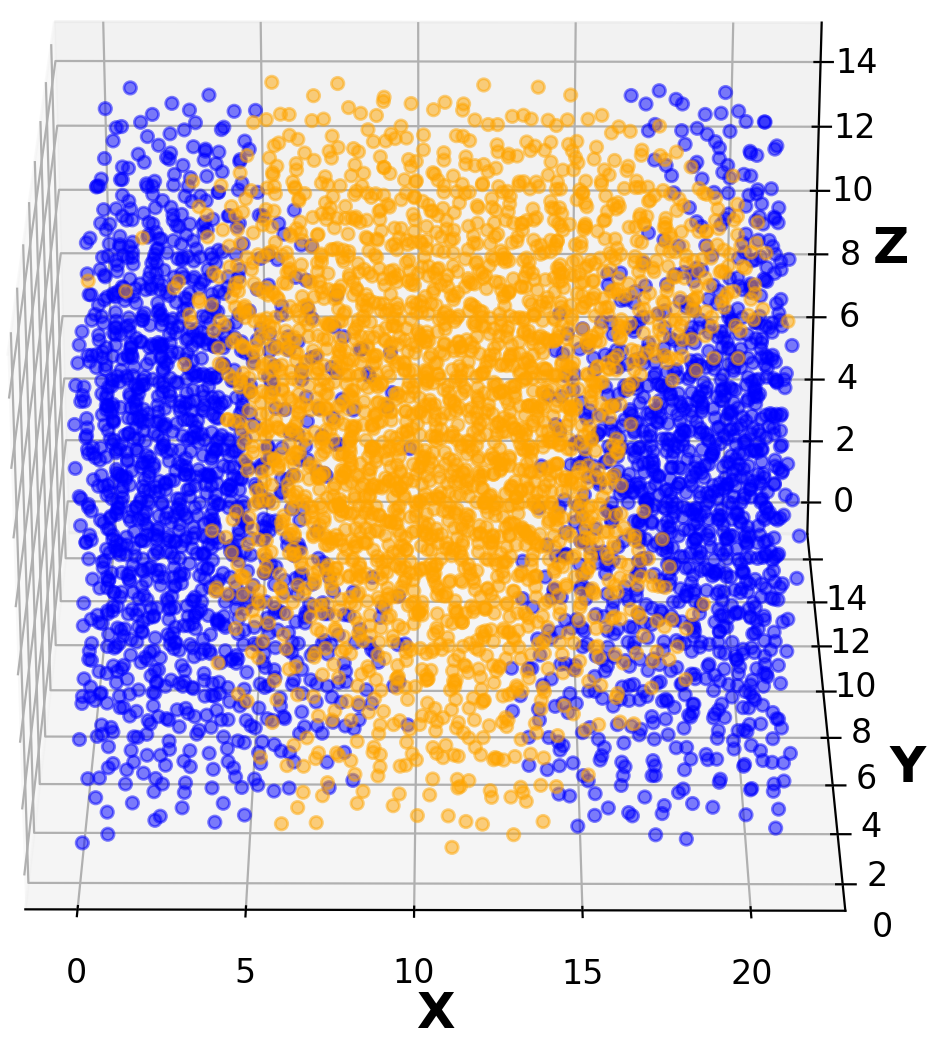}
\caption{$T^*$ = 0.3, Snapshot 3}
   \label{fig:T0.3_20ns}
\end{subfigure}
\hfill
}
\resizebox{\columnwidth}{!}
{
\begin{subfigure}{0.35\textwidth}
\includegraphics[width=2.1in]{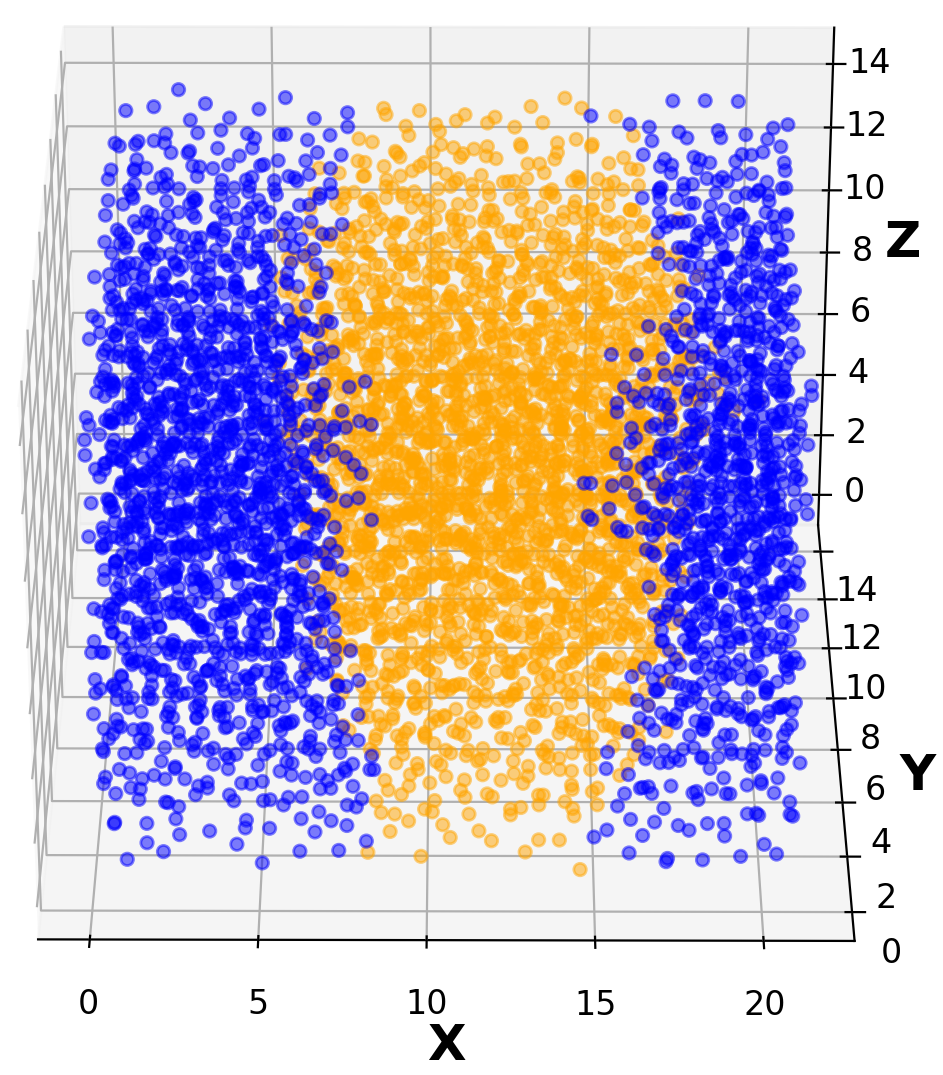}
\caption{$T^*$ = 0.2, Snapshot 1}
   \label{fig:T0.2_0ns}
\end{subfigure}
\hfill
\begin{subfigure}{0.35\textwidth}
\includegraphics[width=2.1in]{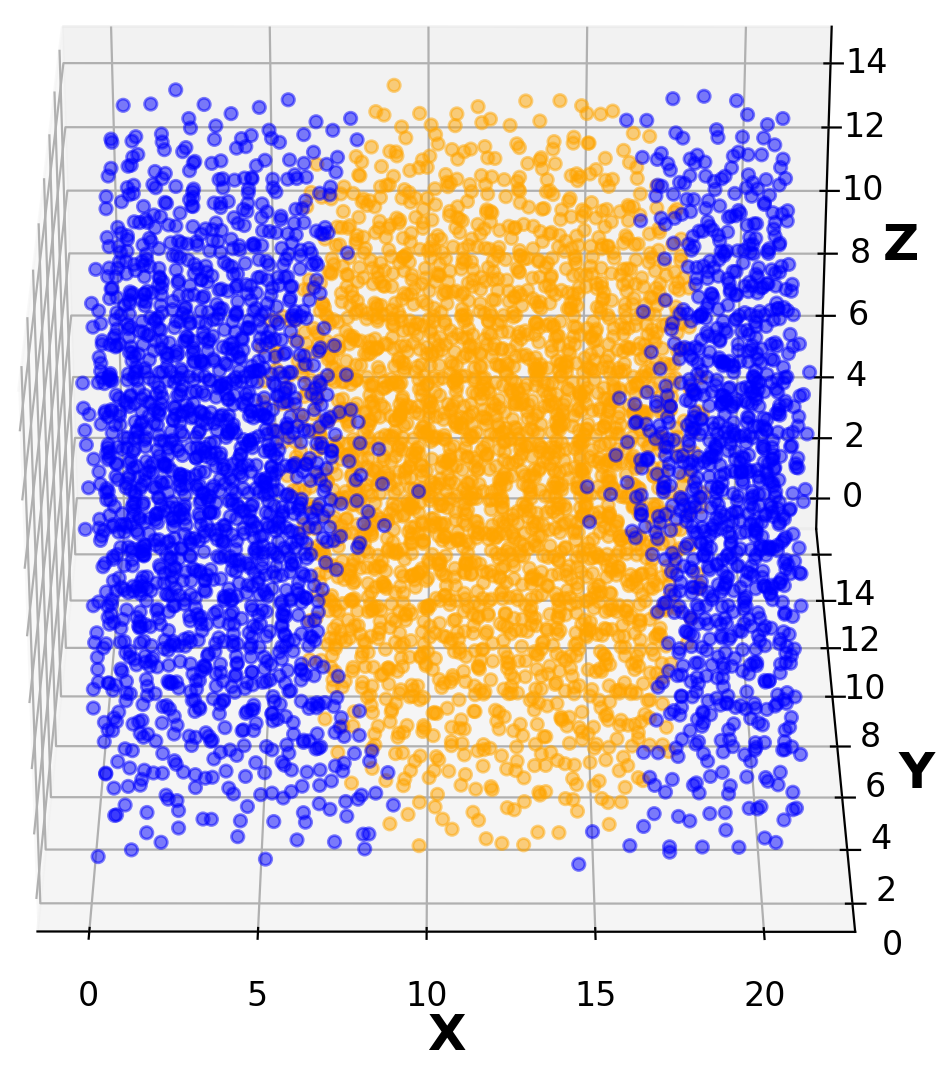}
\caption{$T^*$ = 0.2, Snapshot 2}
   \label{fig:T0.2_10ns}
\end{subfigure}
\hfill
\begin{subfigure}{0.35\textwidth}
\includegraphics[width=2.1in]{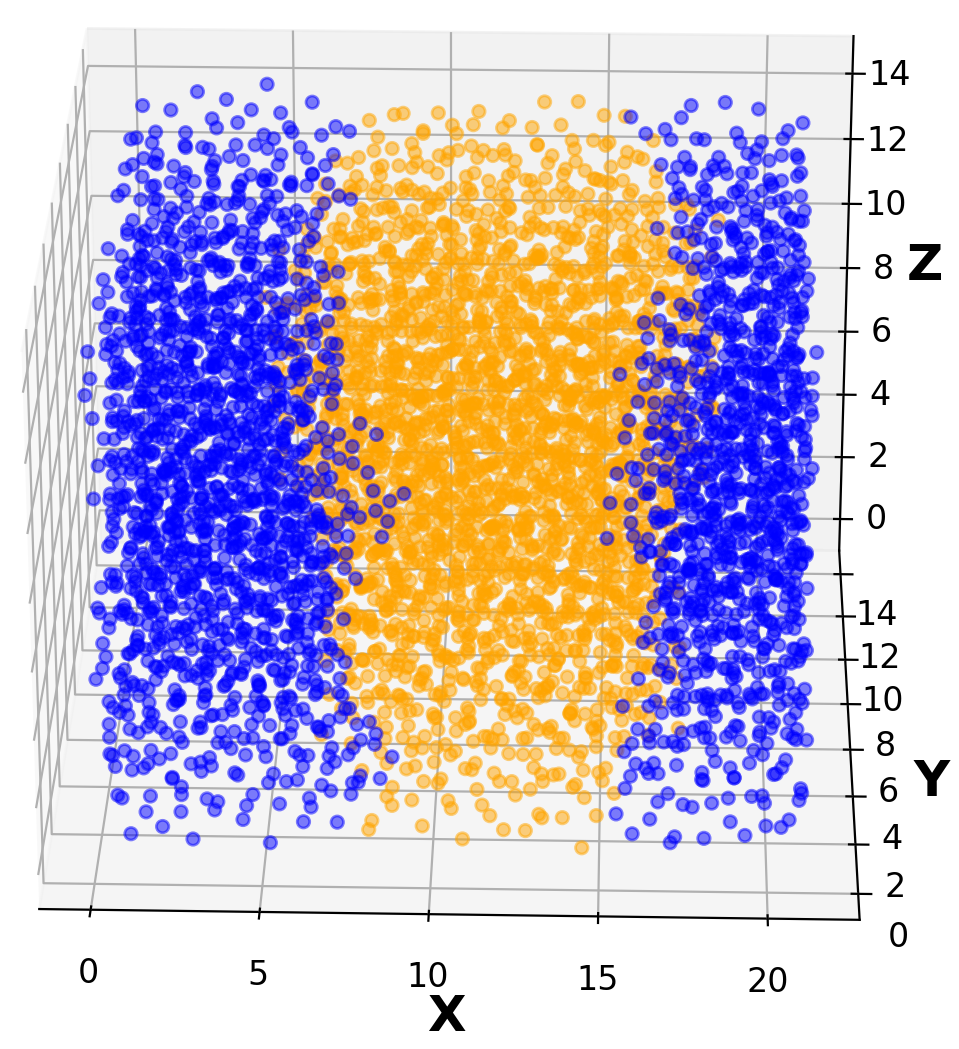}
\caption{$T^*$ = 0.2, Snapshot 3}
   \label{fig:T0.2_20ns}
\end{subfigure}
\hfill
}
\caption{ 3D snapshots of meso-states at different temperatures. The snapshots are at every 10ns lag time.}  
\label{fig:snapshots}
\end{figure}

\begin{figure}
\resizebox{\columnwidth}{!}
{
\begin{subfigure}{0.35\textwidth}
\includegraphics[width=2.1in]{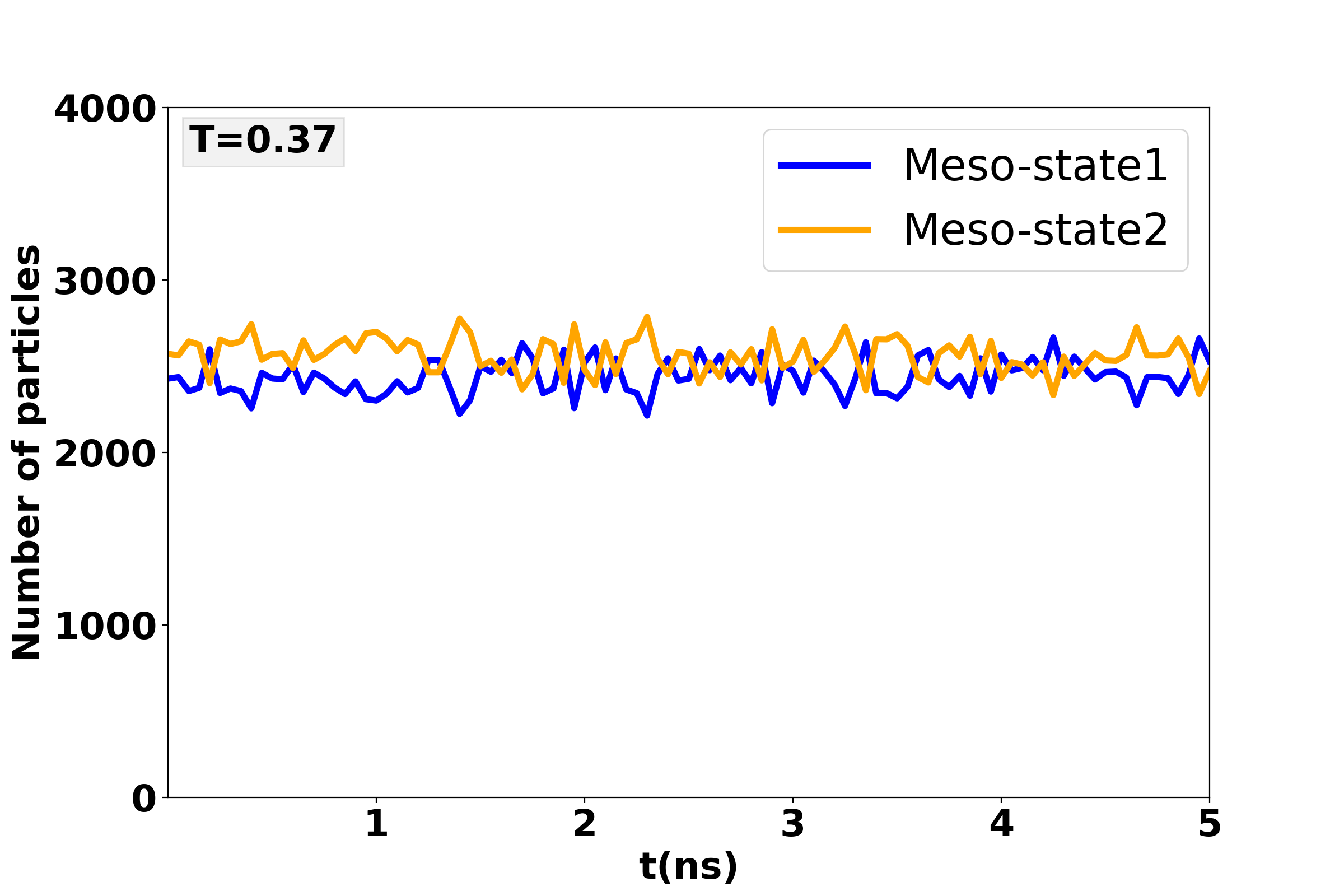}
\caption{$T^*$ = 0.37}
   \label{fig:T0.37_flucs}
\end{subfigure}
\hfill
\begin{subfigure}{0.35\textwidth}
\includegraphics[width=2.1in]{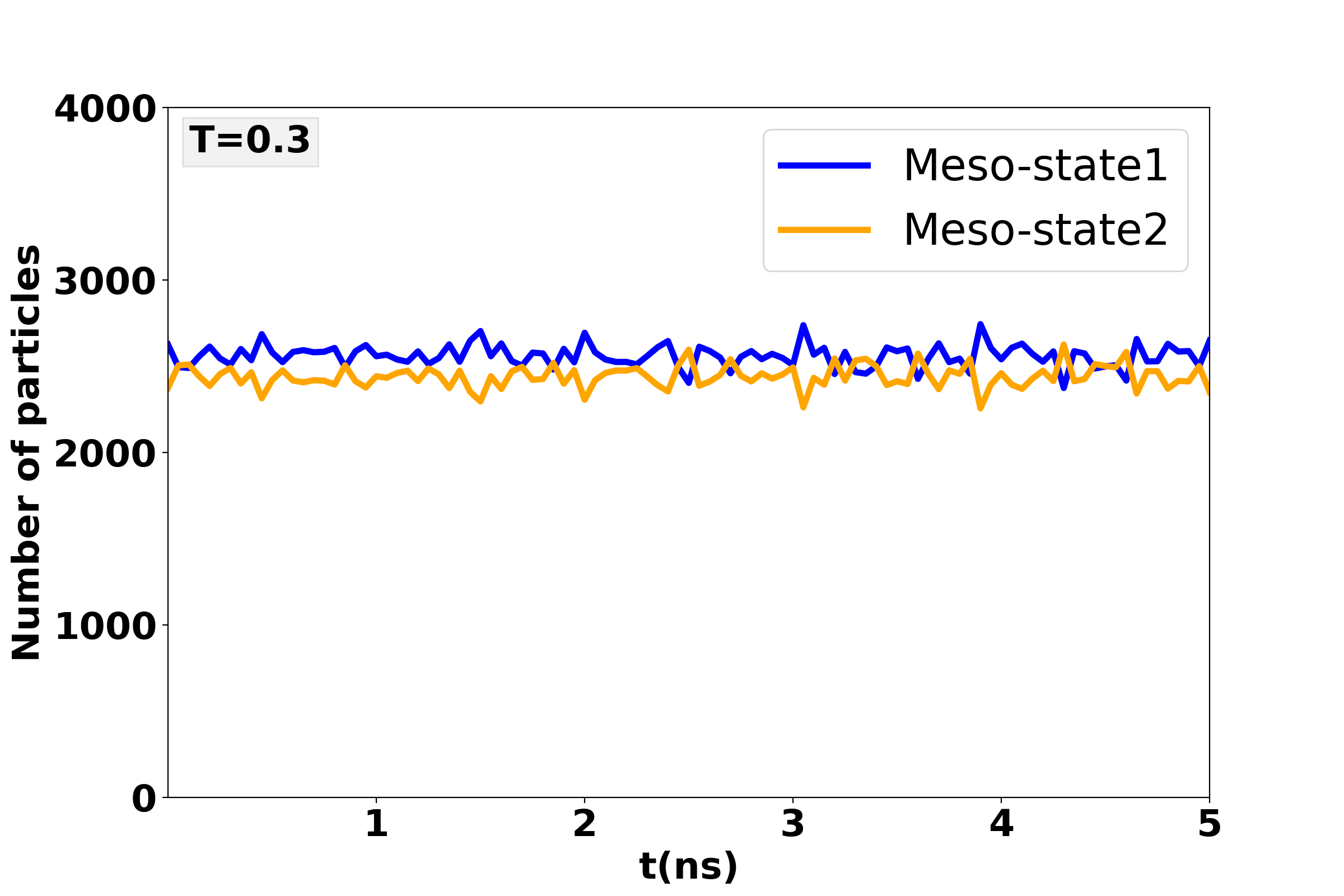}
\caption{$T^*$ = 0.3}
   \label{fig:T0.3_flucs}
\end{subfigure}
\hfill
}
\resizebox{\columnwidth}{!}
{
\begin{subfigure}{0.35\textwidth}
\includegraphics[width=2.1in]{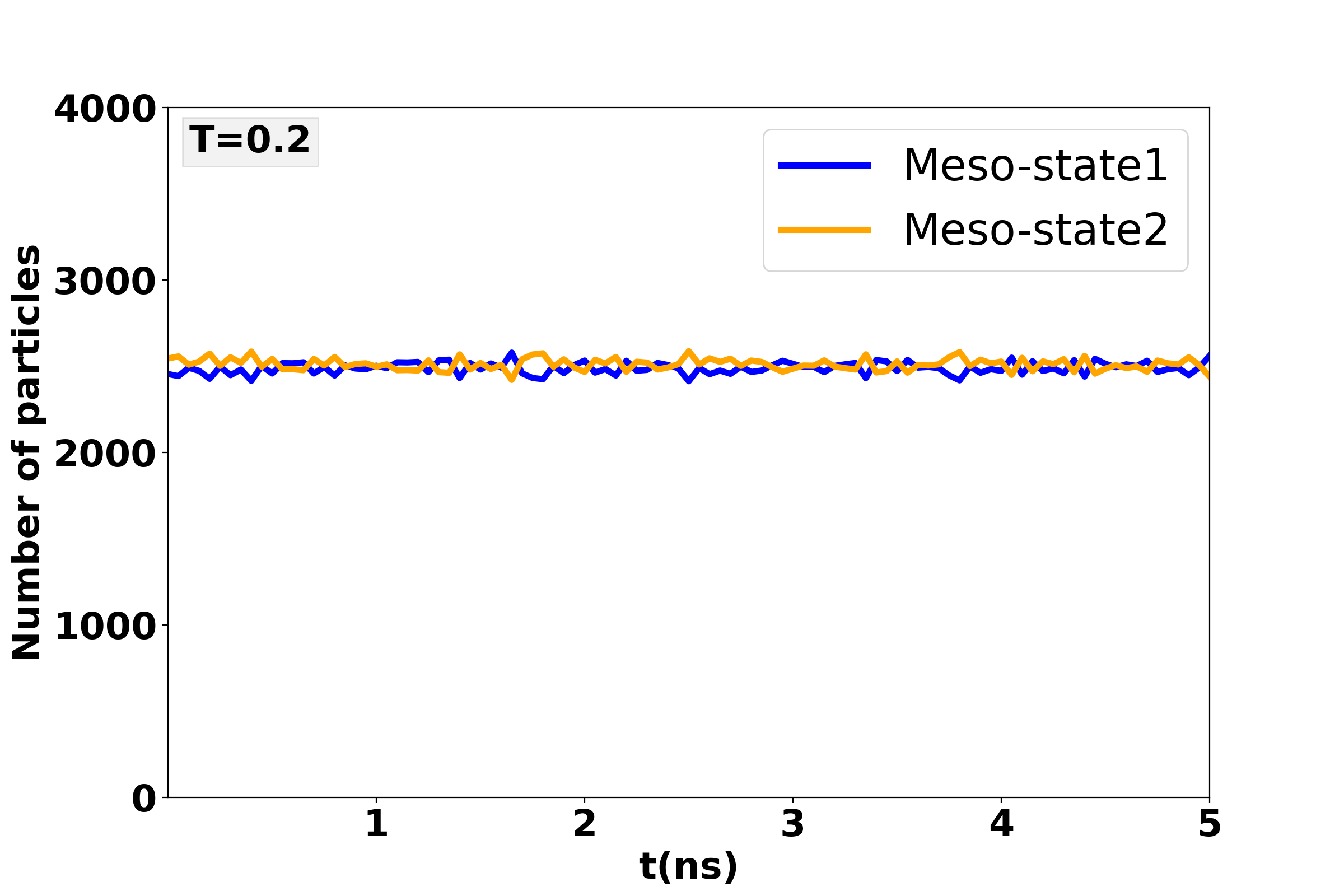}
\caption{$T^*$ = 0.2}
   \label{fig:T0.2_flucs}
\end{subfigure}
\hfill
\begin{subfigure}{0.35\textwidth}
\includegraphics[width=2.1in]{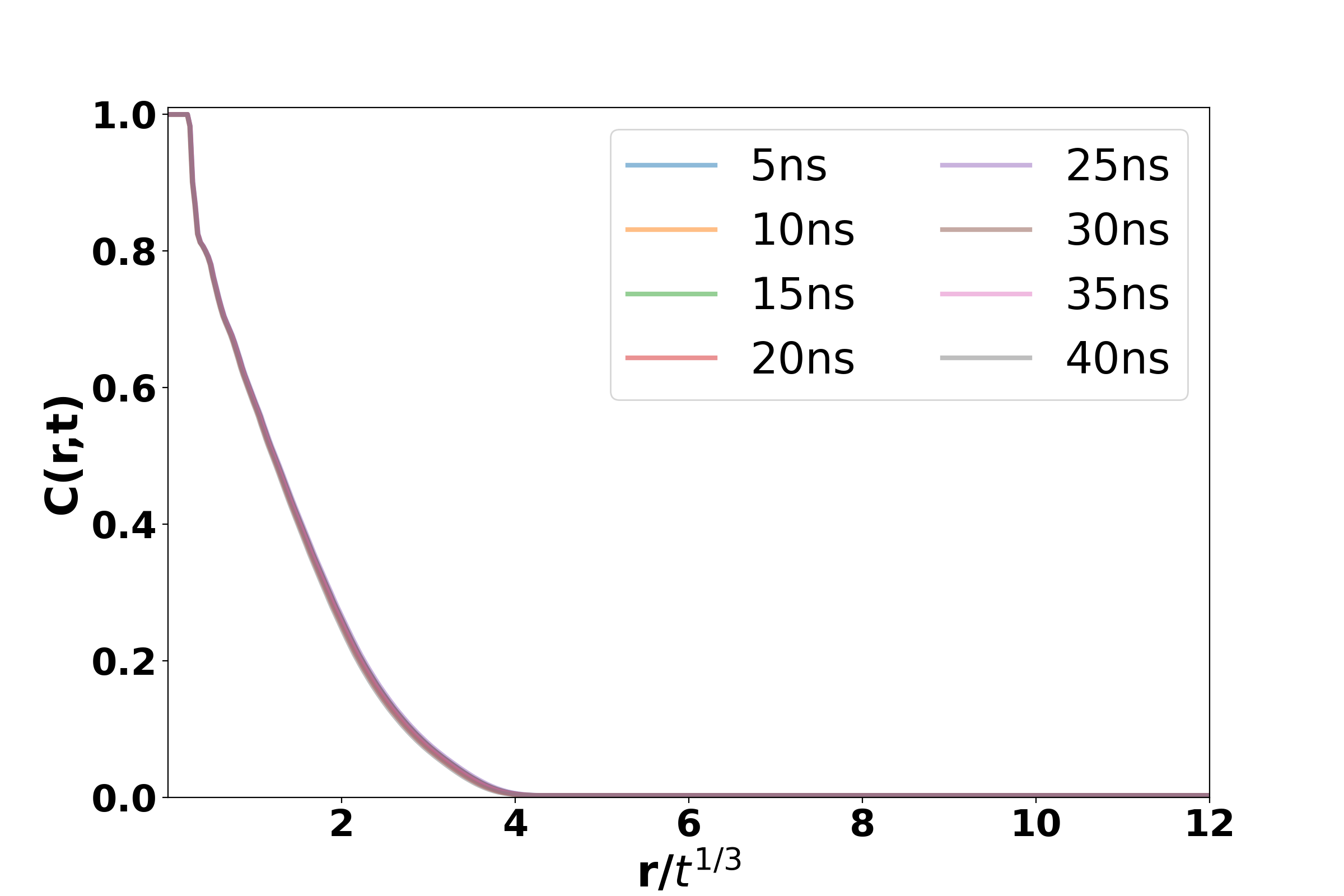}
\caption{$C(r,t)$}
   \label{fig:Crt}
\end{subfigure}
\hfill
}
\caption{ a-c) Particle number fluctuations of meso-states at different temperatures over 5ns;  d) Equal-time correlation $C(r,t)$ at $T^*$ = 0.2}  
\label{fig:fluctuations}
\end{figure}

In the Section \ref{sec:rdf}, the stability of nano-domains is associated with the onset of cage-breaking processes which is reflected as a plateau in the self intermediate coherent function $F(k,t)$ or in diffusion dynamics via MSD(\ref{fig:msd}). The timescale of the cage processes depends on temperature, a quantitative study will interesting, but some qualitative observations can still be made.
Given the glass transition temperature being $T_g^*$ = 0.25 for this system, different configurational snapshots can be used to qualitatively examine timescale of nano-domains. \ref{fig:snapshots} shows three different 10ns lag time snapshots  which are taken at three different temperatures. At $T^*$ = 0.2 which is below $T_g^*$, almost all of particles are immobile and freeze at their local domains, so lifetime of nano-domains are indefinitely long. Meanwhile, as temperature goes up, more particles are able to escape out the cage as illustrated from $T^*$ = 0.3 to $T^*$ = 0.37 in \ref{fig:snapshots}, hence nano-domain shapes are changing relatively quickly and become less static. These phenomena are confirmed by quantitatively evaluating particles fluctuation of the domains as shown in \ref{fig:fluctuations}a-c. The magnitude of particles fluctuations increases as the temperature increases because of higher number of mobile particles. 

Another signature of the liquid-liquid phase separation that follows the quenching from a high temperature equilibrium (normal liquid) state to a super-cooled state is the scaling law of the domain size growth~\cite{Aranson,Binder1975,Humayun}. In this case, the two meso-states are the equilibrium thermodynamic states with domains formed either via 
  spinodal decomposition or nucleation such as shown in \ref{fig: iter3_xyz}~\cite{viet,jayme}.  It is well-established that the growth of characteristic domain size follows an algebraic growth law in time ~\cite{Aranson,Binder1975,Humayun} $L(t) \sim t^{1/3}$ for conserved scalar order parameters (even though the A/B particles are treated as the same in our classficaiton, but the growth dynamics still follows the conserved order parameter scaling law as the real dynamics is still constrained by the swapping of  A/B identity) ~\cite{A.J.Bray,MazenkoGeneF,Mazenko,FongLiu,Bray} which can be tested by the calculation of equal-time correlation function $C(r,t)$ from our classification. Considering a scalar order parameter $\psi$, the equal-time correlation is: $C(r,t) = N^{-1}\sum_{i} \psi_i(t)\psi_{i+r}(t)$ where N is the number of particles, $i +r$ indicates a neighboring particle displaced by a distance $r$ relative to the reference particle $i$ with $\psi_i$ = +1 for particles identities of state 1 and $\psi_{i}$ = -1 for particles identities of state 2, hence the product of $\psi_i(t)\psi_{i+r}(t)$ will be +1 between pair of particles from the same state and will be -1 otherwise. Since the domain identities for each particle are known from the classification scheme, the result of $C(r,t)$ indeed confirms the scaling law of domains growth $L(t) \sim t^{1/3}$ as shown in \ref{fig:Crt}.

\begin{figure}
\resizebox{\columnwidth}{!}
{
\begin{subfigure}{0.35\textwidth}
\includegraphics[width=2.1in]{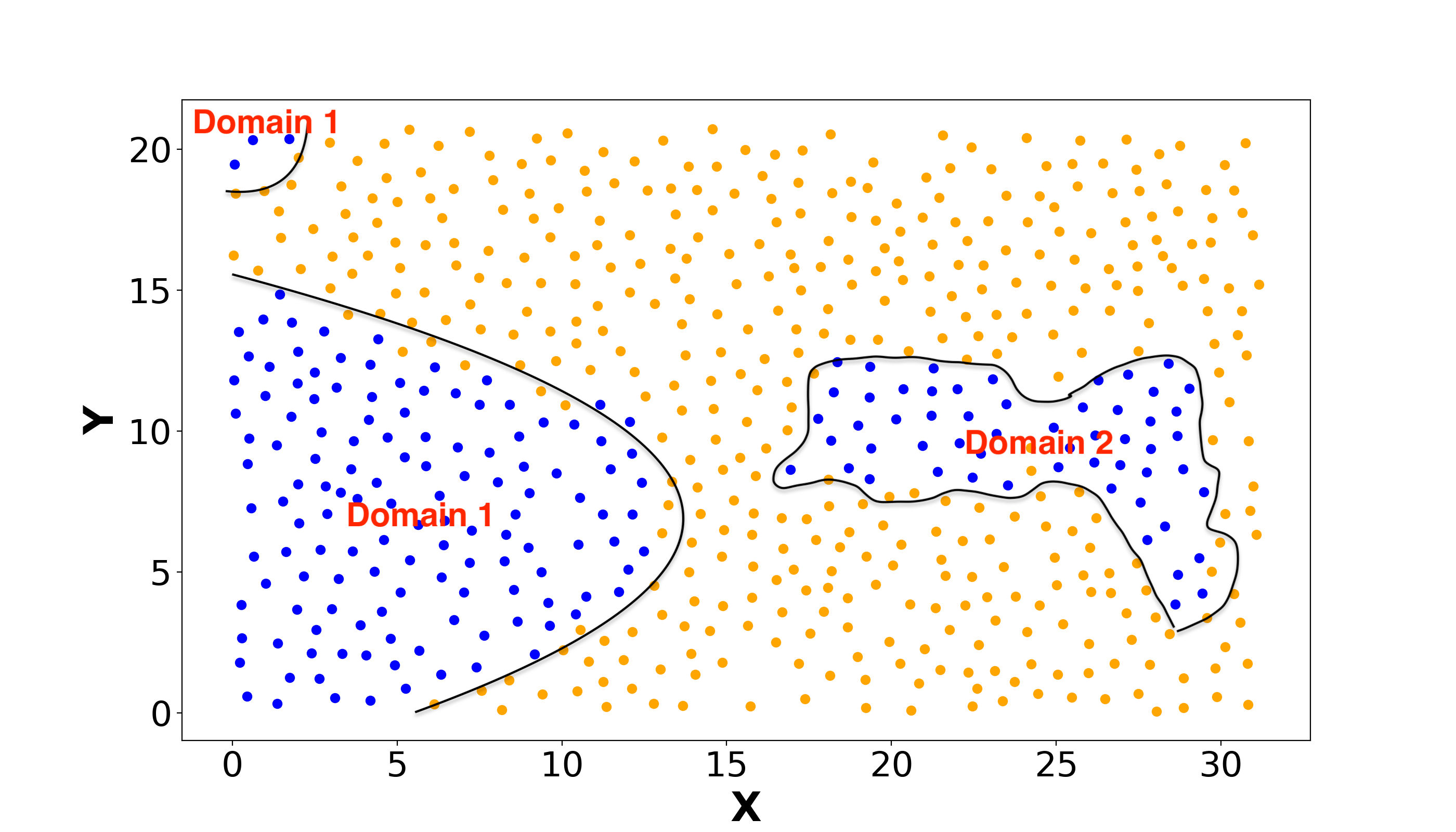}
\caption{Snapshot 1, Cross section 1}
   \label{fig:26ns_cs2}
\end{subfigure}
\hfill
\begin{subfigure}{0.35\textwidth}
\includegraphics[width=2.1in]{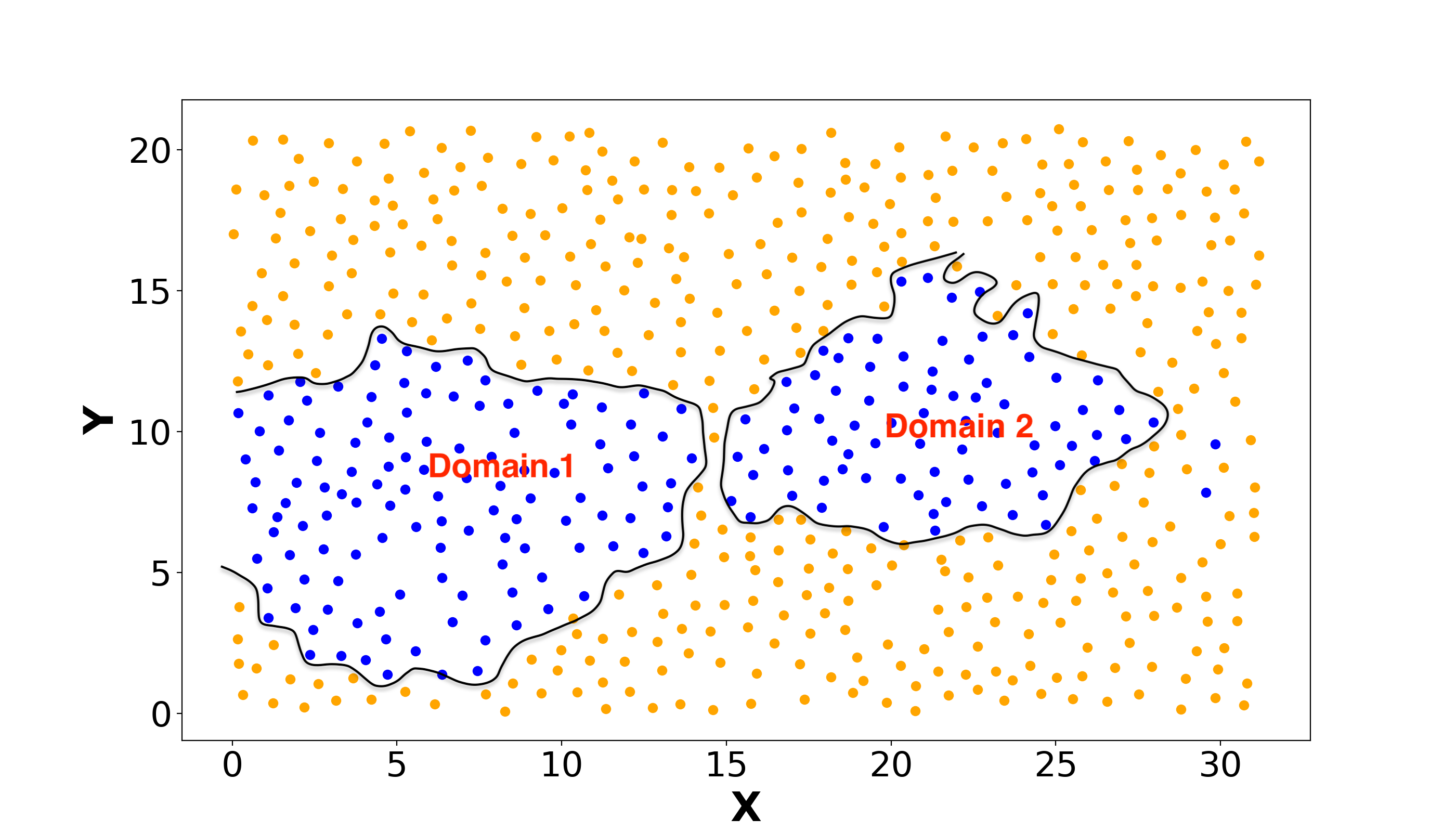}
\caption{Snapshot 2, Cross section 1}
   \label{fig:28ns_cs2}
\end{subfigure}
\hfill
}
\resizebox{\columnwidth}{!}
{
\begin{subfigure}{0.35\textwidth}
\includegraphics[width=2.1in]{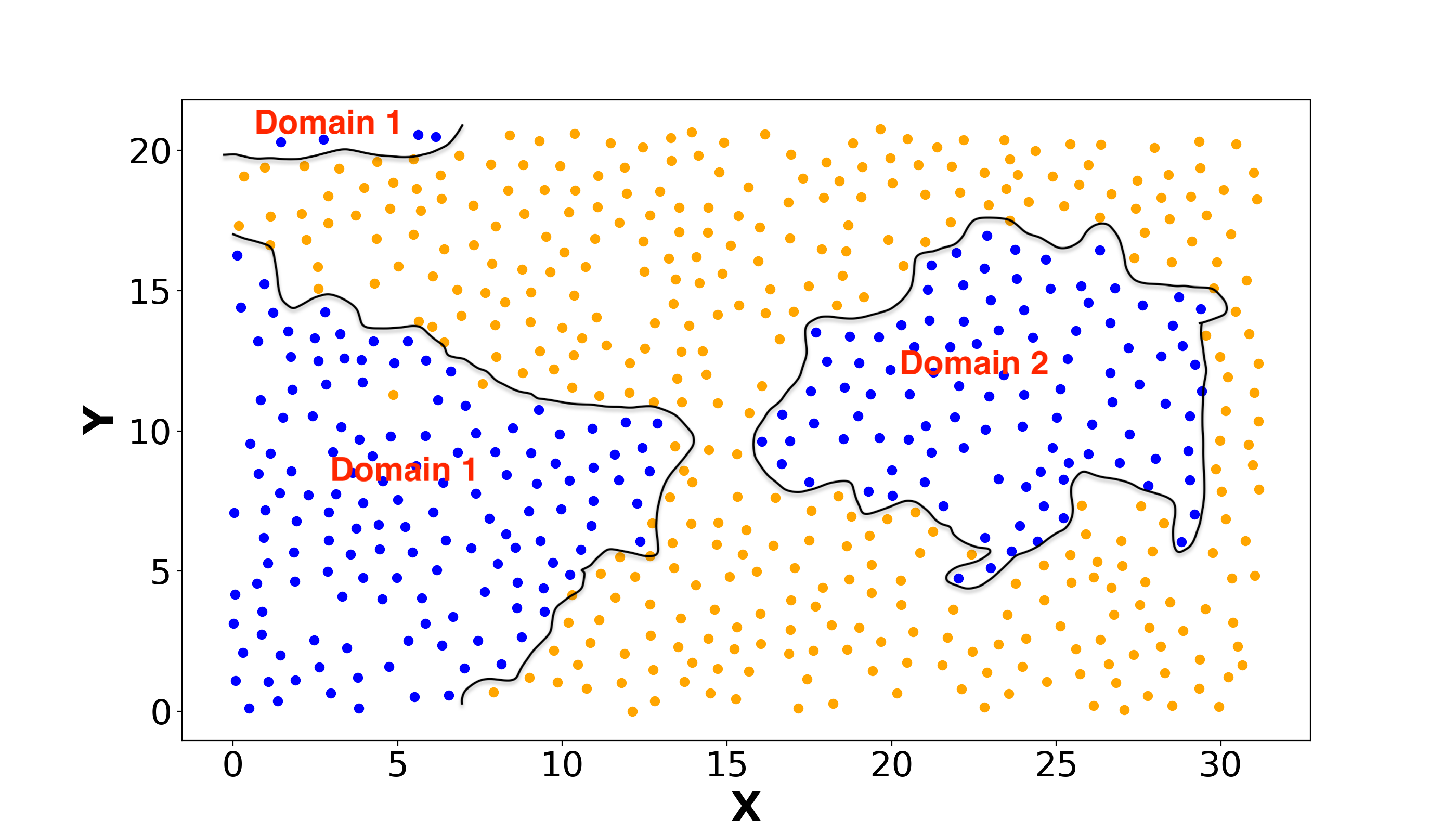}
\caption{Snapshot 1, Cross section 2}
   \label{fig:26ns_cs7}
\end{subfigure}
\hfill
\begin{subfigure}{0.35\textwidth}
\includegraphics[width=2.1in]{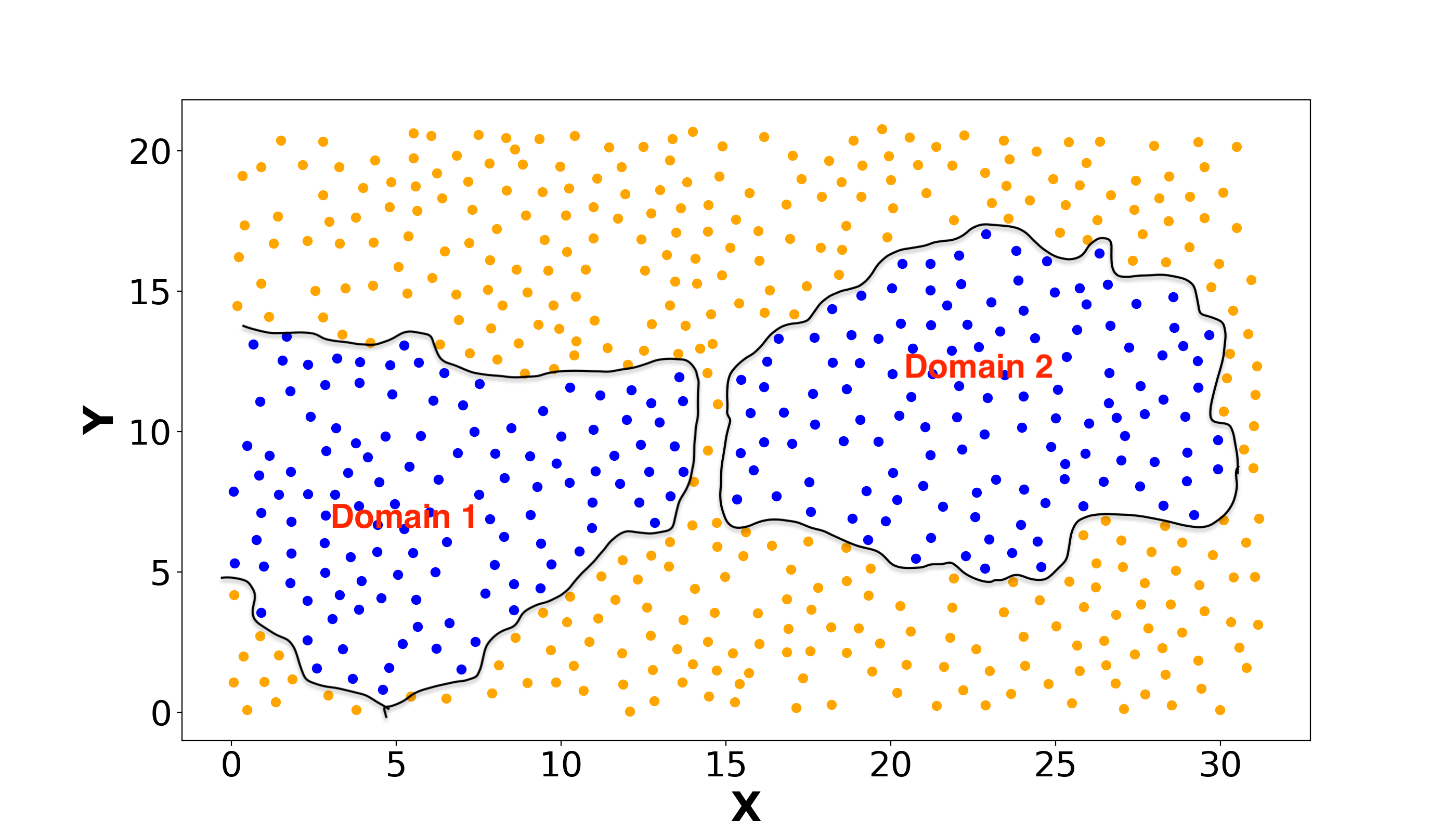}
\caption{Snapshot 2, , Cross section 2}
   \label{fig:28ns_cs7}
\end{subfigure}
\hfill
}
\caption{ 2D representation of bifurcated domains for meso-states of 16000 particles system at $T^*=0.2$, Panels a,c and b,d are two different cross section along Z-direction of two different snapshots. The two snapshots are at 2ns lag time. The green lines are an guide to the eye for the boundaries among bifurcated domains and the numbers represent the various domains of a specific meso-state.}  
\label{fig:finite_size}
\end{figure}

\begin{figure}
\resizebox{\columnwidth}{!}
{
\begin{subfigure}{0.35\textwidth}
\includegraphics[width=2.1in]{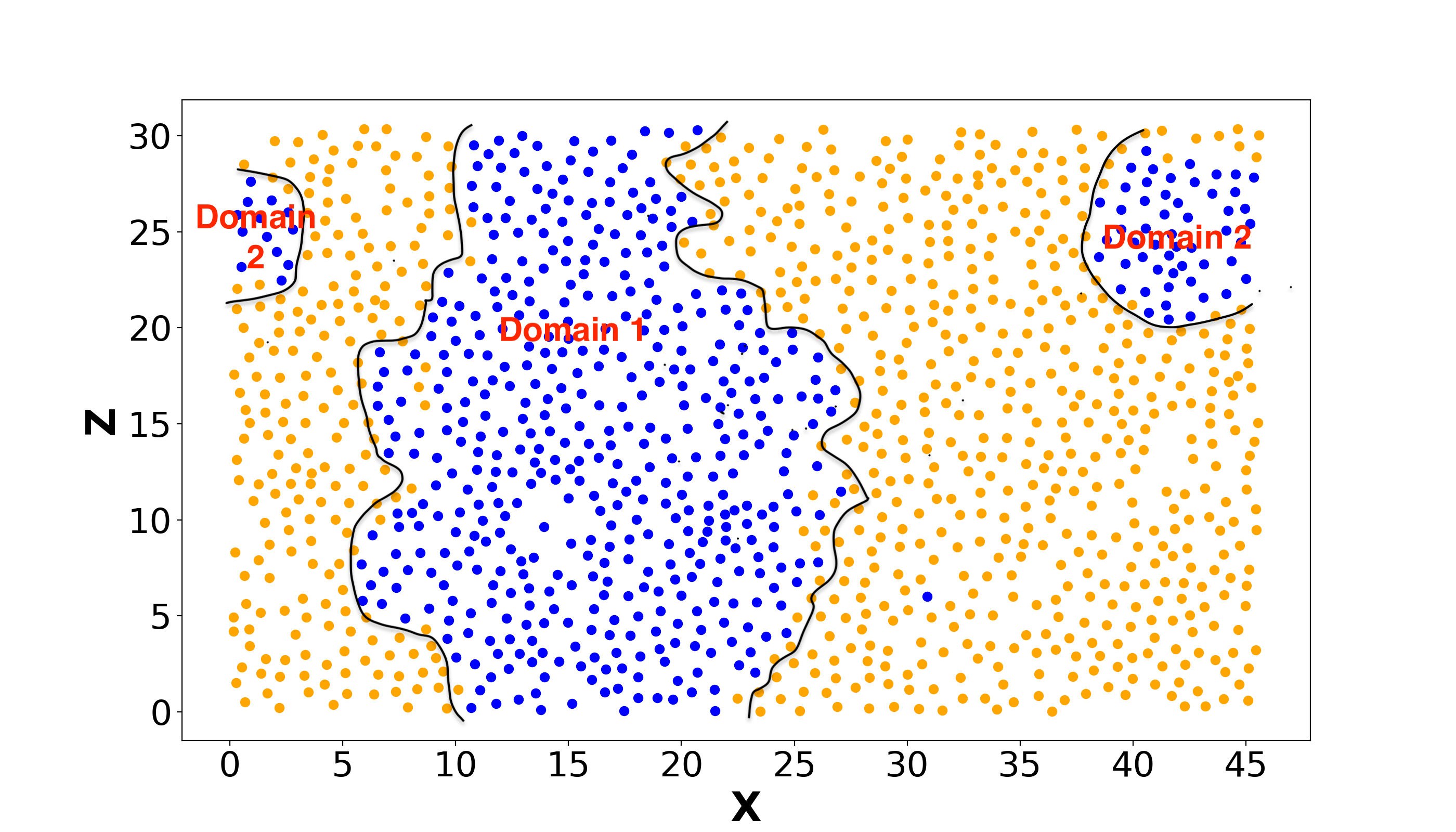}
\caption{Cross section 1}
   \label{fig:50000_cs0}
\end{subfigure}
\hfill
\begin{subfigure}{0.35\textwidth}
\includegraphics[width=2.1in]{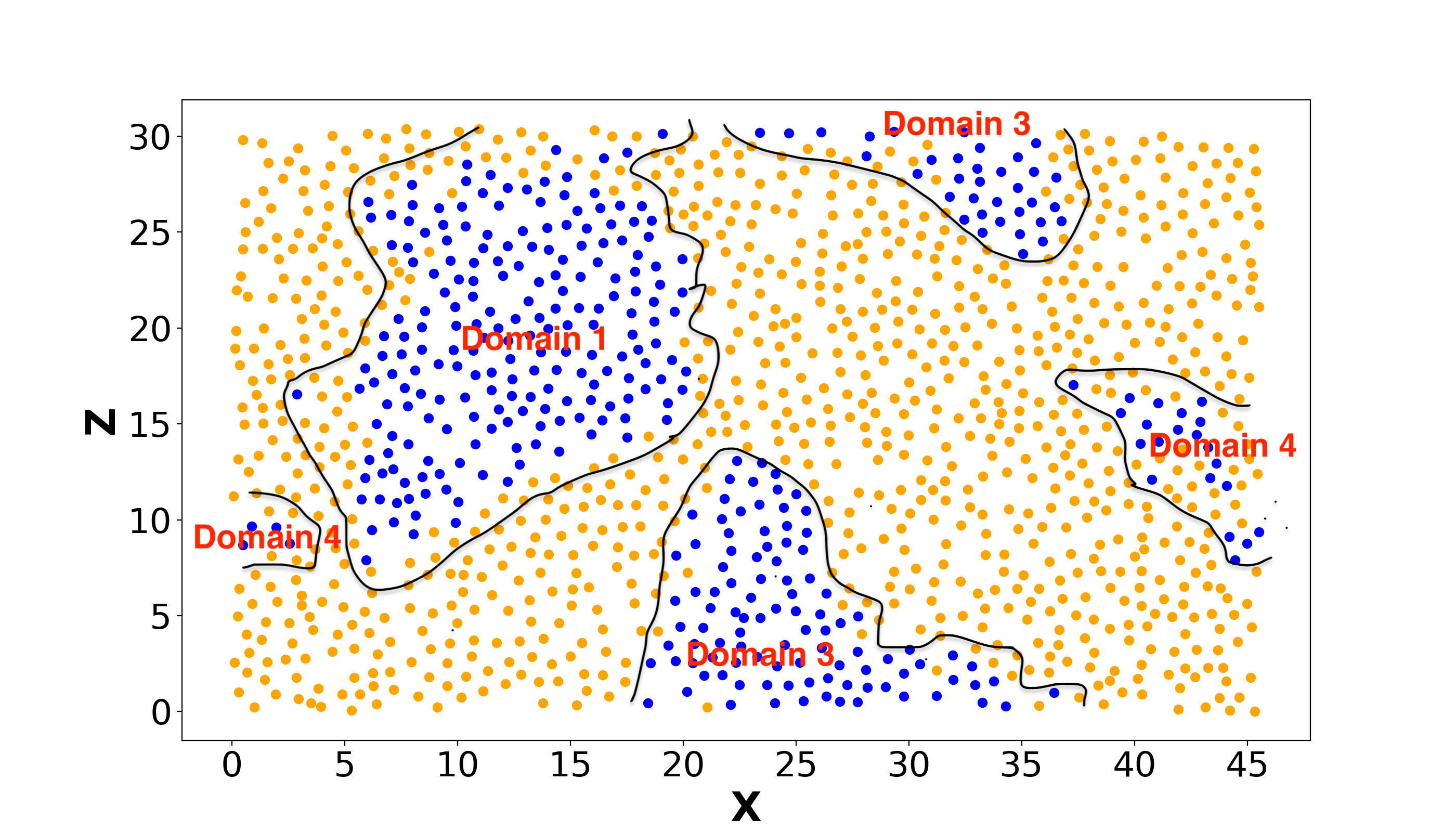}
\caption{Cross section 2}
   \label{fig:50000_cs12}
\end{subfigure}
\hfill
}
\resizebox{\columnwidth}{!}
{
\begin{subfigure}{0.35\textwidth}
\includegraphics[width=2.1in]{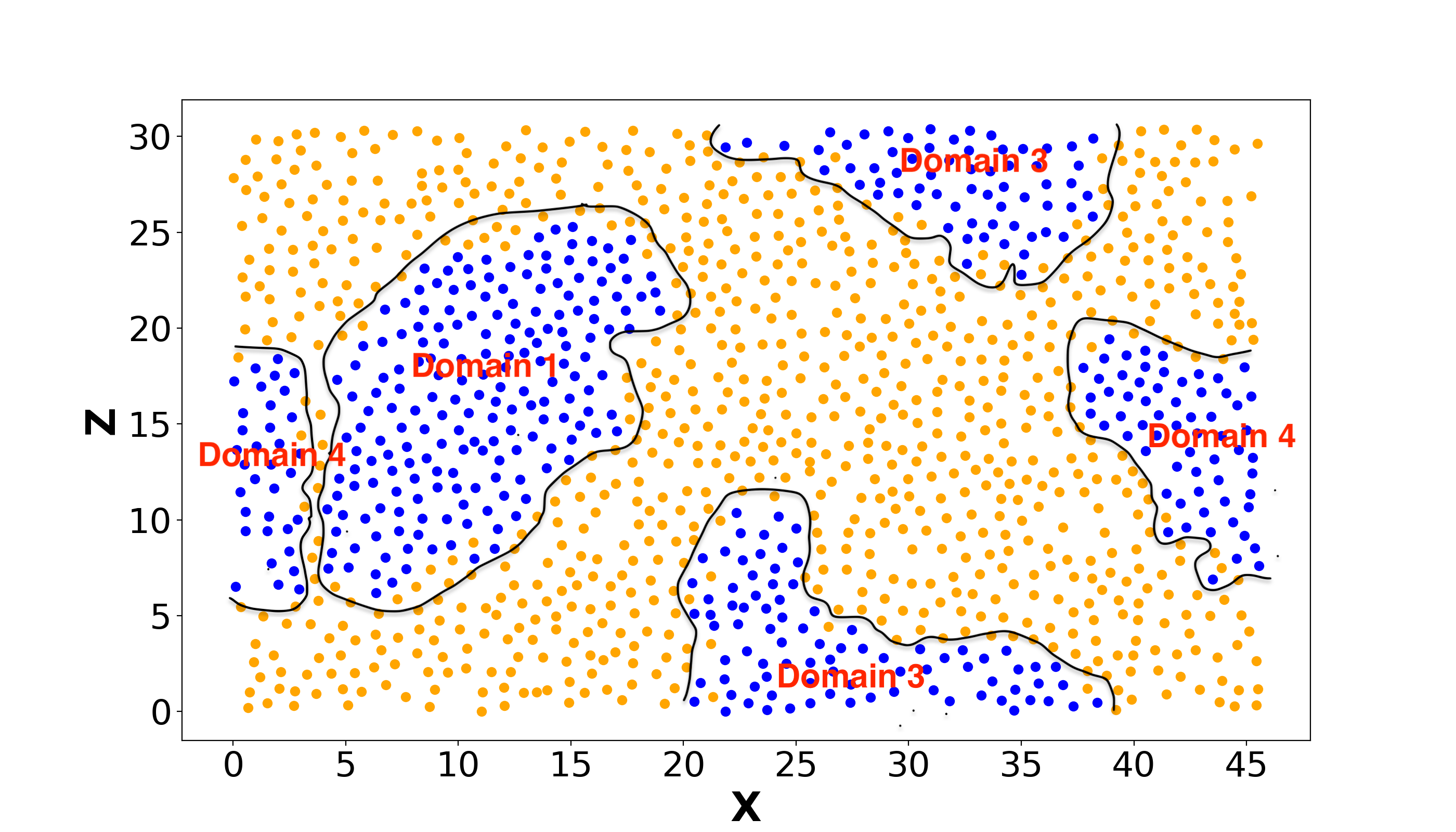}
\caption{Cross section 3}
   \label{fig:50000_cs15}
\end{subfigure}
\hfill
\begin{subfigure}{0.35\textwidth}
\includegraphics[width=2.1in]{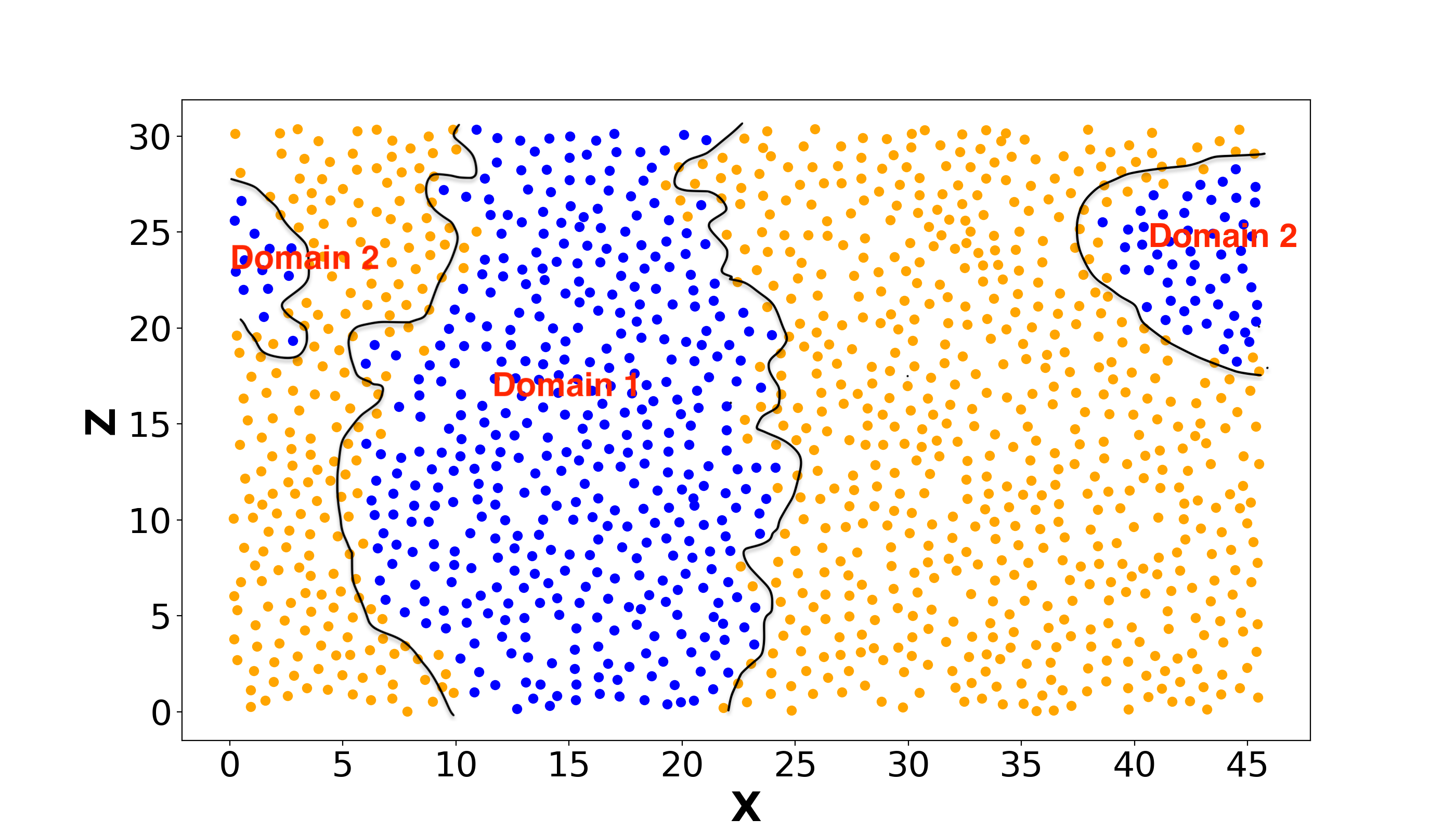}
\caption{Cross section 4 }
   \label{fig:50000_cs37}
\end{subfigure}
}
\caption{2D cross sections along Y-axis to represent bifurcated domains of meso-states of 50000 particles system at $T^*=0.2$. The green lines are an guide to the eye for the boundaries among bifurcated domains and the numbers represent the various domains of a specific meso-state. }  
\label{fig:50000_cs}
\end{figure}

Finally, similar to the 2D case~\cite{viet}, the number of domains belonging to the same meso-state will increase to tile up the whole system  as the system size increases. However, 3D domains are hard to visualize, so their spatial structures are illustrated by taking different 2D cross sections at different configurations. \ref{fig:26ns_cs2} and \ref{fig:26ns_cs7} shows two different cross sections of two domains that belong to the same meso-state1. These two domains are  also shown by taking a second snapshot at 2ns later in \ref{fig:28ns_cs2} and \ref{fig:26ns_cs7}. As the system size continues to increase, the meso-state1 will split into more domains as shown in \ref{fig:50000_cs} for 50000 particles system. It should be emphasized that the structure of a domain might disappear in some regions of the space along different cross sections as happened to the bifurcated domain 2,3,4 to illustrate the finiteness of each domain in 3D configurational space.

\section{Concluding Remarks}

ML methods are used to develop a scheme to identify spatially co-existing  meso-states or nano-domains both structurally and configurationally.  The physical interpretation of these meso-states are explicitly demonstrated by the observation of bimodality of  $\overline{WCNs}$ distribution along each solvation shell, the corresponding construction of weighted partial $\it g(r)$; the formation of interfaces from calculations of the pressure and density profiles. Given the classification, heterogeneous dynamics of these nano-domains are captured by the difference in the collective distribution of diffusion constants; spatial characterization of these nano-domains is used to evaluate their lifetimes to understand of cage effect for longer relaxation dynamics. Furthermore, kinetic domain growth scaling law calculation presents a direct evidence to indicate that such domains are the result of liquid-liquid phase separation when the system is at supercooled condition from quenching. 

Using the classification scheme developed in this report, the L-L phase separation behaviors can be studied in details. The observed domain structures provide a natural molecular realization of the Adam-Gibbs' Cooperative Rearranging Regions or the mosaic picture of ROFT. These domain structures naturally lead to two types of relaxation dynamics, 
 the intra-domain relaxation is largely due to diffusion inside a domain and the inter-domain relaxation which is related to the coarsening kinetics of the first-order phase transitions.
Therefore, the classification scheme provides a platform for further extensive statistical mechanics analysis of supercooled liquids.

\section{Acknowledgement}

This work is supported by the Division of Chemical and Biological Sciences, Office of Basic Energy Sciences, U.S. Department of Energy, under Contact No. DE-AC02-07CH11358 with Iowa State University. 

\appendix

\section{Selection of K = 2} \label{sec:kmeans}
\renewcommand{\thefigure}{A\arabic{figure}}
\setcounter{figure}{0}

\begin{figure}[h!] 
\resizebox{\columnwidth}{!}
{
\begin{subfigure}{0.35\textwidth}
\includegraphics[width=1.5in]{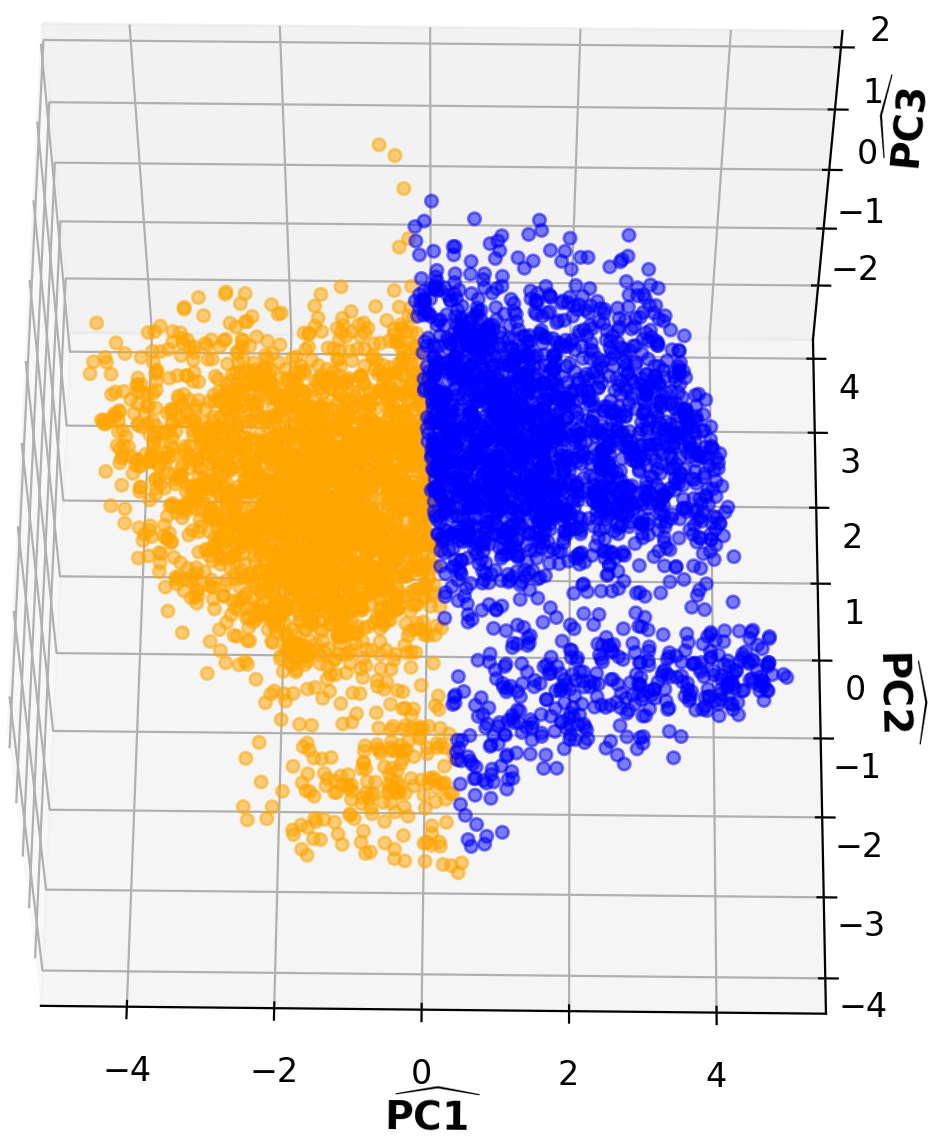}
\caption{K-means, K = 2}
   \label{fig:pca_dmk2}
\end{subfigure}
\hfill
\begin{subfigure}{0.35\textwidth}
\includegraphics[width=1.5in]{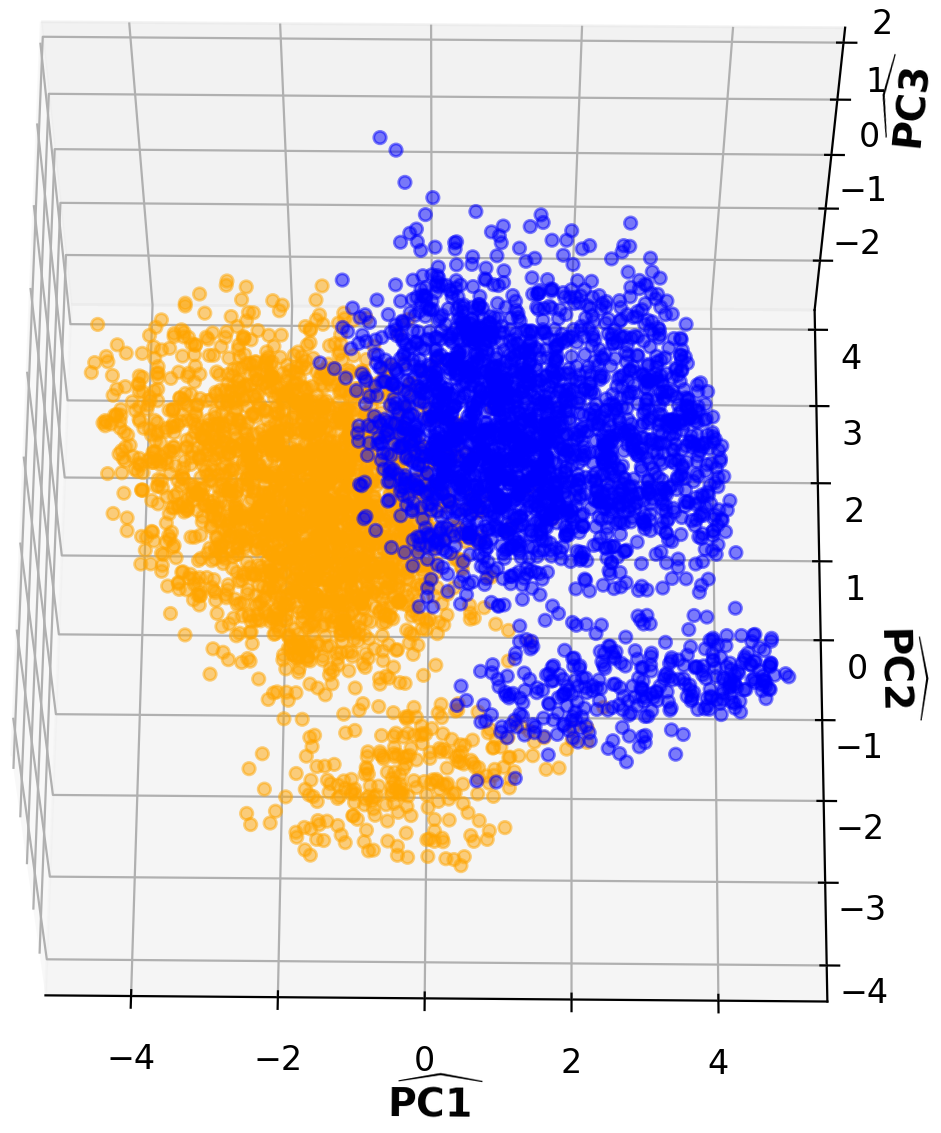}
\caption{Converged iteration, K = 2}
   \label{fig:pca_ck2}
\end{subfigure}
\hfill
}
\resizebox{\columnwidth}{!}
{
\begin{subfigure}{0.35\textwidth}
\includegraphics[width=1.5in]{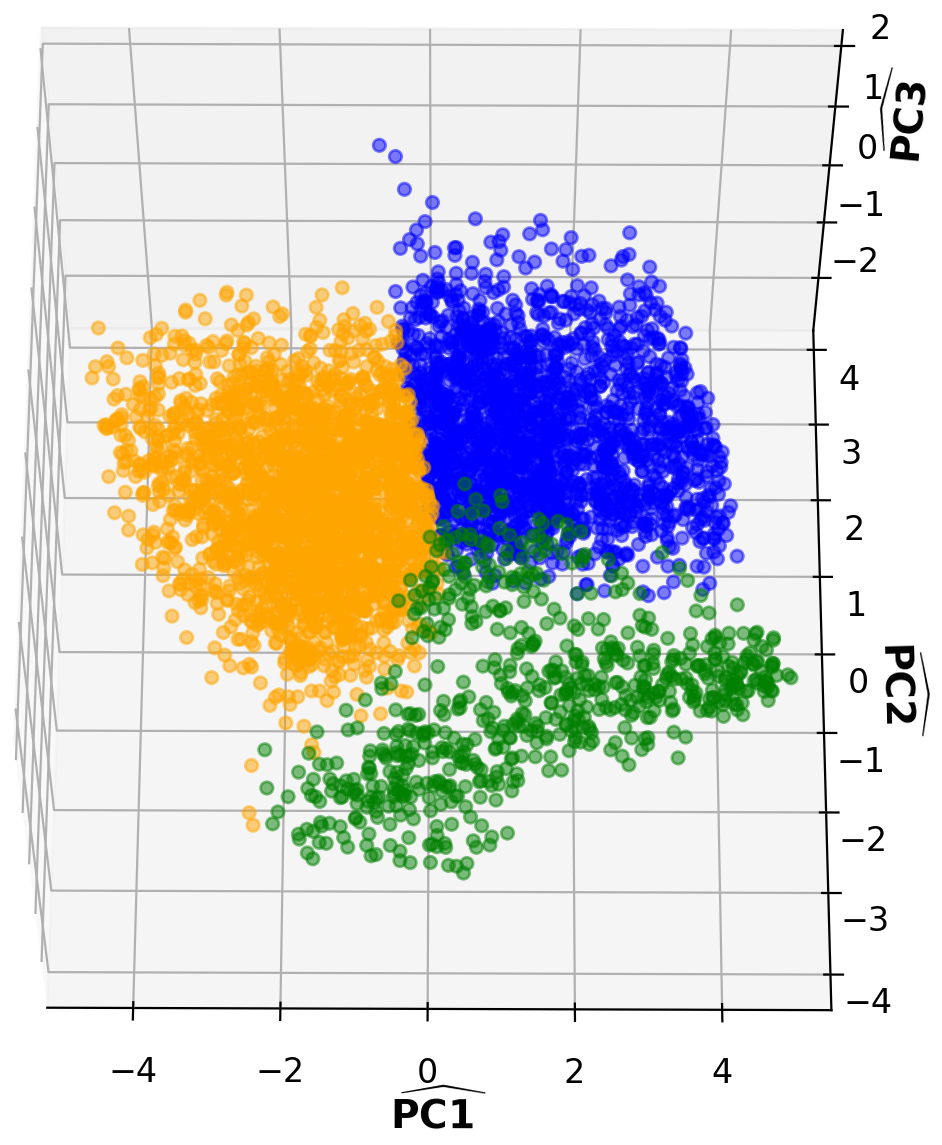}
\caption{K-means, K = 3}
   \label{fig:pca_dmk3}
\end{subfigure}
\hfill
\begin{subfigure}{0.35\textwidth}
\includegraphics[width=1.5in]{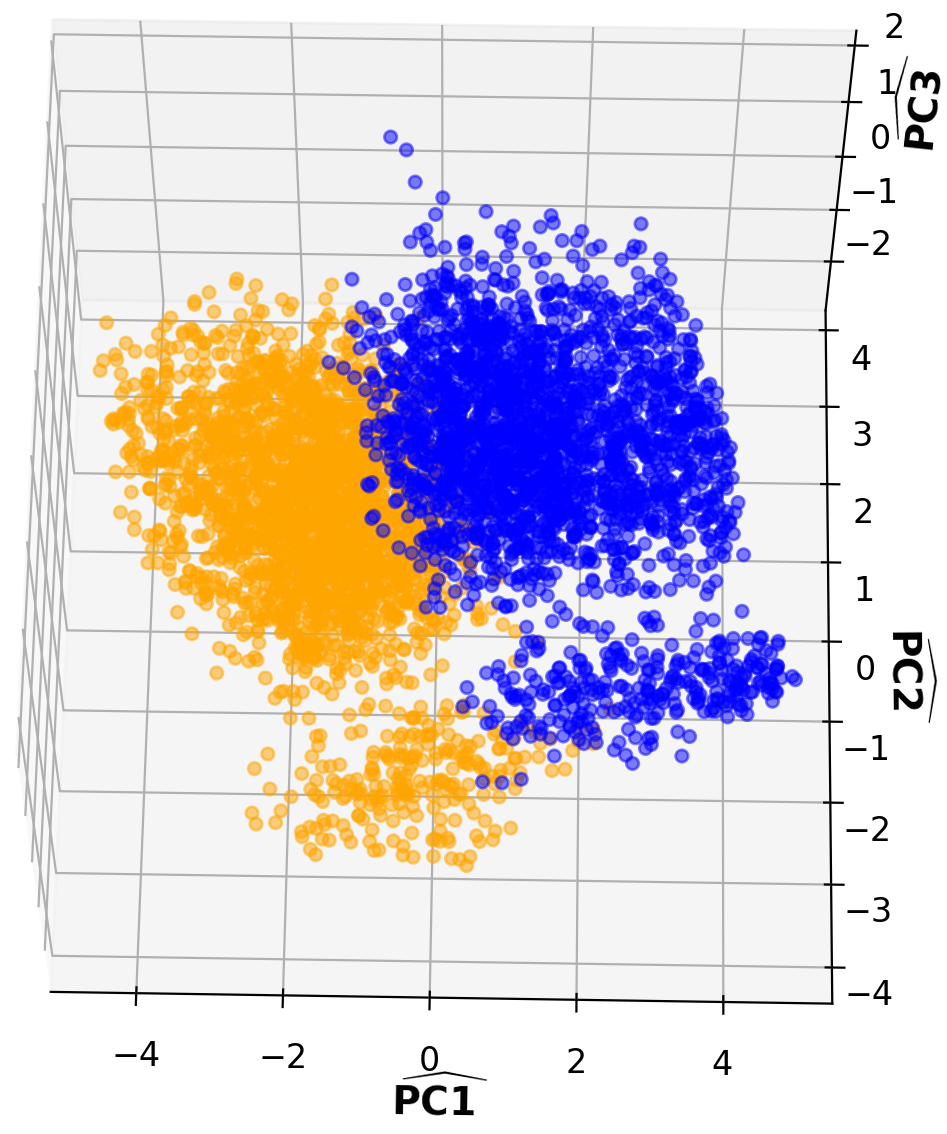}
\caption{Converged iteration, K = 3}
   \label{fig:pca_ck3}
\end{subfigure}
\hfill
}
\resizebox{\columnwidth}{!}
{
\begin{subfigure}{0.35\textwidth}
\includegraphics[width=1.5in]{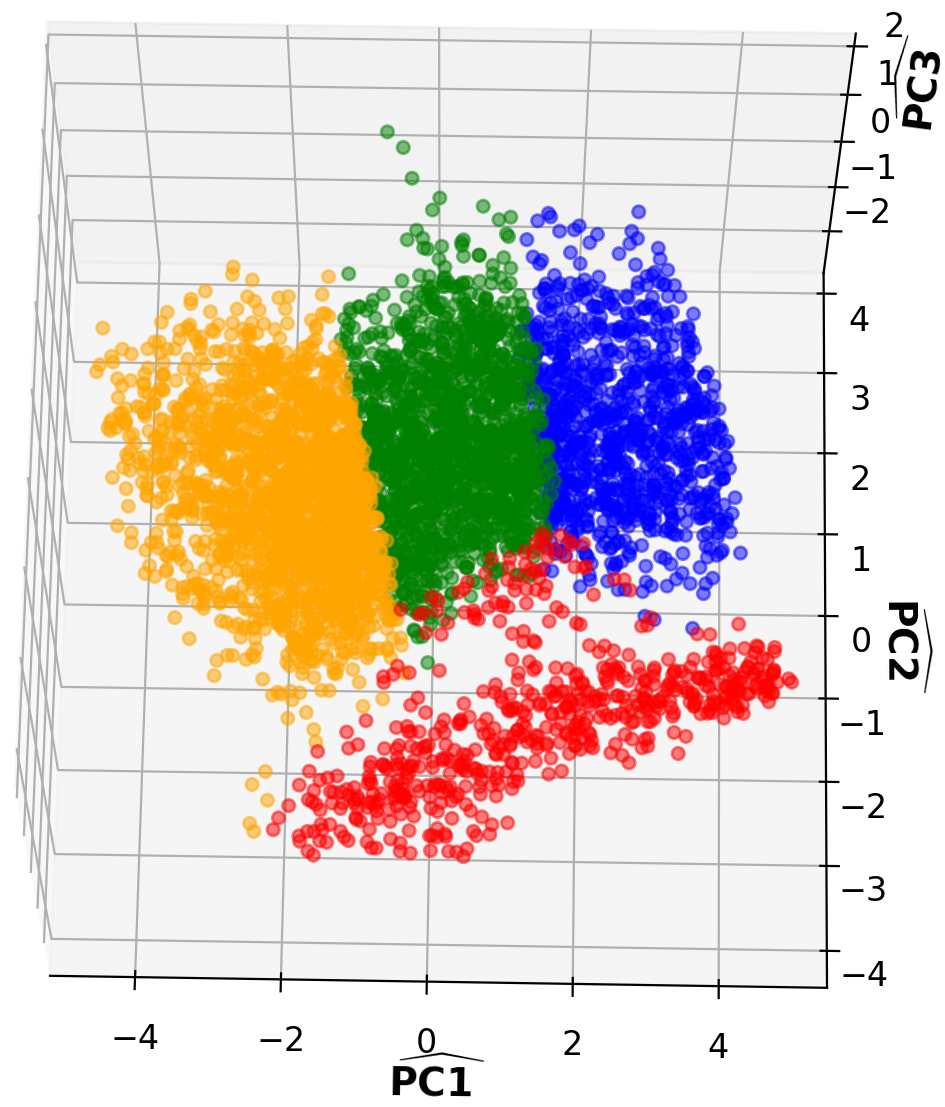}
\caption{K-means, K = 4}
   \label{fig:pca_dmk4}
\end{subfigure}
\hfill
\begin{subfigure}{0.35\textwidth}
\includegraphics[width=1.5in]{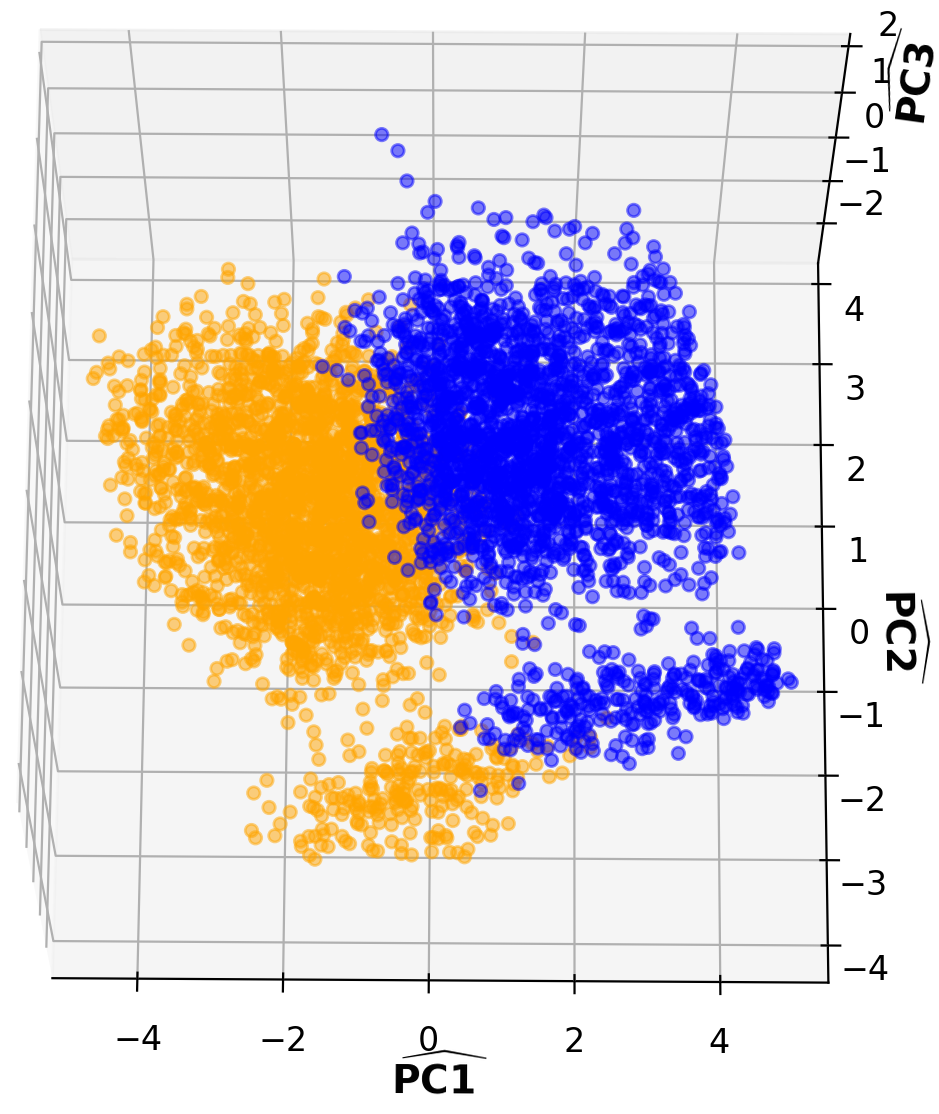}
\caption{Converged iteration, K = 4}
   \label{fig:pca_ck4}
\end{subfigure}
\hfill
}
\caption{ Comparision for different K values for initial K-means and after converged iteration of 5000 particles in PC-space at $T^*=0.2$}  
\label{fig:snapshots_si}
\end{figure}

\begin{figure}[h!] 
\resizebox{\columnwidth}{!}
{
\begin{subfigure}{0.35\textwidth}
\includegraphics[width=1.5in]{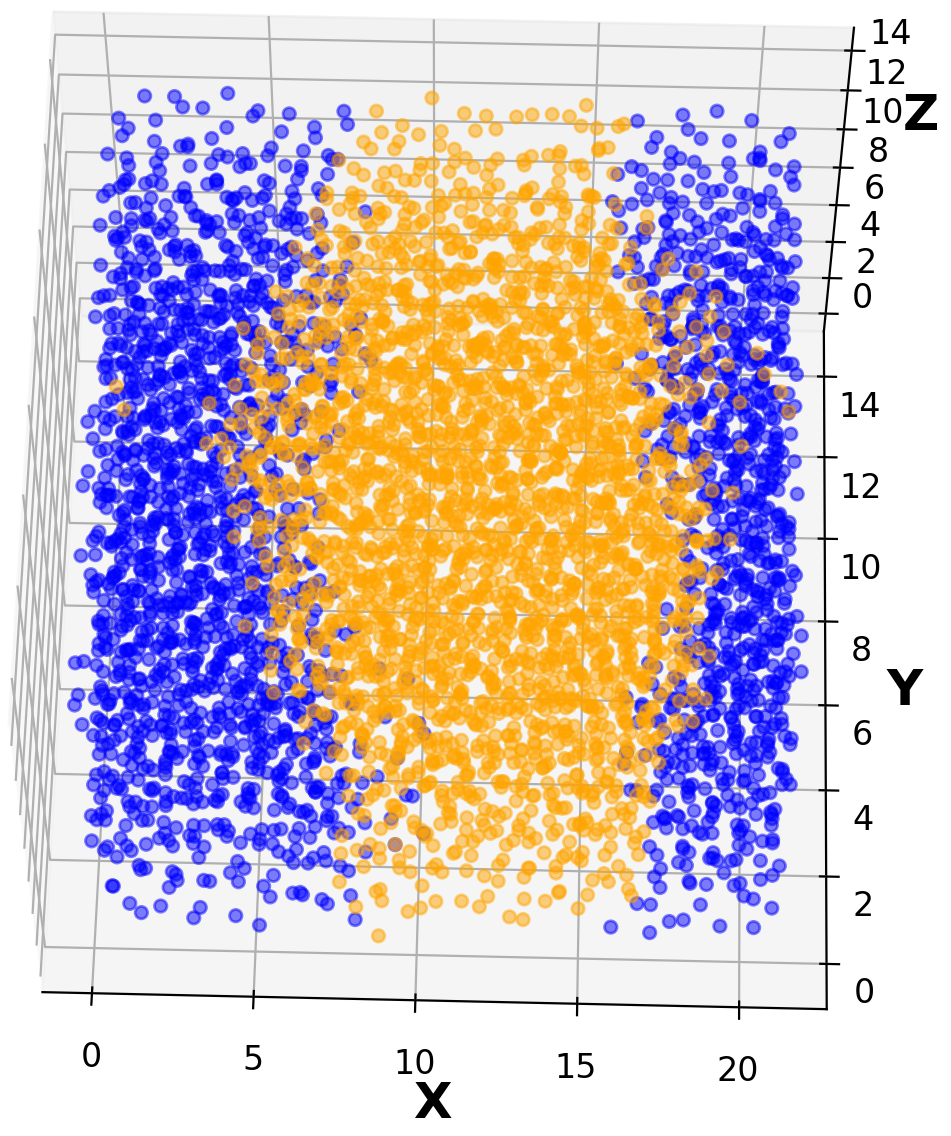}
\caption{K-means, K = 2}
   \label{fig:xyz_dmk2}
\end{subfigure}
\hfill
\begin{subfigure}{0.35\textwidth}
\includegraphics[width=1.5in]{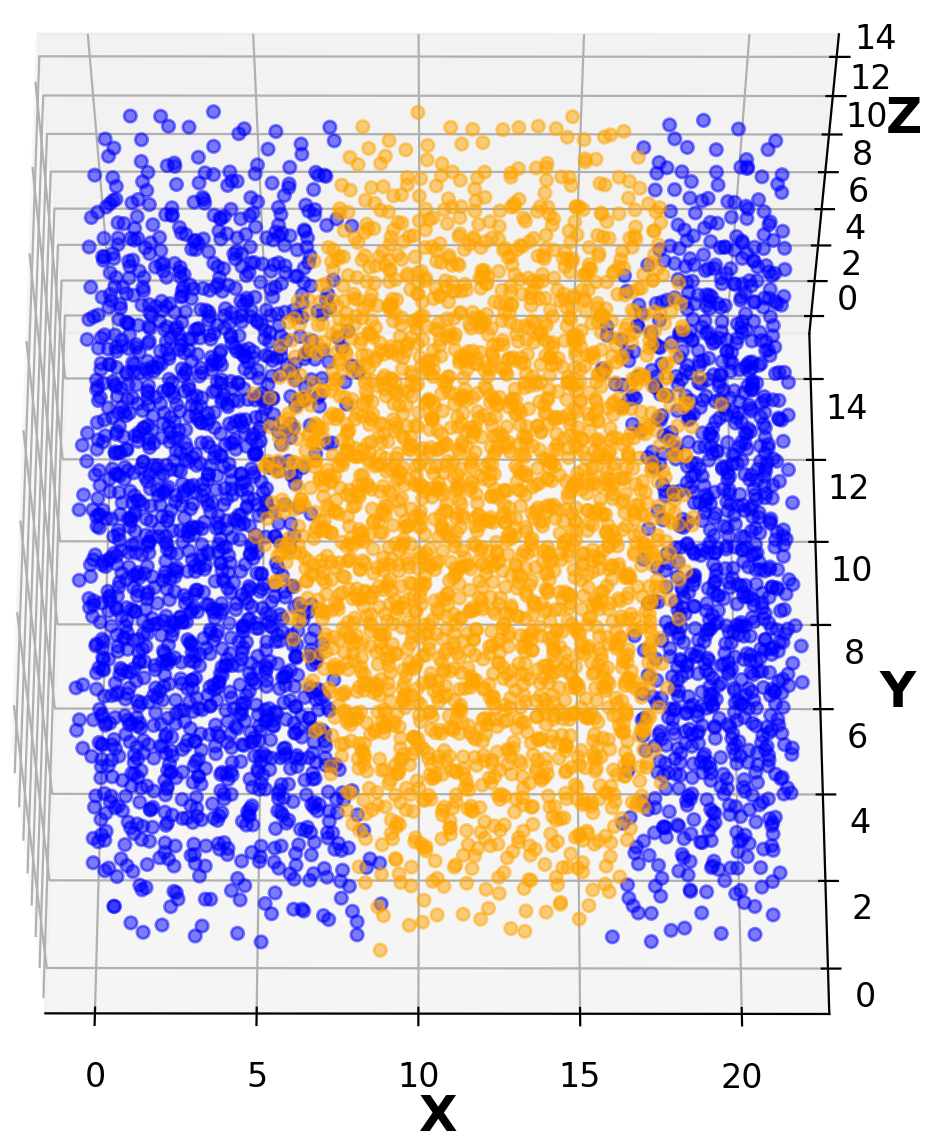}
\caption{Converged iteration, K = 2}
   \label{fig:xyz_ck2}
\end{subfigure}
\hfill
}
\resizebox{\columnwidth}{!}
{
\begin{subfigure}{0.35\textwidth}
\includegraphics[width=1.5in]{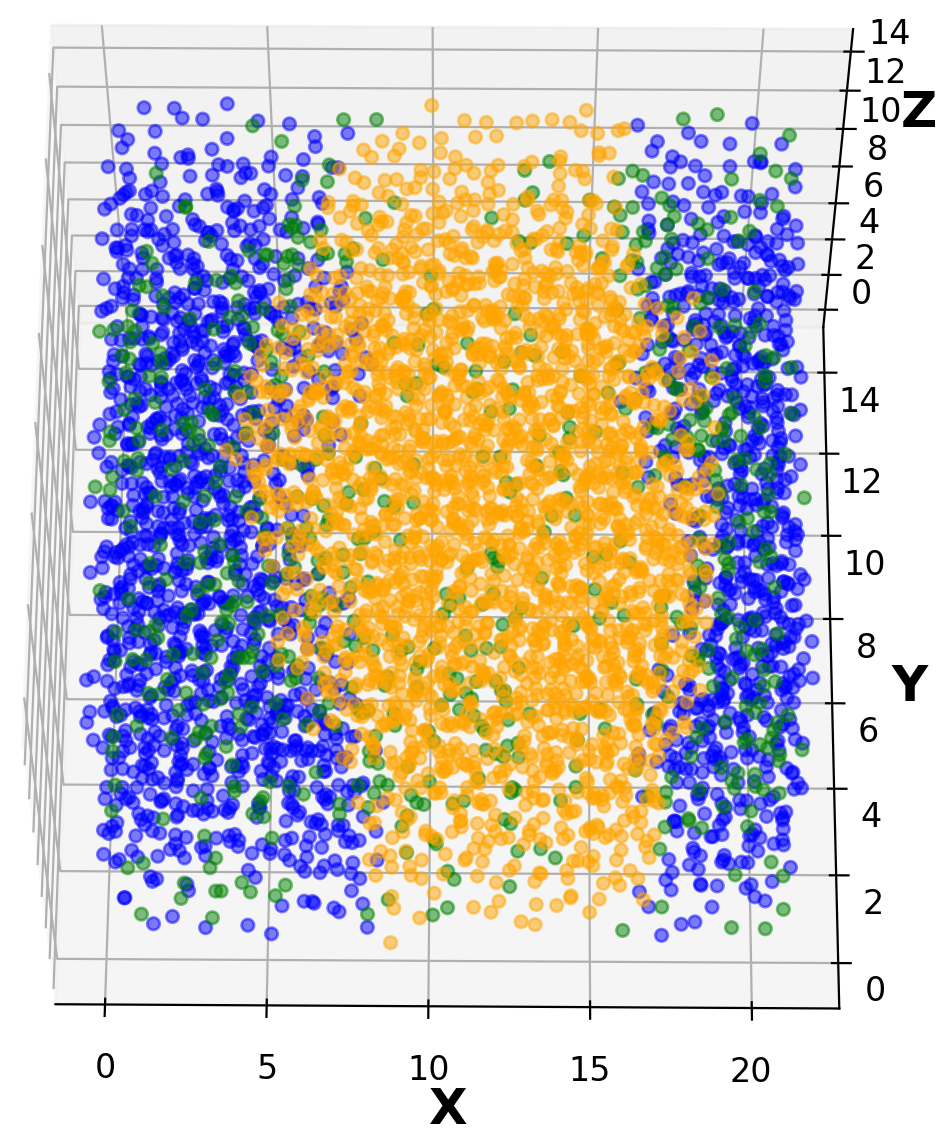}
\caption{K-means, K = 3}
   \label{fig:xyz_dmk3}
\end{subfigure}
\hfill
\begin{subfigure}{0.35\textwidth}
\includegraphics[width=1.5in]{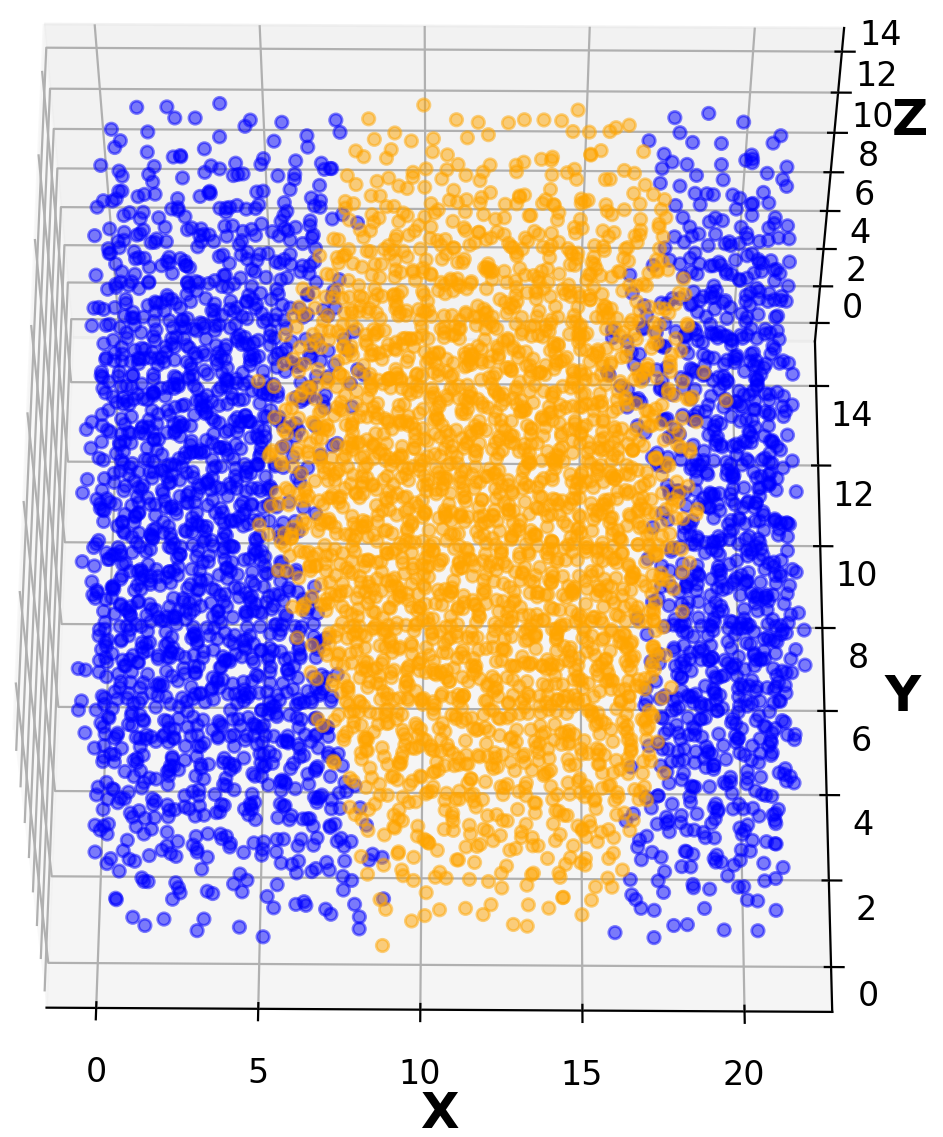}
\caption{Converged iteration, K = 3}
   \label{fig:xyz_ck3}
\end{subfigure}
\hfill
}
\resizebox{\columnwidth}{!}
{
\begin{subfigure}{0.35\textwidth}
\includegraphics[width=1.5in]{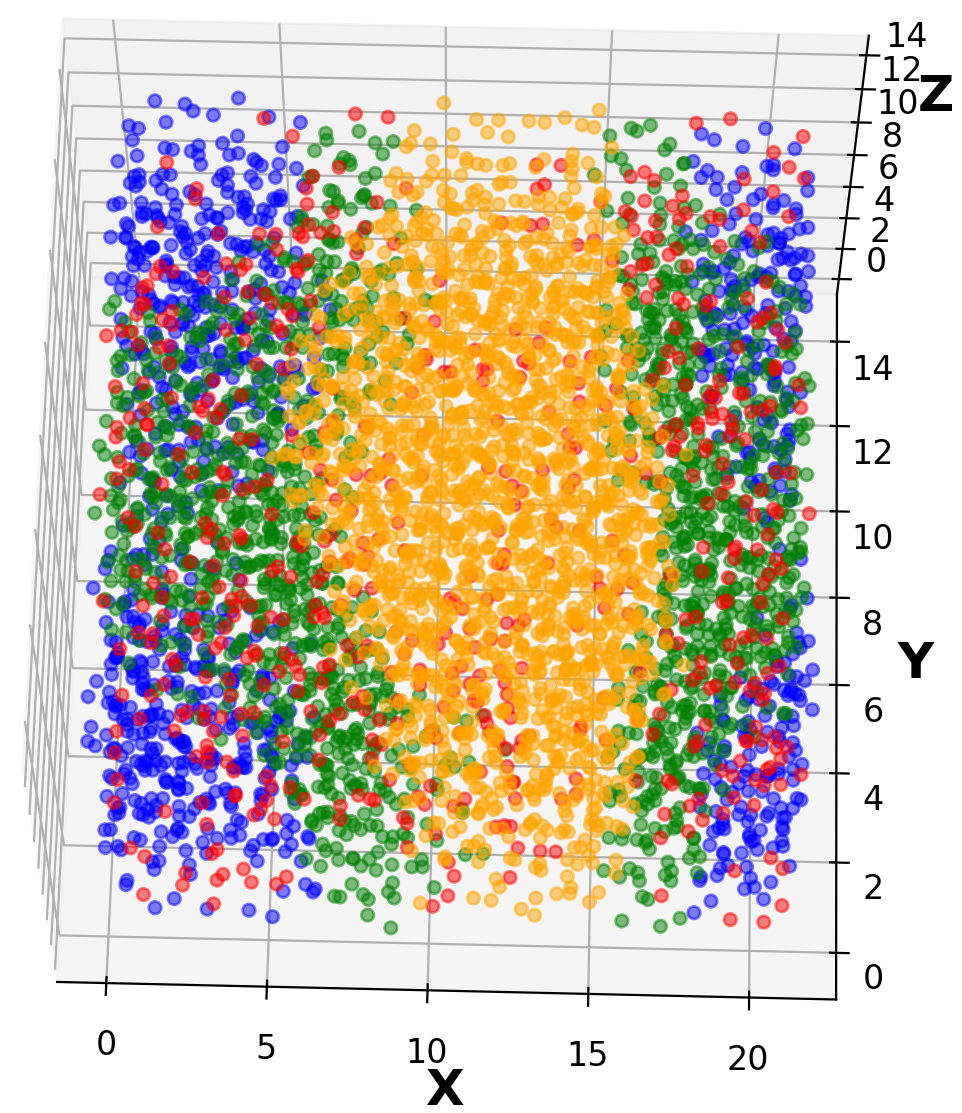}
\caption{K-means, K = 4}
   \label{fig:xyz_dmk4}
\end{subfigure}
\hfill
\begin{subfigure}{0.35\textwidth}
\includegraphics[width=1.5in]{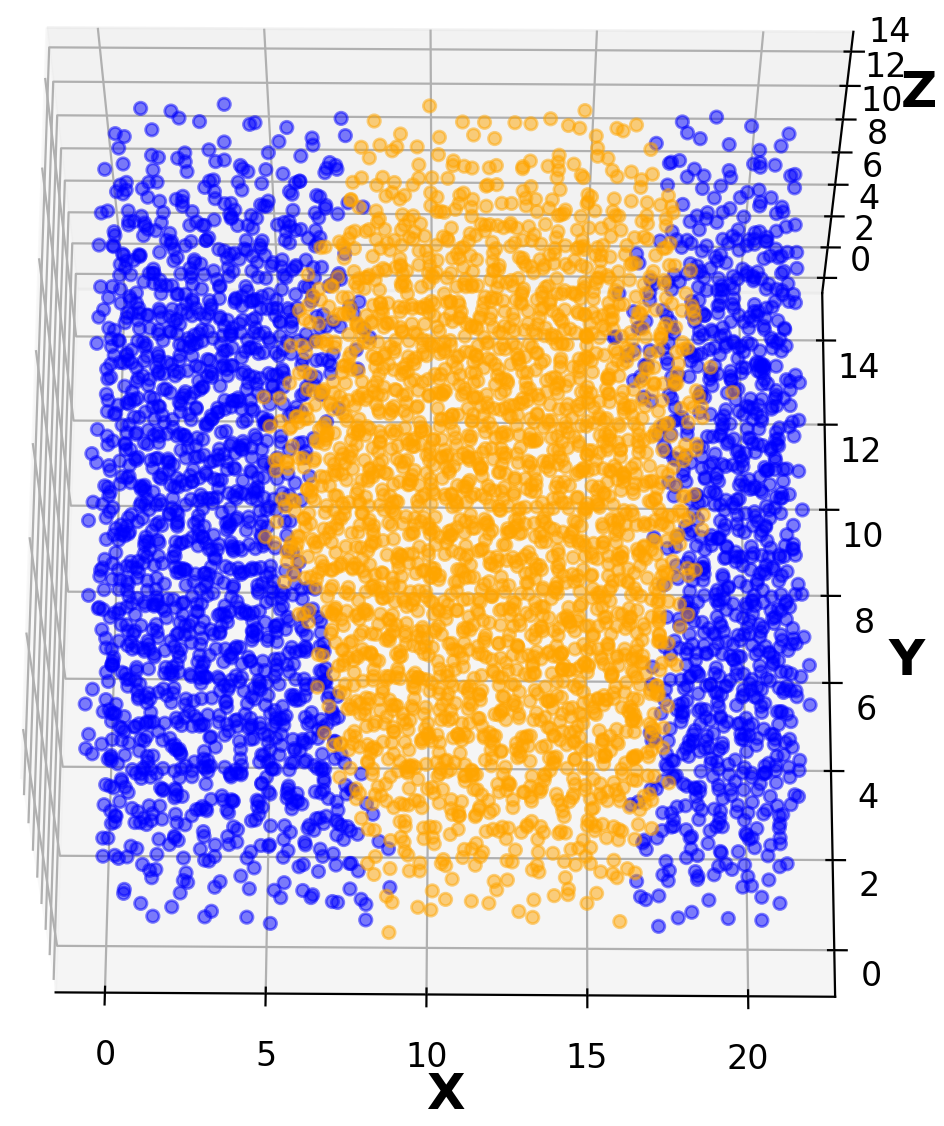}
\caption{Converged iteration, K = 4}
   \label{fig:xyz_ck4}
\end{subfigure}
\hfill
}
\caption{ Comparision for different K values for initial K-means and after converged iteration of 5000 particles in real space at $T^*=0.2$}  
\label{fig:si_kmeans}
\end{figure}

In the main text, the Elbow test can not  determine an effective K value for initial K-means clustering but a possible range of K clusters. For a selection of K = 2 as described in the main text, we first constructed K-means clustering models with various numbers of K (from 2 to 4) in the PC-space as shown in \ref{fig:pca_dmk2},\ref{fig:pca_dmk3},\ref{fig:pca_dmk4} and in the  configurational space by direct mapping in \ref{fig:xyz_dmk2},\ref{fig:xyz_dmk3},\ref{fig:xyz_dmk4}. For all models of K-means clustering in the previous step, we then performed the co-learning strategy for each K = 2,3,4 to sort out the one K that all models of K-means converge in both the PC and real space.  Final results presented in both PC (\ref{fig:pca_ck2},\ref{fig:pca_ck3},\ref{fig:pca_ck4} ) and configurational space (\ref{fig:xyz_ck2},\ref{fig:xyz_ck3},\ref{fig:xyz_ck4}) show convergence for K = 2 for all K-means models, thus support our K=2 choice. Physically for an one-component system the the Gibbs phase rule will lead to K=2 as well.  

\section{Angle Projection to visualize the meso-state structure} \label{sec:angles}
\renewcommand{\thefigure}{B\arabic{figure}}
\setcounter{figure}{0}

 In addition to the figures in the main text, different angle projections of \ref{fig:init_PCA} and \ref{fig:init_xyz} are presented here to provide different view for the domain structure.  

\begin{figure}
\resizebox{\columnwidth}{!}
{
\begin{subfigure}{0.35\textwidth}
\includegraphics[width=2.1in]{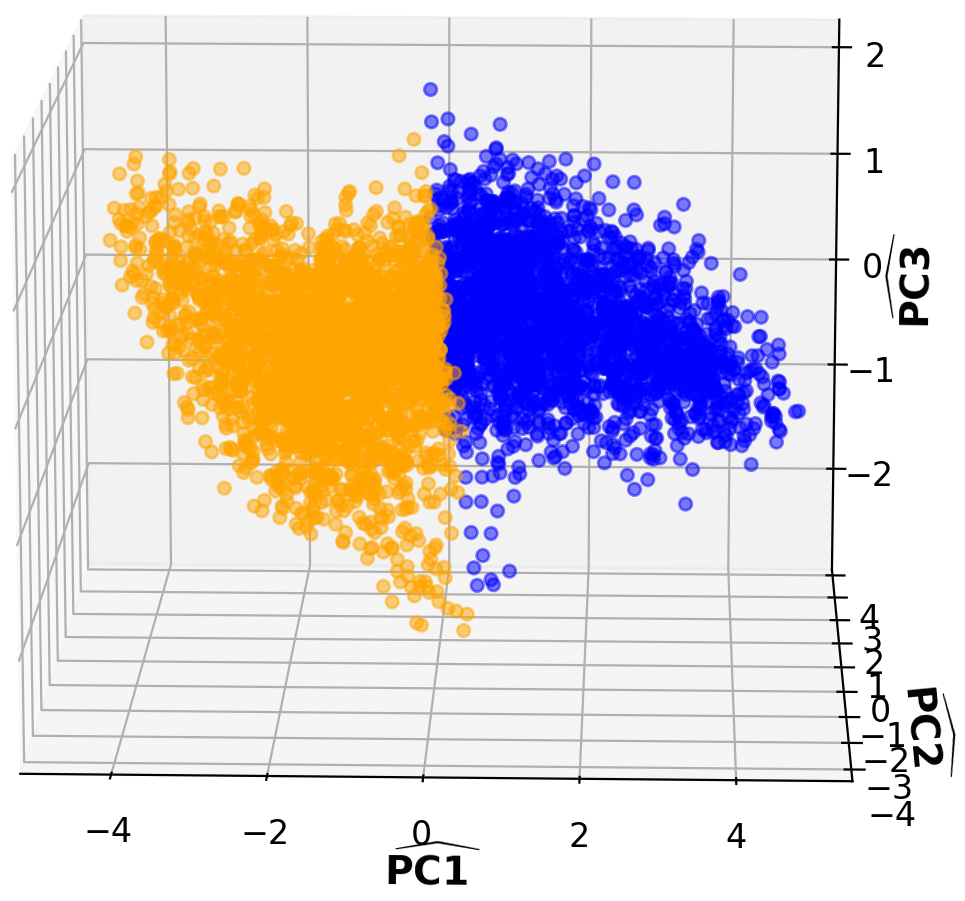}
\caption{}
   \label{}
\end{subfigure}
\hfill
\begin{subfigure}{0.35\textwidth}
\includegraphics[width=2.1in]{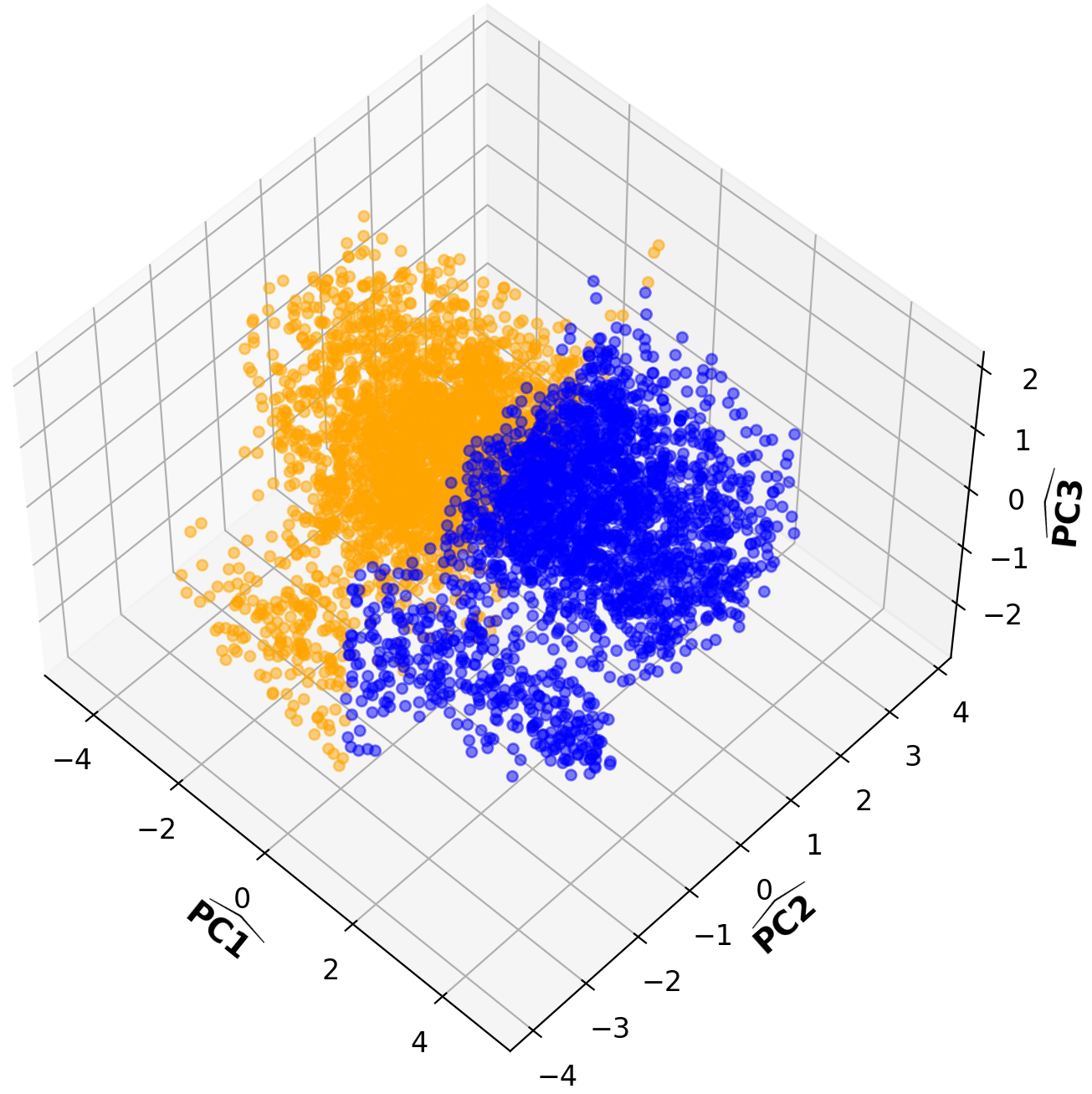}
\caption{}
   \label{}
\end{subfigure}
\hfill
\begin{subfigure}{0.35\textwidth}
\includegraphics[width=2.1in]{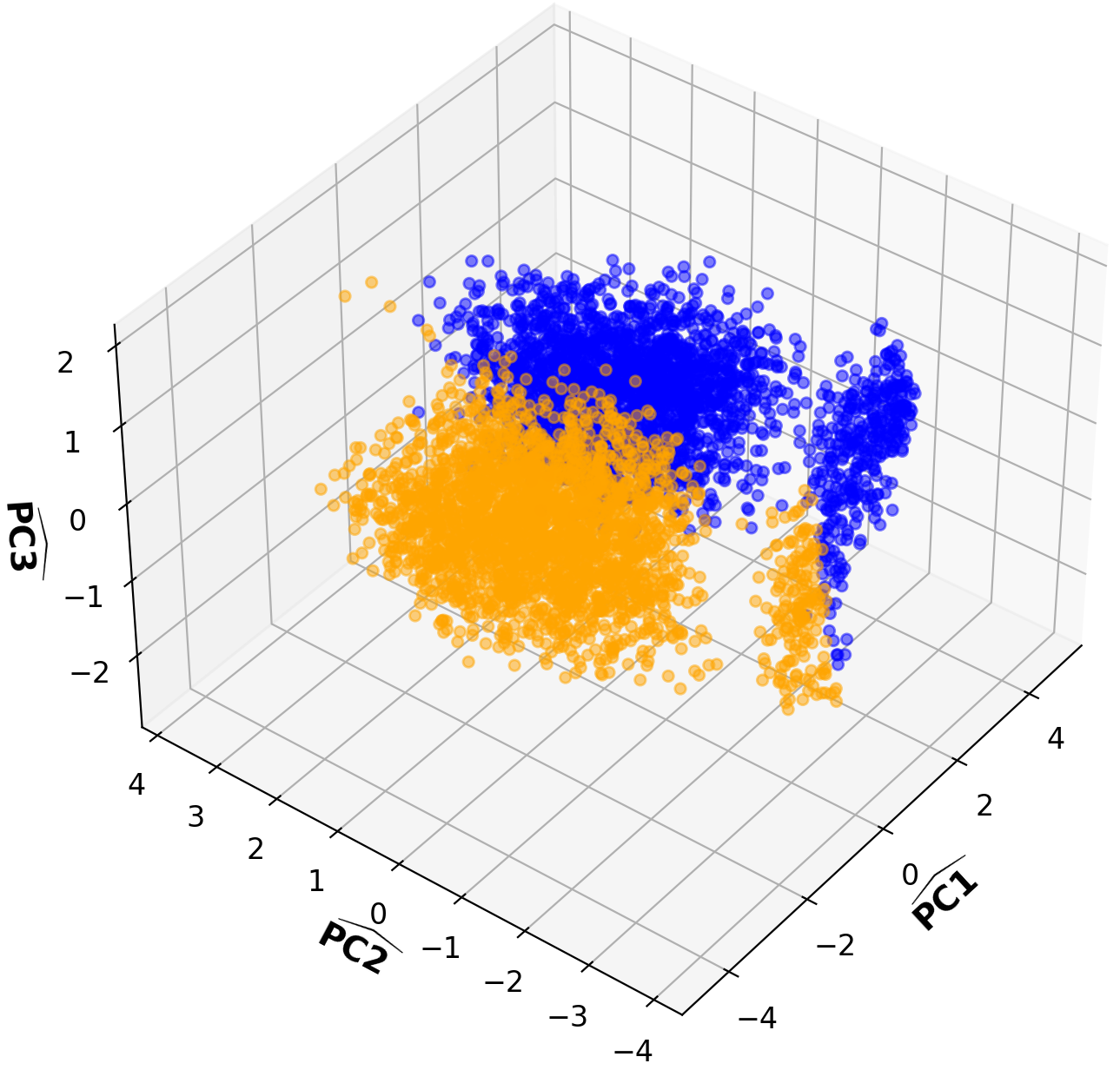}
\caption{}
   \label{}
\end{subfigure}
\hfill
}
\resizebox{\columnwidth}{!}
{
\begin{subfigure}{0.35\textwidth}
\includegraphics[width=2.1in]{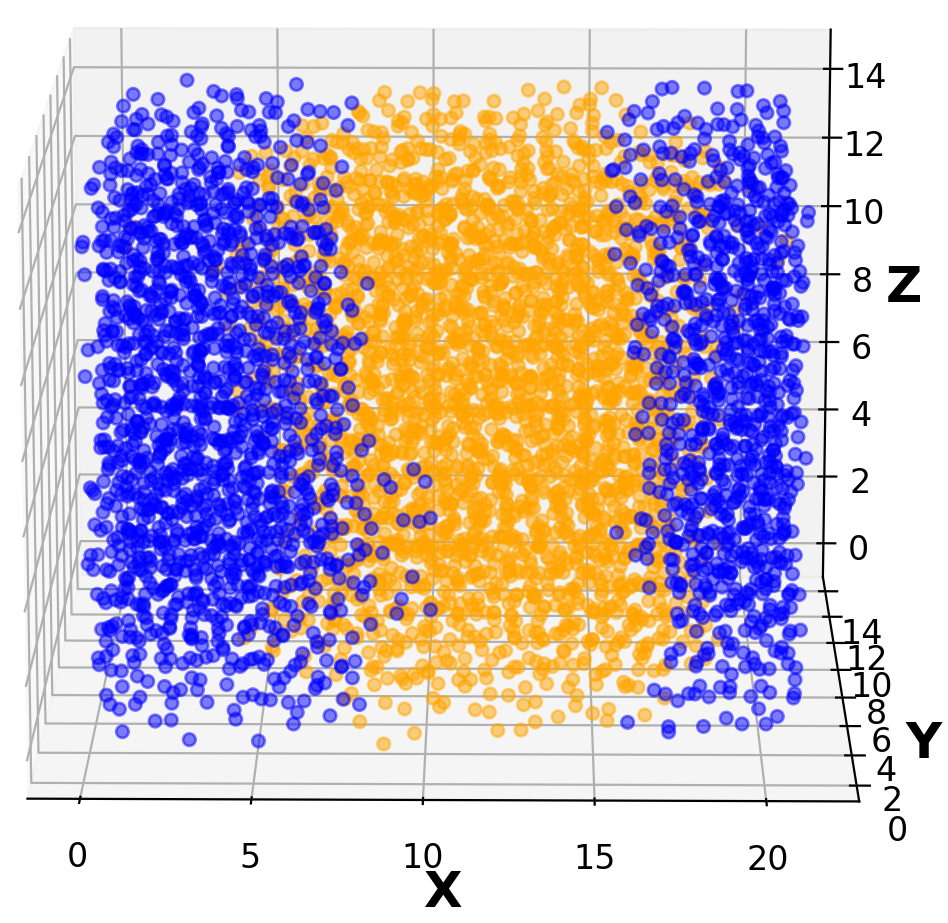}
\caption{}
   \label{}
\end{subfigure}
\hfill
\begin{subfigure}{0.35\textwidth}
\includegraphics[width=2.1in]{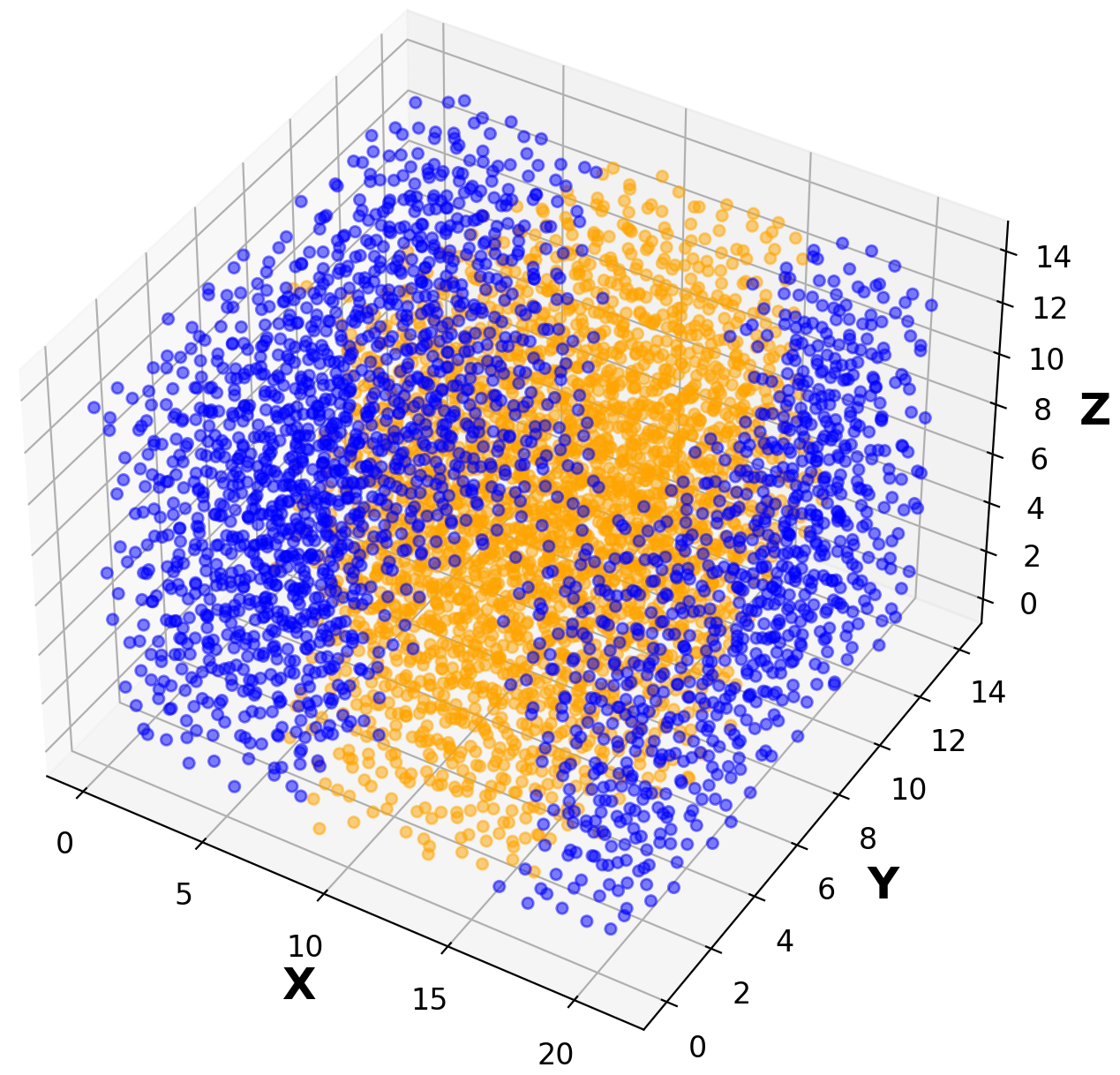}
\caption{}
   \label{}
\end{subfigure}
\hfill
\begin{subfigure}{0.35\textwidth}
\includegraphics[width=2.1in]{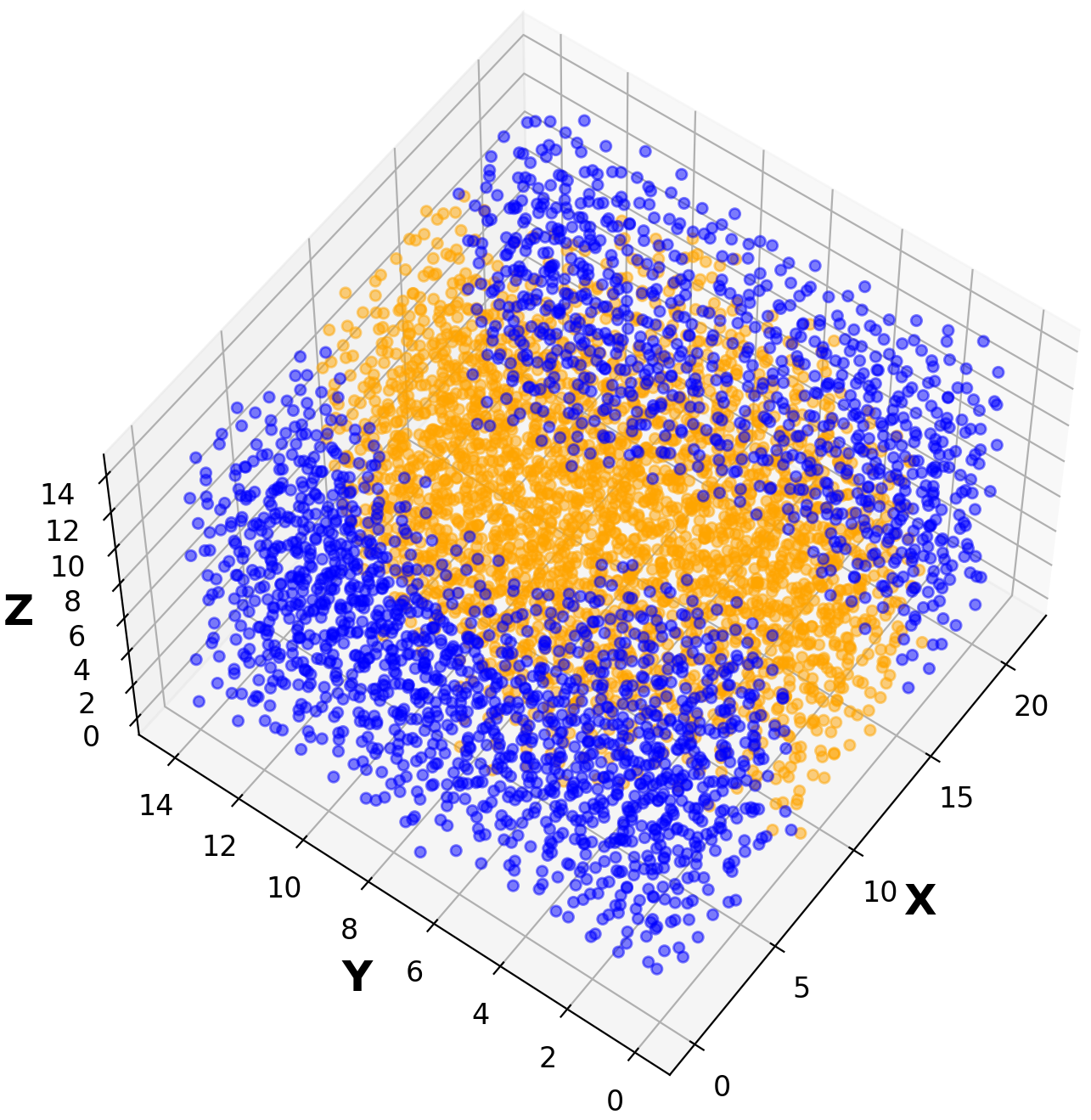}
\caption{}
   \label{}
\end{subfigure}
\hfill
}
\caption{ Different angles projection to display the local meso-states structure in PC and configurational spaces from the \ref{fig:init_PCA} and \ref{fig:init_xyz} in the main text. Panels a,b,c are for PC-space and d,e,f are for the configurational space }  
\label{fig:angle_projected}
\end{figure}

\section{Classification of A/B particles type} \label{sec:AB}
\renewcommand{\thefigure}{C\arabic{figure}}
\setcounter{figure}{0}

\begin{figure}
\resizebox{\columnwidth}{!}
{
\begin{subfigure}{0.35\textwidth}
\includegraphics[width=1.5in]{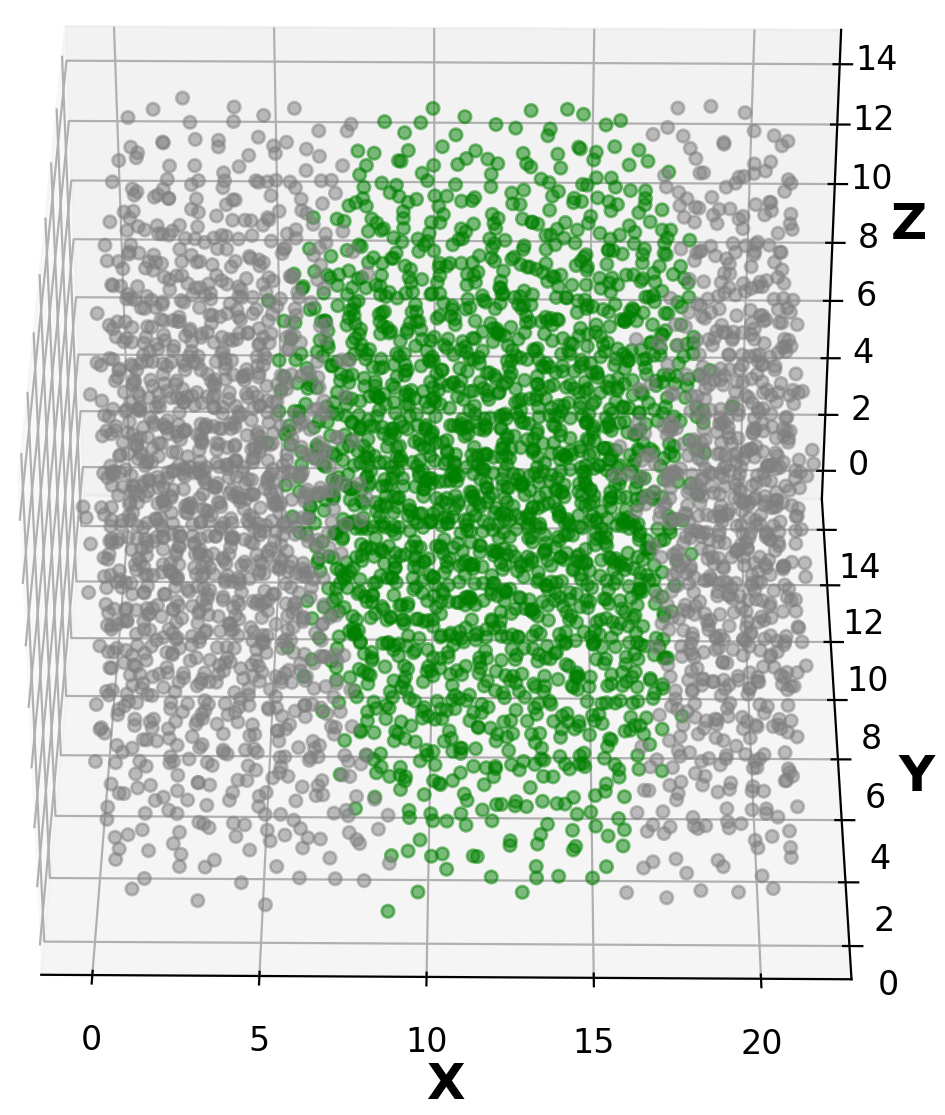}
\caption{ A particles in both meso-states}
   \label{fig:A_2states}
\end{subfigure}
\hfill
\begin{subfigure}{0.35\textwidth}
\includegraphics[width=1.5in]{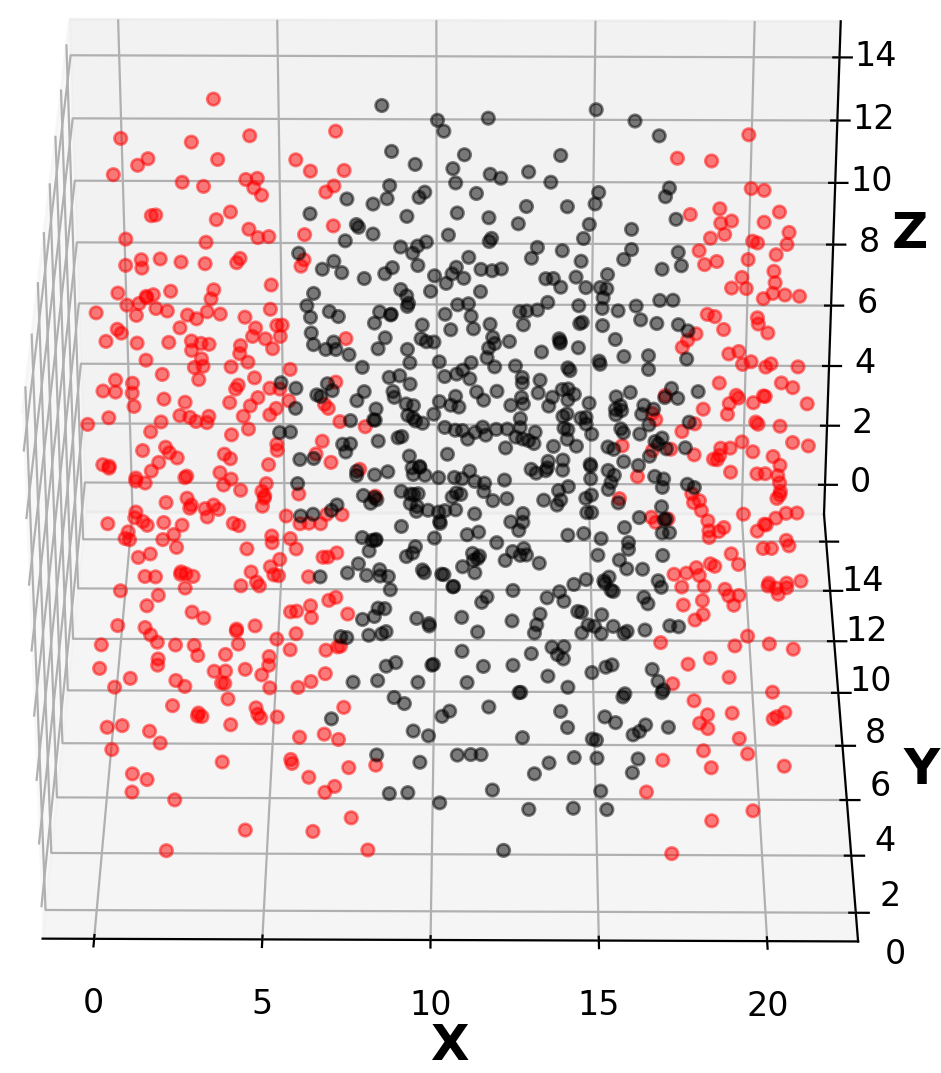}
\caption{B particles in both meso-states}
   \label{fig:B_2states}
\end{subfigure}
\hfill
}

\resizebox{\columnwidth}{!}
{
\begin{subfigure}{0.35\textwidth}
\includegraphics[width=1.5in]{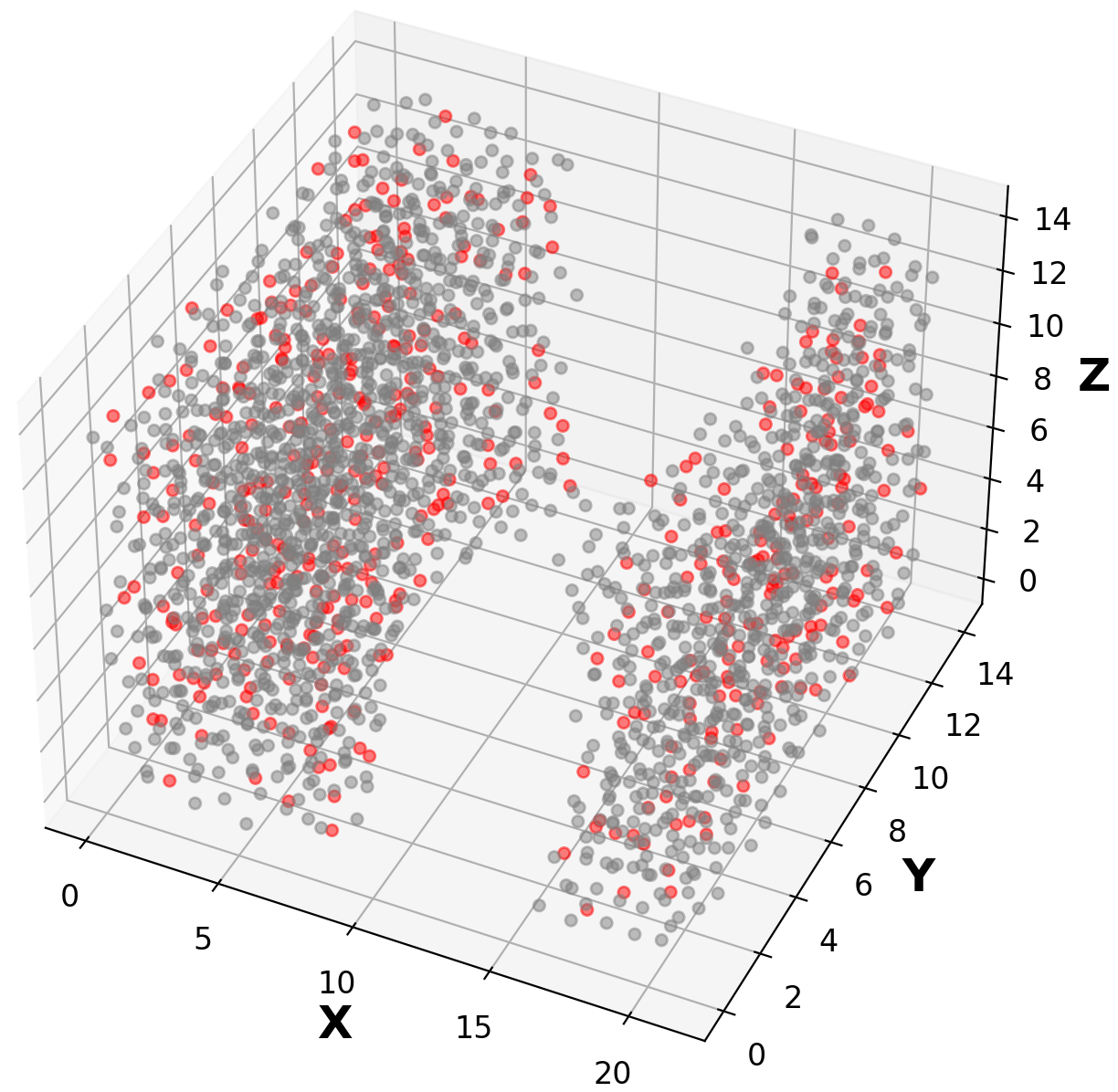}
\caption{A and B particles of meso-state 1 }
   \label{fig:AB_state0}
\end{subfigure}
\hfill
\begin{subfigure}{0.35\textwidth}
\includegraphics[width=1.5in]{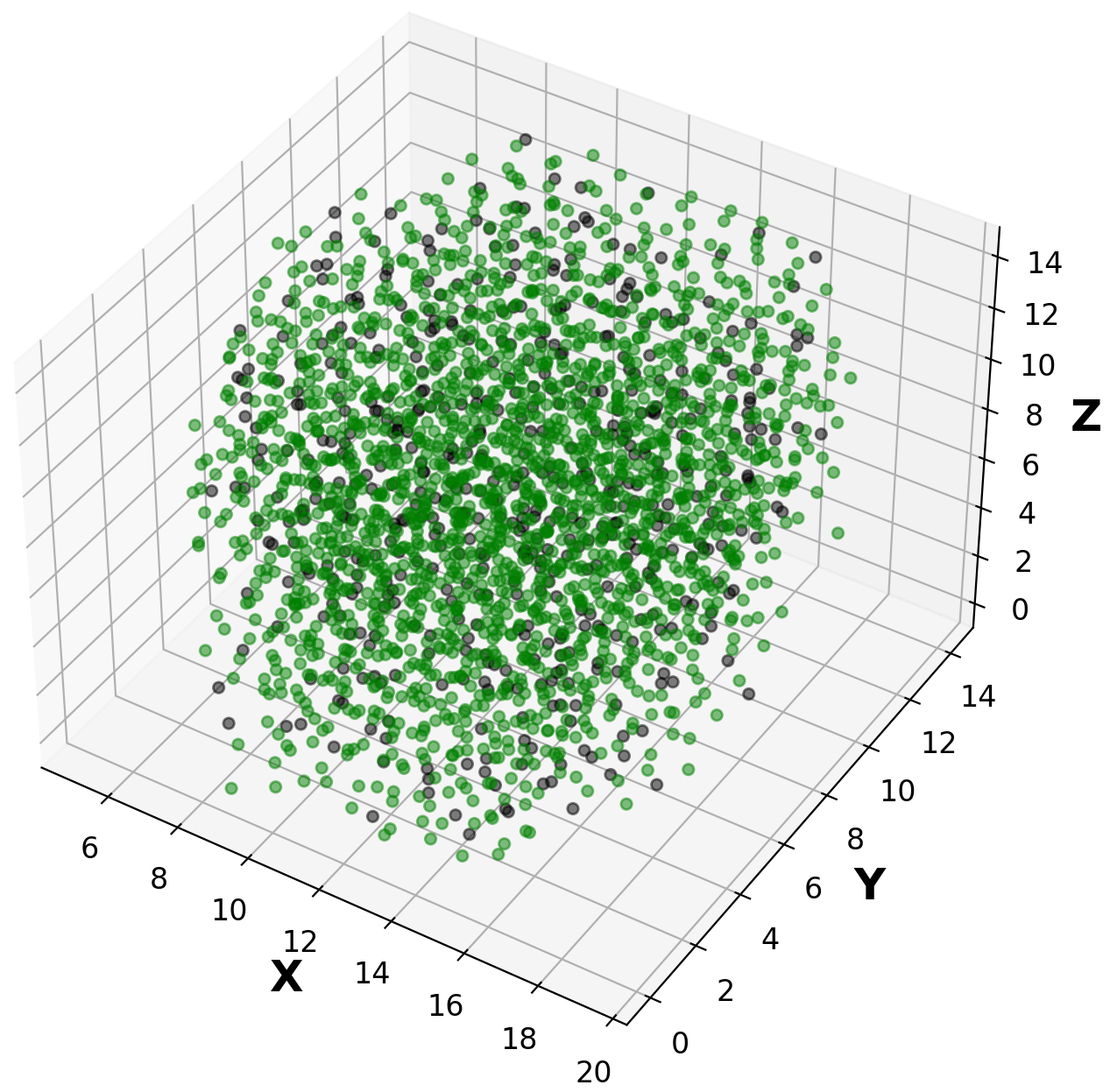}
\caption{A and B particles of meso-state 2}
   \label{fig:AB_state1}
\end{subfigure}
}
\caption{ A 3D projection of classified A/B particles in configurational space of 5000 particles system at $T^*=0.2$. Panels a and b are classified A or B particles in both meso-states while panel c and d are A and B particles of a particular meso-state.  Grey and red are A and B particles belonging to meso-state 1, green and black are A and B particles of meso-state 2. Meso-state 1 has 2052 A particles and 459 B particles while the meso-state 2 has 1948 A particles and 541 B particles, so the ratio of A/B particles in each meso-state is roughly 4.47 and 3.60, respectively, which is different from the bulk one.}  
\label{fig:AB_classified}
\end{figure}

In the main text, the identity of particles' type (A/B) is ignored when collecting $\overline{WCNs}$ for the classification of particles into meso-states. Indeed  each meso-state consists of a mixture of A and B particles shown in Figure \ref{fig:AB_classified}

\bibliographystyle{apsrev4-1}
\bibliography{nano_domain_2} 
\end{document}